\documentclass[12pt]{article}
\usepackage{epsfig}
\usepackage{amsmath}
\usepackage{hhline}
\usepackage{amssymb}
\usepackage{times}

\renewcommand{\arraystretch}{1.75}
\newlength{\dinwidth}
\newlength{\dinmargin}
\newcommand{\GeV}{\textrm{\,GeV}}
\newcommand{\gev}{\textrm{\,GeV}}
\newcommand{\mev}{\textrm{\,MeV}}

\newcommand{\pbinv}{\textrm{\,pb}^{-1}}
\newcommand{\cm}{\textrm{\,cm}}

\newcommand{\ksf}{\ensuremath{K^0_s}}
\newcommand{\lsfa}{\ensuremath{\Lambda(\bar{\Lambda})}}
\newcommand{\lsf}{\ensuremath{\Lambda}}
\newcommand{\lsa}{\ensuremath{\bar{\Lambda}}}
\newcommand{\lambdas}{\ensuremath{\lambda_s}}

\newcommand{\GeVSq}{{\rm\,GeV^2}}
\newcommand{\qsq}{\ensuremath{Q^2} }

\newcommand{\bars}{\ensuremath{\bar{s}}}
\newcommand{\baru}{\ensuremath{\bar{u}}}
\newcommand{\bard}{\ensuremath{\bar{d}}}

\def\EJC{{\em Eur. Phys. J.} {\bf C}}

\begin{document}

\begin{titlepage}

\begin{flushleft}

DESY 08-095  \hfill    ISSN 0418-9833\\
{\tt December 2008}
\end{flushleft}

\vspace{2cm}
\noindent

\begin{center}
\begin{Large}

{\bf Strangeness Production at low ${\bf Q^2}$ \\
in Deep-Inelastic \boldmath{${ep}$} Scattering at HERA }

\vspace{2cm}

H1 Collaboration

\end{Large}
\end{center}

\vspace{2cm}

\begin{abstract}

\noindent
The production of  neutral strange hadrons
is investigated using deep-inelastic 
scattering events measured with the H1 detector at HERA.
The measurements are made in the phase space defined 
by the negative 
four-momentum transfer squared of the photon 
$2 < Q^2 < 100$ $\GeVSq$ and the inelasticity $0.1 < y < 0.6$.
The \ksf\ and \lsfa\ production cross sections and
their ratios are  determined.  
 \ksf\ production is  compared 
to the production  of charged particles in the same region of phase space. 
The $\Lambda - \bar{\Lambda}$ asymmetry is 
also measured and
found to be consistent with zero. 
Predictions of leading order Monte Carlo 
programs are compared to the data.

\end{abstract}

\vspace{1.5cm}

\begin{center}
Accepted by \EJC
\end{center}

\end{titlepage}

\begin{flushleft}

F.D.~Aaron$^{5,49}$,           
C.~Alexa$^{5}$,                
V.~Andreev$^{25}$,             
B.~Antunovic$^{11}$,           
S.~Aplin$^{11}$,               
A.~Asmone$^{33}$,              
A.~Astvatsatourov$^{4}$,       
A.~Bacchetta$^{11}$,           
S.~Backovic$^{30}$,            
A.~Baghdasaryan$^{38}$
% ...

\bigskip{\it
% ...
$ ^{4}$ Inter-University Institute for High Energies ULB-VUB, Brussels;
          Universiteit Antwerpen, Antwerpen; Belgium$^{ c}$ \\
 $ ^{5}$ National Institute for Physics and Nuclear Engineering (NIPNE) ,
          Bucharest, Romania \\
% ...
$ ^{11}$ DESY, Hamburg, Germany \\
% ...
$ ^{25}$ Lebedev Physical Institute, Moscow, Russia$^{ e}$ \\
% ...
$ ^{30}$ Faculty of Science, University of Montenegro,
          Podgorica, Montenegro$^{ e}$ \\
% ...
$ ^{33}$ Dipartimento di Fisica Universit\`a di Roma Tre
          and INFN Roma~3, Roma, Italy \\
% ...
$ ^{38}$ Yerevan Physics Institute, Yerevan, Armenia \\
% ...
$ ^{49}$ Also at Faculty of Physics, University of Bucharest,
          Bucharest, Romania \\
% ...
$ ^c$ Supported by FNRS-FWO-Vlaanderen, IISN-IIKW and IWT
      and  by Interuniversity
Attraction Poles Programme,
      Belgian Science Policy \\
% ...
$ ^e$ Supported by the Deutsche Forschungsgemeinschaft \\
% ...
}
\end{flushleft}
%
%

\newpage

\section{Introduction}
\label{sec:Intro}


The production of strange hadrons in high energy particle collisions
allows the investigation of strong interactions in the perturbative and
non-perturbative regimes.  
Strange quarks  are 
created in the non-perturbative process of colour string
fragmentation, which constitutes
the dominant production mechanism of strange hadrons.
In deep-inelastic scattering (DIS), strange quarks also originate from the strange sea in the 
nucleon, boson-gluon fusion and heavy quark decays.
Measurements of strangeness production have been used
to investigate the suppression of strangeness relative to lighter flavours 
in fragmentation.
The universality of fragmentation in different processes can be studied by  comparing
differential cross sections of the production of \ksf\ and \lsfa\ 
hadrons in various regions of phase space.
Further information is gained by studying the ratios of production rates of \lsfa\ to \ksf\ and
of \ksf\ to charged hadrons ($h^{\pm}$) as some model
dependencies are expected to cancel.

\par
The baryon production mechanism was studied in 
$e^+e^-$ annihilation~\cite{TASSO,OPAL,SLD,ALEPH-TUNE,DELPHI-TUNE},
 a process without incident baryons. Data involving a baryon in the initial state, like $ep$ collisions
 at HERA,
 provide additional information.
In particular, data on the $\Lambda$ - \lsa ~production asymmetry from  HERA 
are of  interest as an 
experimental constraint for theories  of  baryon number transfer \cite{TRANSPORT}.
Fixed target data have shown~\cite{EMC} that the \lsf\ production rate substantially
exceeds that of the \lsa\ in the so-called remnant region because the baryon number 
of the target is conserved.

This paper presents a measurement of neutral  strange particle (\ksf and \lsf ) 
production in DIS at negative 
four momentum transfer squared
$ 2 < \qsq < 100 \GeVSq$ and at low values of Bjorken $x$. 
The study is based
on data collected with the H1 detector at HERA
at a centre-of-mass energy of $319 \gev$ in the years $1999$ and $2000$.
This data sample is 40 times larger than that used in the previous H1
publication \cite{H1-K0} and covers a wider kinematic range.
Measurements of \ksf\ and \lsf\ production in different kinematic ranges
have also been reported by the ZEUS collaboration~\cite{ZEUS-K0-2}.
The differential cross sections of \ksf\ mesons, \lsfa\ baryons and their
ratio as well as  the ratio of \ksf\ to charged hadrons
are presented  as a function of various kinematic variables,
both in the laboratory frame and in the Breit frame. 
The results are compared with predictions 
obtained from leading order Monte Carlo calculations,
based on matrix elements, with parton shower simulations.
%
The main feature of the data is a
suppression of strange quark production relative to lighter quarks;
this is discussed within the context of
the framework of the LUND~\cite{LUND} fragmentation model.


\section{Phenomenology}
\subsection{Production of Strange Hadrons }

\begin{figure}[!hbt]
\centering
\includegraphics[width=55mm]{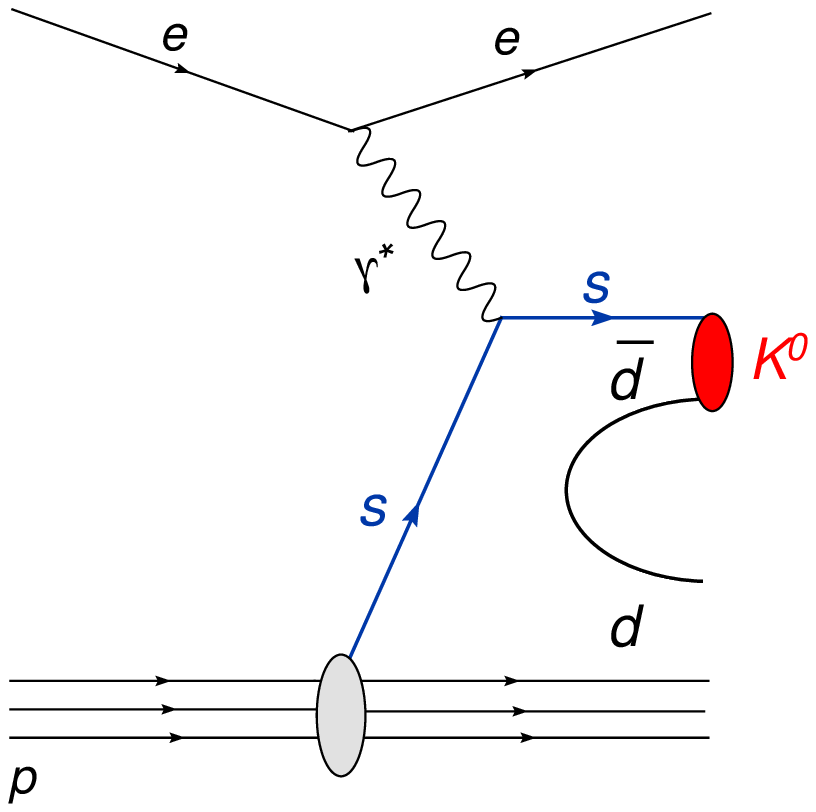} 
\hspace{1.5cm}
\includegraphics[width=55mm]{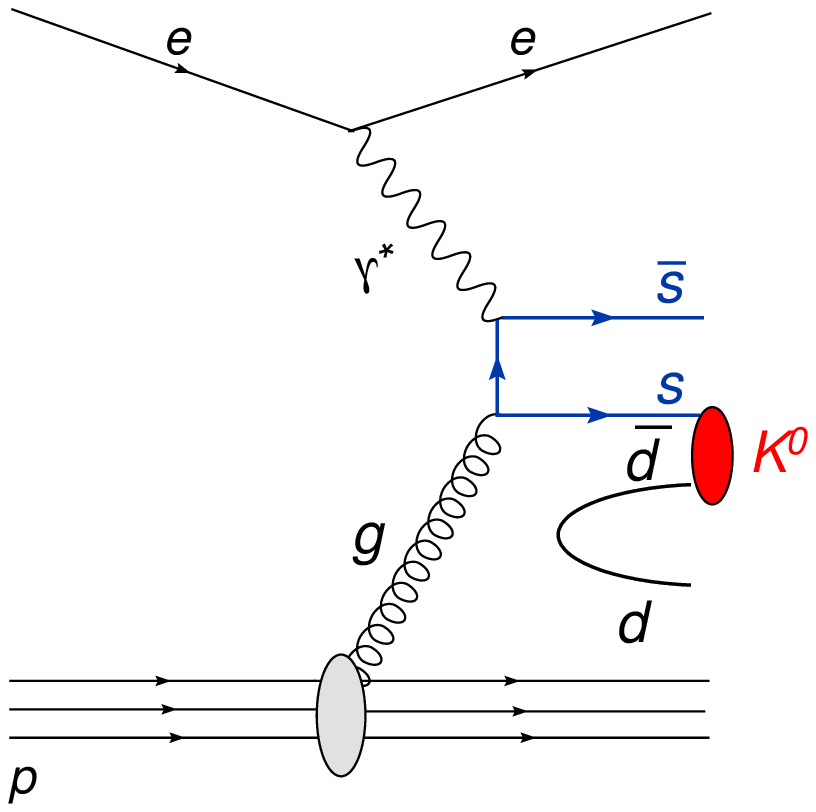}
\\
\vspace{1cm}
\includegraphics[width=55mm]{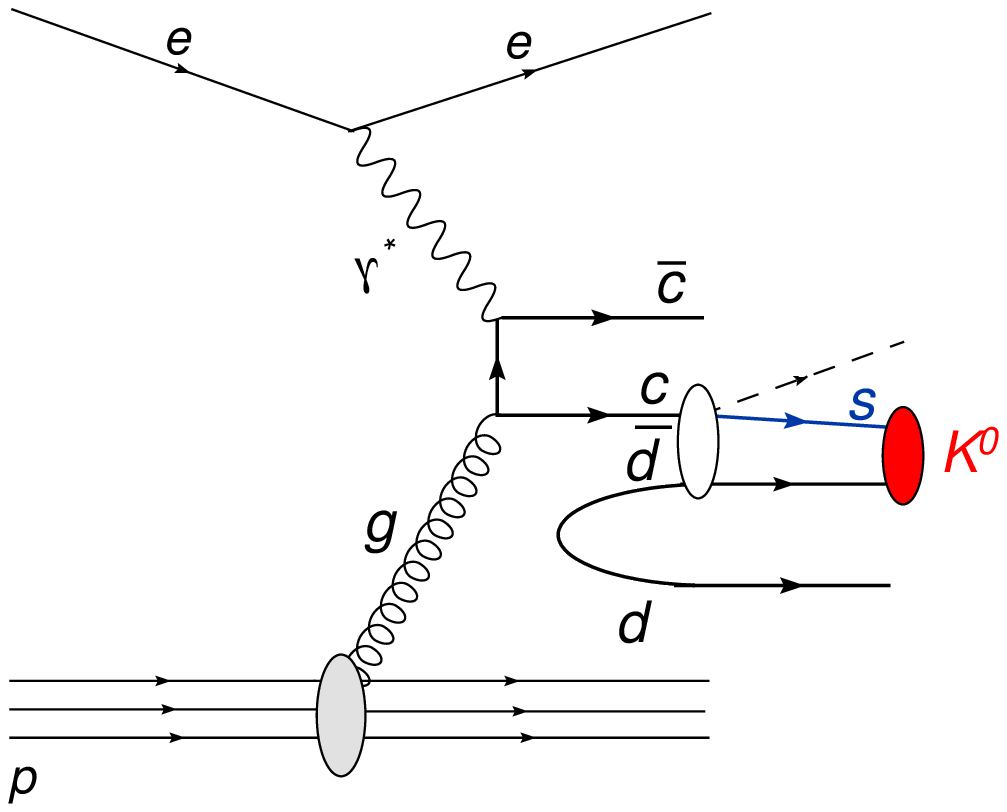} 
\hspace{1.5cm}
\includegraphics[width=55mm]{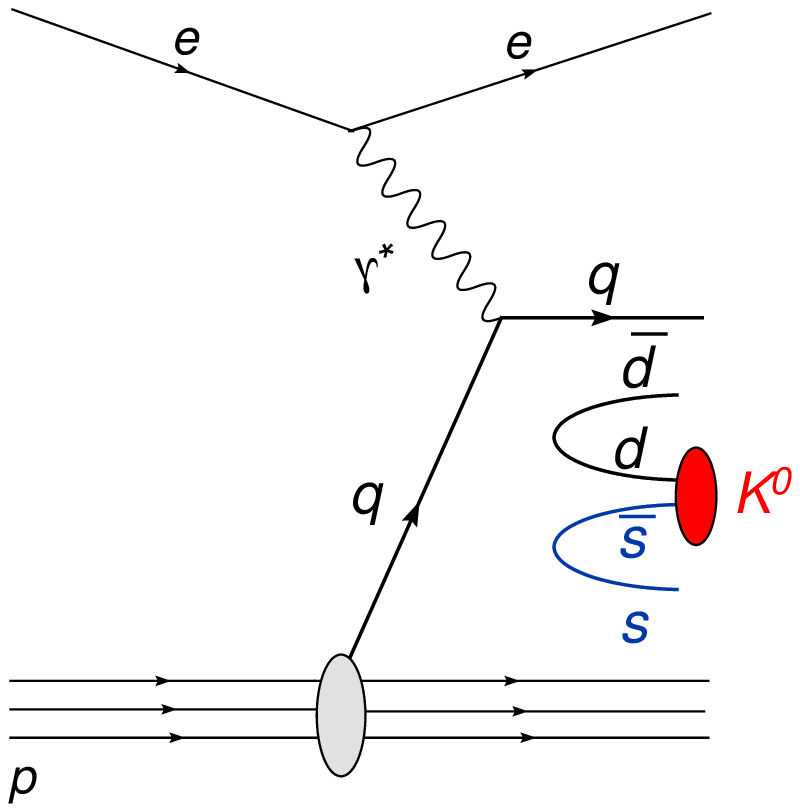}
\caption[Diagrams for the different processes contributing to strangeness production] 
{Schematic diagrams of the different processes contributing to strangeness production:
a) direct production in the QPM, b) BGF, c) 
decays of heavy quarks and d) hadronisation.}
\begin{picture}(0,0)
   \put(-63,120){\bfseries a)}
   \put(9,120){\bfseries b)}
   \put(-63,60){\bfseries c)}
   \put(9,60){\bfseries d)}
\end{picture}
\label{fig:sProduction}
\end{figure}

Particles with strangeness can be produced in DIS
in the hard sub-process 
and in the hadronisation of the colour field, as
illustrated schematically in figure~\ref{fig:sProduction}.
 
Figure~\ref{fig:sProduction}a) 
shows strangeness production 
within the quark parton model (QPM),
where a strange quark $s$ from the nucleon sea participates in the hard interaction.  
Figure~\ref{fig:sProduction}b) illustrates $s$ production in a boson-gluon fusion (BGF) 
process, where a gluon emitted from the nucleon splits
into a $s\bar{s}$ quark pair. 
Figure~\ref{fig:sProduction}c) depicts heavy 
 quark (charm $c$ and beauty $b$) production by boson-gluon fusion (BGF)
with subsequent weak decay into $s(\bar{s})$ quarks. This process is  suppressed 
at low \qsq\ due to
the masses of the heavy quarks.
These production mechanisms (figures~\ref{fig:sProduction}a, b, c) are characterised 
by a hard scale allowing for
a perturbative treatment.
The relative rate of the BGF processes depends strongly on 
the Bjorken scaling variable 
$x$ due to the strong rise of the gluon density at low $x$.
In the kinematic region studied in this
analysis (low $x$) the BGF contributions are expected to be
significant.
According to the Monte Carlo predictions described below,
roughly  $25\,\%$   
of the strange hadrons originate from strange quarks produced in the hard interaction 
either directly (figures~\ref{fig:sProduction}a and b) or through heavy quark production 
in BGF processes (figure~\ref{fig:sProduction}c). 
In regions of phase space where the quark masses are not relevant with respect
to the process scales (e.g. at very high \qsq) this rate can reach up to $50\%$.

The largest contribution to strange quark production is due to the 
colour field  fragmentation processes, as
illustrated in figure \ref{fig:sProduction}d).
As these processes occur at large distances they cannot be treated perturbatively
and thus phenomenological models, such as the LUND string model~\cite{LUND}, are required 
for their description. 
\par
Frames of reference customarily used to study particle production are the laboratory
and the Breit frame~\cite{Breit}.
In the Breit frame of reference the virtual space-like photon has momentum $Q$ but no energy.
The photon direction defines the negative $z$-axis with the proton
moving in the $+z$ direction. The transverse momentum in the Breit frame
$p_T^{Breit}$ is computed with respect to this axis.
 Particles from the proton remnant are almost collinear
to the incoming proton direction, therefore the hemisphere defined by $p^{Breit}_z > 0$ is labelled as 
the target hemisphere. 
Equally, in the QPM the struck quark only populates
the current hemisphere ($p^{Breit}_z < 0$). 
Higher order processes modify  this simple picture 
as they generate transverse momentum in the final
 state and may lead to particles from the hard subprocess 
propagating into the target hemisphere. 
%

In the current 
hemisphere, the mechanism of particle production should 
in principle resemble that of collisions  without an incident proton
like $e^+e^-$.
In analogy with $e^+e^-$ collisions the fragmentation variable 
$x_p^{Breit}=2 | {\vec{p}} \, |/Q$ is defined,
 where $\vec{p}$ is the momentum of the particle in the Breit frame;
$x_p^{Breit}$ corresponds to $x_p=p/p_{beam}$ in   $e^+e^-$ collider experiments.
Strange quarks produced directly in the hard interaction 
are expected to preferentially populate the current hemisphere, which is
less sensitive to non-perturbative strangeness contributions.
In the case of baryon production the hemisphere separation is useful 
to study also baryon transfer, which is expected to be relevant at high 
$x_p^{Breit}$ in the target frame.

   
\subsection{ Monte Carlo Simulation }
\label{models}

The deep-inelastic $ep$ interactions are 
simulated using the DJANGOH program~\cite{DJANGOH} .
It generates hard partonic processes at Born level and at leading order in $\alpha_S$
(e.g. $\gamma^* q \rightarrow q$, $\gamma^* q \rightarrow q g$,
$\gamma^* g \rightarrow q\bar{q}$~), 
convoluted with the  parton distribution 
function (PDF) for the proton,
chosen herein to be CTEQ6L~\cite{CTEQ6}.
The factorisation and renormalisation scales are set
to $\mu_f^2=\mu_r^2=Q^2$. 
Within DJANGOH, higher order QCD effects are accounted for using either 
the parton shower
approach as implemented in LEPTO~\cite{LEPTO} (referred to as MEPS) 
or by the so-called colour dipole model
approach available within ARIADNE~\cite{ARIADNE} (referred to
as CDM~\cite{CDM}). 
In LEPTO, the parton showers are ordered in the transverse momenta
($k_T$) of emissions, 
according to the leading log($Q^2$) approximation. 
In the ARIADNE program, the partons are generated by colour dipoles
spanned between the partons in the cascade;
since the dipoles radiate independently, there is no $k_T$ ordering. 

The hadronisation process
is modelled according to  the LUND colour string fragmentation model \cite{LUND},
as implemented in the JETSET~\cite{JETSET} program.
Within this model,  the strange quark suppression is predominantly
described by the (constant) factor $\lambda_s=P_s/P_q$, where
$P_s$ and $P_q$ are the probabilities for creating strange 
($s$) or light ($q=u$ or $d$) 
quarks in a non-perturbative process from the colour field during 
the fragmentation process. 
Further important parameters of this model are the diquark 
suppression factor $\lambda_{qq}=P_{qq}/P_{q}$, 
i.e. the probability of producing 
a light diquark pair $qq\bar{q}\bar{q}$ from the vacuum with respect to  
a light $q\bar{q}$ pair, and
the strange diquark suppression factor $\lambda_{sq}=(P_{sq}/P_{qq})/(P_s/P_q)$,
which models the relative production of strange diquark pairs. These are the two
 most relevant factors for the description of baryon production.
The $s\bars$ pair production rate is primarily dominated by
$\lambda_s$,  i.e. 
 $u(\baru):d(\bard):s(\bars)=1:1:\lambda_s$.
The values tuned to hadron production measurements
by the ALEPH collaboration~\cite{ALEPH-TUNE}
($\lambda_s =0.286, \lambda_{qq}=0.108$, and $\lambda_{sq}=0.690$)
are taken herein as default values 
for the simulation of hadronisation within JETSET.
 
Previously published H1 and ZEUS data~\cite{H1-K0,ZEUS-K0-1} 
are better described by 
a lower value $\lambda_s = 0.2$. 
A recent ZEUS analysis~\cite{ZEUS-K0-2}
favours $\lambda_s = 0.3$ from cross section results
and $\lambda_s = 0.22$  from measurements of the strange mesons
to charged hadrons ratio. 
The same theoretical framework is also used in $e^+e^-$ analyses and
thus allows for tests of strangeness suppression universality.

Monte Carlo event 
samples generated with DJANGOH are used for the acceptance and efficiency 
correction of the data.  All generated events are passed through
the full  GEANT \cite{geant} based simulation of the H1 apparatus and are reconstructed
and analysed using the same programs as for the data. 
%


%
\section{Experimental Procedure}
\label{method}
\subsection{H1 Detector}
\label{detector}
A detailed description of the H1 detector can be found in \cite{h1det}. 
In the following, only those detector components important 
for the present analysis are described.
H1 uses a right handed Cartesian
coordinate system with the origin at the nominal $ep$
interaction point.
The proton beam direction defines the positive $z$-axis of the laboratory frame and
transverse momenta are measured in the  $x-y$ plane. 
The polar angle $\theta$ is measured with respect
to this axis and 
the pseudorapidity $\eta$ 
is given by $\eta=-\ln {\tan {\frac{\theta}{2}}}$.

Charged particles are measured in the Central Tracking Detector (CTD) 
in the range $-1.75 < \eta < 1.75$.
%
The CTD comprises two  cylindrical Central Jet
Chambers (CJCs),
arranged concentrically around the beam-line, complemented by a silicon vertex
detector (CST)~\cite{Pitzl:2000wz},
two $z$-drift chambers and two multi-wire proportional 
chambers for triggering purposes, all within a solenoidal magnetic field of
strength $1.16$ $\textrm{T}$. 
The transverse momentum resolution is 
$\sigma(p_T) / p_T $ $\simeq 0.006 \, p_T \, /\GeV \, \oplus 0.015$~\cite{Kleinwort}.
In each event 
the tracks are used in a common fit procedure to determine the $ep$ interaction
vertex. 

The tracking detectors are surrounded by a Liquid Argon calorimeter (LAr) in 
the forward and central region ($-1.5 < \eta < 3.4$) and by a lead-scintillating 
fibre calorimeter (SpaCal) in the 
backward region~\cite{Appuhn:1996na} ($-4 < \eta < -1.4$)\@. 
The SpaCal is designed for the detection of scattered positrons 
in the DIS kinematic range considered here
and has an electromagnetic energy resolution of
$\sigma_E / E \simeq 7 \% /\sqrt{E / \textrm{ GeV}} \, \oplus 1\%$.
The backward  drift chamber (BDC), positioned in front of the SpaCal, 
improves the measurement of the positron polar angle and
is used to reject neutral particle background.
The DIS events studied in this paper are triggered 
by an energy deposition in the SpaCal,
complemented by signals in the CJCs and in the multi-wire proportional 
chambers. 

The luminosity is determined from the rate of the Bethe-Heitler process, 
$ep \rightarrow ep\gamma$, measured using a calorimeter located
close to the beam pipe at $z=-103~\textrm{m}$.

%
\subsection{Selection of DIS Events}
\label{selection}
The analysis is based on a data sample  
corresponding to an integrated 
luminosity of ${\cal L}=49.9 \pbinv$, recorded 
when HERA collided positrons at an energy $E_{e}$ 
=  $27.6\gev$ with protons at $920\gev$ in the years
$1999$ and $2000$. 

The selection of DIS events  is based on the identification of
the scattered positron as a compact calorimetric deposit in the SpaCal.
The cluster radius is  required to be less than 
$3.5\,\cm$, consistent with an electromagnetic energy deposition.
The cluster centre must be
geometrically associated with a charged track candidate in the BDC. 
These conditions reduce background from photoproduction processes.
%

At fixed centre of mass energies $\sqrt{s}$
the kinematics of the scattering process are described using
the Lorentz invariant variables $Q^2$, $y$ and $x$. 
These variables can be expressed as a function of the scattered 
positron energy $E_{e}^{\prime}$ and its scattering angle $\theta_{e}$
in the laboratory frame:
\begin{equation}
Q^2  = 4 E_e E_{e}^{\prime} \cos^2 \left( \frac{\theta_{e}}{2} \right)  ,
  \textrm{ \quad  } 
y = 1 - \frac{E_{e}^{\prime}}{E_e} \sin ^2
\left( \frac{\theta_{e}}{2} \right),  \;\; \quad
x = \frac{Q^2}{ys}.
\label{eq:kine}
\end{equation}

The negative four-momentum transfer squared $Q^2$  and the 
\mbox{inelasticity  $y$} 
are required to lie in the ranges 
$2 < Q^2 < 100\GeVSq$ and
$0.1 < y < 0.6$.
Background from events at low $Q^2$, in which the electron escapes 
undetected down the beam pipe and a hadron fakes the electron signature, is suppressed
by the requirement that the difference $\Sigma(E-p_z)$ between the total
energy and the longitudinal momentum must be in the range
$35 < \Sigma(E - p_z) <70 \gev$, where the sum includes all measured hadronic final
state particles and the scattered electron candidate.
Events are accepted if the
$z$-coordinate of the event vertex, reconstructed using
the tracking detectors, lies within $\pm 35{\rm\,\cm}$ of
the mean position for $ep$ interactions. 

%
\subsection{Selection of Hadron Candidates}
\label{lambda}
The neutral strange \ksf\ meson and \lsf\
baryon states\footnote{Unless explicitly mentioned, 
a reference to a state implicitly
includes the charge conjugate of that state.}
are measured by the kinematic reconstruction of their decays 
$\ksf    \to  \pi^+ \pi^-$ and 
$\Lambda \to  p \pi^-$.
The analysis is based on charged particles measured in the central 
region of the H1 detector with a minimum transverse momentum 
 $p_T  \geq 0.12\gev$.
The neutral strange hadrons $K^0_s$ and \lsf\ are identified by 
fitting pairs of oppositely charged tracks
in the $x-y$ plane to their secondary decay vertices,
with the direction of flight of the mother particle constrained to the
primary event vertex.
\ksf\ and \lsf\ candidates are retained if the
fit probability is above $1\,\%$.
In order to reduce background, 
the radial distance $L$ of the secondary vertex to the beam line
is required to be larger than 
$5\ \textrm{mm}$ 
and the vertex separation significance $L/\sigma_L > 4$,
where $\sigma_L$ is the uncertainty of $L$. 
The transverse momentum and the pseudorapidity 
 of the \ksf\ (\lsf) candidates are required to satisfy 
 $0.5 < p_T < 3.5\,\gev$ and $|\eta| < 1.3$. 
A detailed description of the analyses can be found in~\cite{marc,anna}.

\begin{figure}
\begin{center}
\includegraphics[width=99mm]{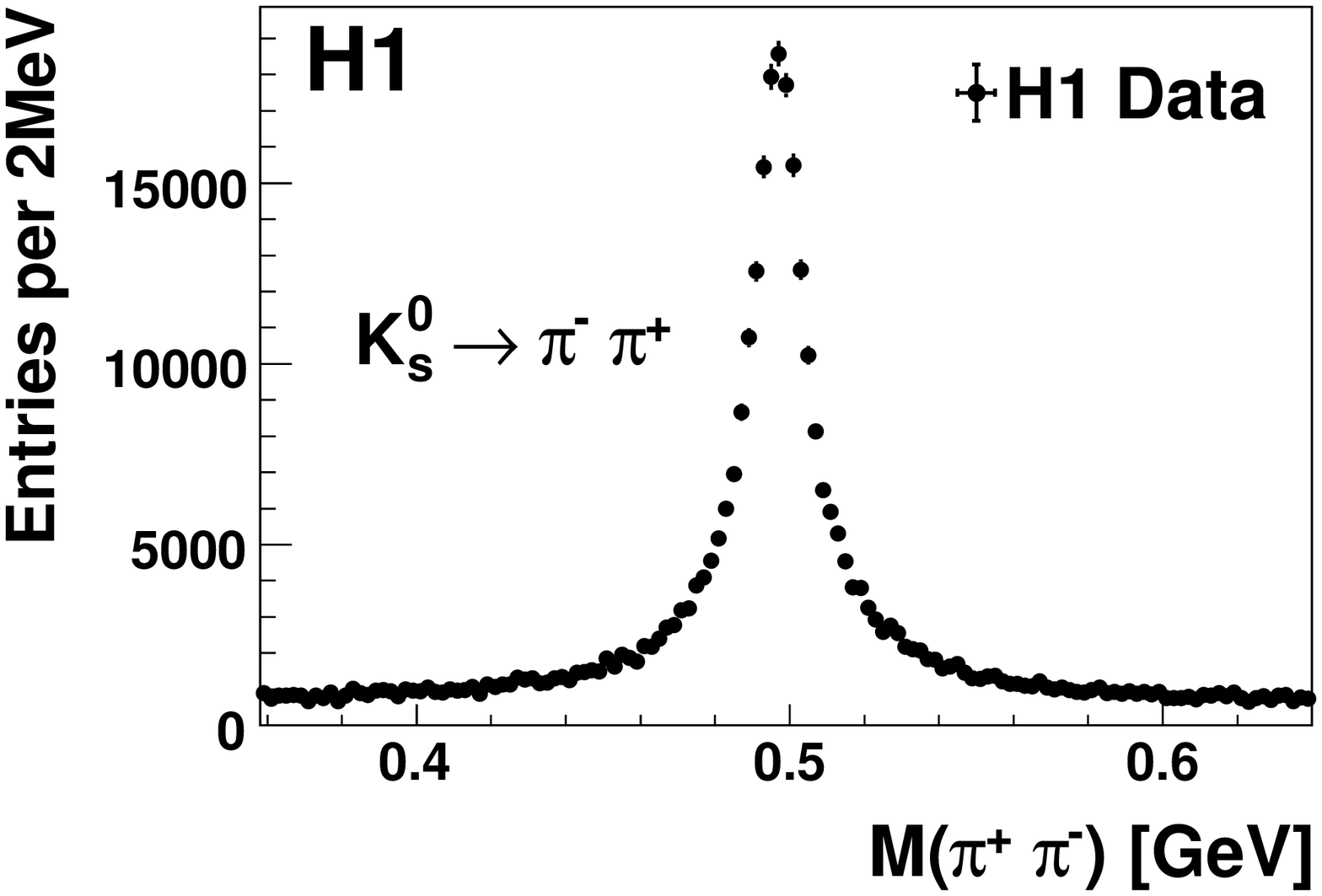}
\setlength{\unitlength}{\textwidth}
\caption{The invariant mass spectrum for 
$\pi^+\pi^{-}$ particle combinations.
The data are shown  with error bars denoting the statistical
uncertainty.
%
} 
\label{fig:kmass}
\end{center}
\end{figure}

\begin{figure}
\begin{center}
\includegraphics[width=79mm]{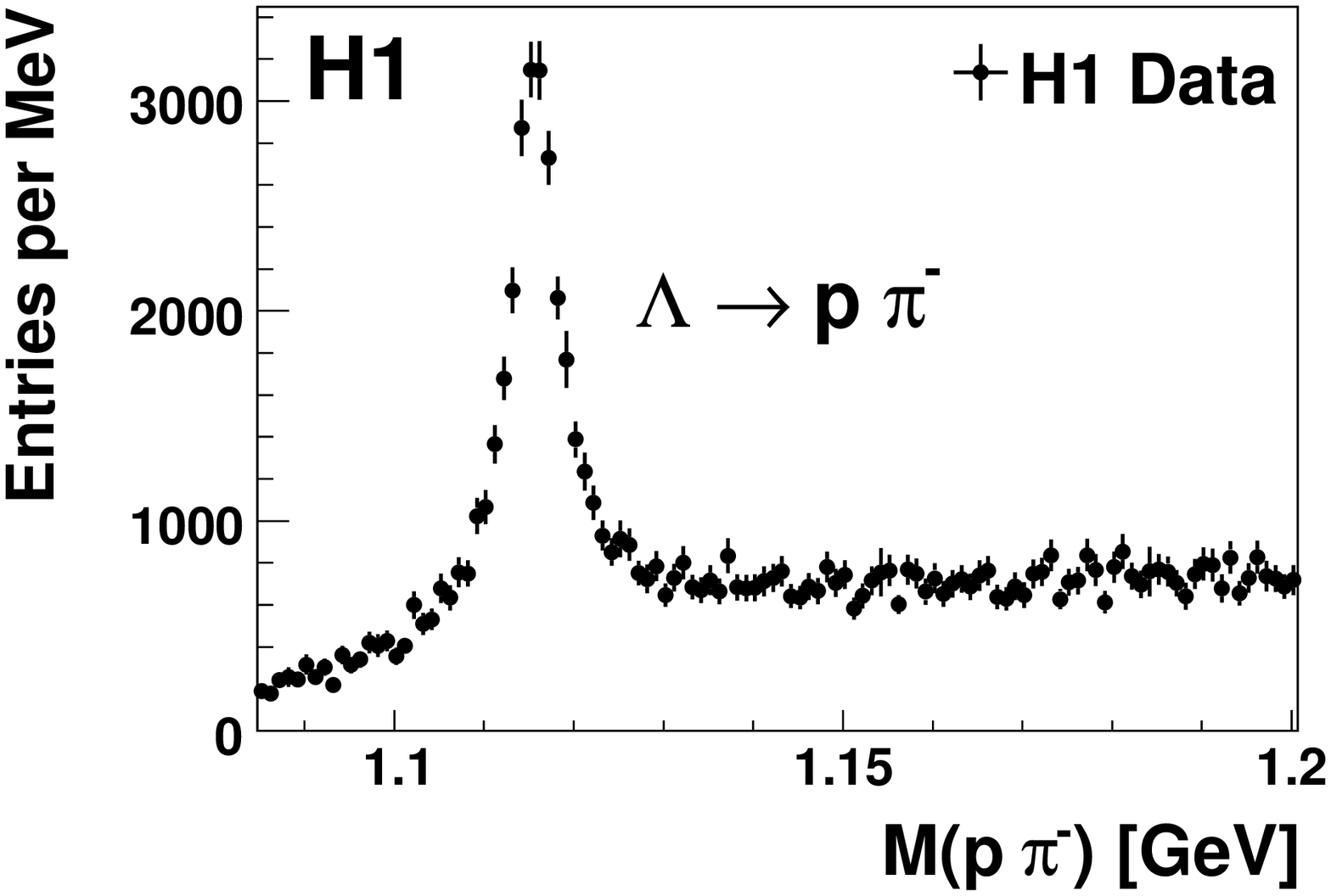}
\includegraphics[width=79mm]{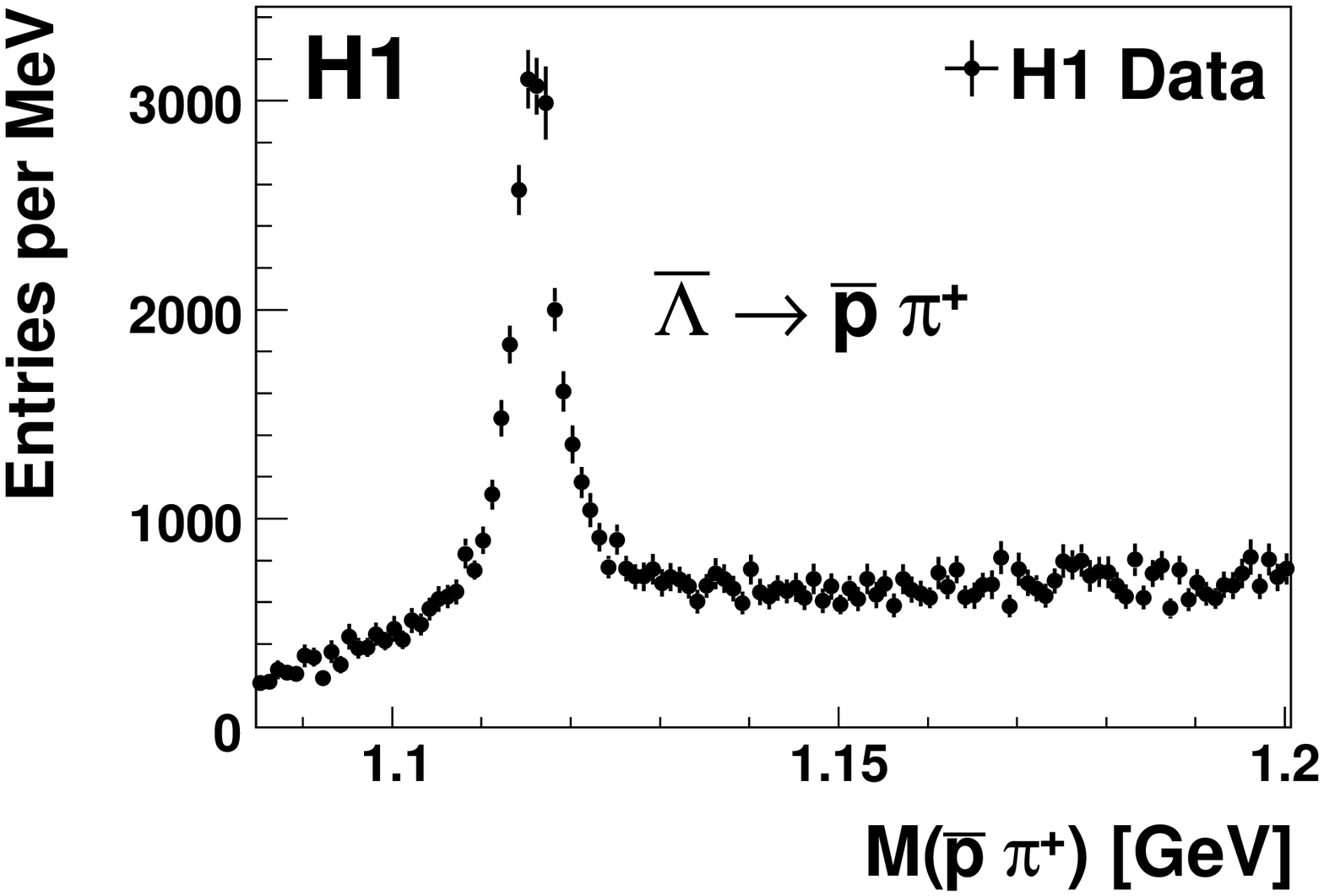}
\setlength{\unitlength}{\textwidth}
\caption{The invariant mass spectra for 
a) \lsf\ $\rightarrow p \pi^{-}$ and b) \lsa\ ${\rightarrow \bar p} \pi^{+}$ 
particle combinations.
The data are shown  with error bars denoting the statistical
uncertainty.
} 
\label{fig:l-mass}
\end{center}
\begin{picture}(0,0)
   \put(18,64){\bfseries a)}
   \put(98,64){\bfseries b)}
\end{picture}
\end{figure}

For \ksf\ candidate reconstruction both tracks are assumed to be pions, while 
for the \lsf\ reconstruction the track with the higher momentum is 
assumed to be the proton and the other track is assumed to be the pion.
The contamination from \lsf\ (\ksf) decays in \ksf (\lsf) candidates  
is suppressed by a rejection of the corresponding invariant mass region: 
$\mid M(\pi p)-m_{\lsf}\mid > 6 \mev$ for the \ksf\ and
$\mid M(\pi\pi)-m_{\ksf}\mid > 10 \mev$ for the \lsf\ selection.
The \lsf\ (\lsa) baryons are tagged by the electrical charge of the decay proton (antiproton).
The invariant mass spectra $M(\pi^{+} \pi^{-})$ and $M(p \pi)$ 
of all candidates passing these criteria are shown in 
figures~\ref{fig:kmass} and \ref{fig:l-mass}, respectively. 

The number of signal particles $N_S$ is obtained by fitting 
the invariant mass spectra with the sum of a signal and a background function.
The signal function $S$ has the same shape for  \ksf\ and \lsf\ and is 
composed of two Gaussian functions 
of identical central value $\mu$ and
of different widths $\sigma_1$ and $\sigma_2$ that account for 
different resolution effects.
The background functions $B_{\ksf}(M)$ and $ B_{\Lambda}(M)$  are chosen with 
different shapes for the \ksf\ and \lsf\ cases.
These functions are defined  according to
\begin{eqnarray}
S(M) &=& P_{0}\cdot G(N_S,\mu,\sigma_1) + (1 - P_{0} )\cdot G(N_S,\mu,\sigma_2),  \\
B_{\ksf}(M) &=& P_{1} + P_{2}\cdot M,  \\
B_{\Lambda}(M) &=& P_{1}\cdot(M-m_{\lsf})^{P_{2}} \cdot e^{(1 + P_{3}\cdot M + P_{4}\cdot M^2)}.
\label{eq:massfits}
\end{eqnarray}

Here, $M$ denotes the $\pi\pi$ and the $p\pi$ invariant mass, respectively, 
and $m_{\lsf}$ the nominal mass of the $\lsf$ baryon \cite{pdg06}.
The normalisation $N_S$, the 
central value $\mu$, the widths $\sigma_1$ and $\sigma_2$ of the Gaussian function $G$ 
and the parameters $P_{i}$ are left free in the fit. 
$P_0$ represents the relative normalisation of the two signal Gaussians.
For the differential distributions the fit is repeated in each of the kinematic bins.

%
The fit yields approximately  
 $213000$ \ksf\ mesons. 
The fitted  mass of $496.9\,\pm\,0.1\,(\textrm{ stat.})\,\mev$
is consistent with the world average \cite{pdg06} and
the measured mean width $13.8\,\pm\,0.4\,(\textrm{ stat.})\,\mev$ is 
described by the simulated detector resolution within 20\%. 
In the case of the \lsf\, the fit yields
approximately $22000$ \lsf\ and  $20000$ \lsa\  baryons.
The fitted mass of $1115.8\,\pm\,0.1\,(\textrm{ stat.})\,\mev$
is also consistent with the world average \cite{pdg06} and
the measured width of $4.3\,\pm\,0.3\,(\textrm{ stat.})\,\mev$ is consistent with 
the detector resolution within 20\%.

Charged hadrons $h^{\pm}$ used for the ratio $R(\ksf/h^{\pm})$ are defined as
long-lived particles with a lifetime $>10^{-8}$~s detected in the same kinematic region as strange particles
($|\eta|<1.3$, $0.5 < p_T < 3.5 \gev$), with the following additional requirements:
each track must point to the primary vertex, 
the number of associated hits in the CJC 
must be greater than eight, the radial track length must be longer than $10\,\cm$ and
the radial distance from the beam line to 
the innermost hit associated with the track must be less than $50\,\cm$.


\section{Results }

\subsection{Determination of Cross Sections}

The total inclusive  cross section $\sigma_{\it vis}$ in the accessible
kinematic region is given by the following expression:
\begin{equation}
\sigma_{vis}(ep \rightarrow e [\ksf, \Lambda, h^{\pm}] X) =
\frac{N}
  { {\cal L} \cdot  \epsilon \cdot BR \cdot (1 + \delta_{rad}) }\qquad , 
\label{eq:sigman}
\end{equation}
where
$N$ represents the observed number of $K^0_s$, the sum of $\lsf$ and $\lsa$ baryons
or the charged hadrons $h^{\pm}$, respectively.
 $\cal{L}$ denotes the integrated luminosity. The branching ratios $BR$ for
the $K^0_s$ and $\Lambda$ decays are taken from \cite{pdg06} and $BR=1$ for charged hadrons.
The number of \ksf\ and \lsf\ particles are determined by fitting the
mass distributions as explained in section~\ref{lambda}.
In the case of differential distributions the same formula is applied in each bin.

The efficiency $\epsilon$ is given by 
$\epsilon = \epsilon_{rec} \cdot \epsilon_{trig}$, 
where $\epsilon_{rec}$ is the reconstruction efficiency
and $\epsilon_{trig}$ is the trigger efficiency. 
The reconstruction efficiency is estimated 
using CDM Monte Carlo event samples for the 
kinematic region and the visible range defined
in sections~\ref{selection} and~\ref{lambda},
and amounts to $33.3\,\%$ 
and $19.5\,\%$ for the $K^0_s$ mesons and the \lsf\ baryons, respectively.
These numbers include the geometric acceptance and the efficiency for track 
and secondary vertex reconstruction. 
The geometric acceptance to find both decay
particles in the CTD is about $80\,\%$ for the  $K^0_s$ mesons
and $70\,\%$  for the \lsf\ baryons, respectively
%

The trigger efficiency is extracted from the 
data using monitor triggers and amounts to 
$81.5\,\%$ and $83.3\,\%$ for the $K^0_s$ and 
the \lsf, respectively.
The radiative correction $\delta_{rad}$ corrects the measured cross
section to the Born level and is calculated using the
program HERACLES~\cite{HERACLES}.
It amounts to $\delta_{rad} = 6.6 (4.3)\,\%$ for the $K^0_s$ (\lsf) 
on average and varies between $-8\%$ and $+19\%$ over the kinematic range considered.
The trigger efficiency and radiative corrections are assumed to be the same for 
particles and antiparticles.

In the case of charged hadrons $h^{\pm}$, the reconstruction efficiency 
$\epsilon_{rec}$ is defined such that it includes
corrections for \ksf\ and \lsf\ decays, secondary interactions, photon 
conversions and the track reconstruction efficiency. 
The total correction $\epsilon(1+\delta_{rad})$  amounts to  $81.1\%$.
	
\subsection{Systematic Uncertainties}

The systematic uncertainties are studied using the CDM Monte Carlo simulation,
unless otherwise stated. 
For the inclusive cross sections, the resulting systematic uncertainties
are summarised in table \ref{table:Systematics}. 
For the differential cross sections, the
systematic uncertainties are estimated in each bin. 
The following contributions are considered:
\begin{itemize}
\item The energy scale  in the Spacal measurements is known to 
$1\,\%$, except for the lowest $Q^2$ bin ($2 < Q^2 < 2.5\, \GeVSq $) 
where the uncertainty on the energy measurement is $2.5\,\%$. 

\item The uncertainty of the measurement of the polar angle
of the scattered positron is $1\,\textrm{mrad}$.

\item The uncertainty on the overall number of reconstructed $K^0_s$ and \lsf\ particles
is determined from data by comparing the numbers obtained from the
fit of the mass spectra with the number obtained by simply
counting the events within $\pm 6\,\sigma$ of the nominal mass after subtracting the 
expected background. The number of background events is estimated
by integrating the background function described in 
equation~\ref{eq:massfits}
over the corresponding interval ($0.42 - 0.58 \gev$ for \ksf\ and $1.085 - 1.2$ for \lsf).
The procedure 
is cross checked by performing the fit in different mass ranges. 

\item The uncertainty  of
 the reconstruction efficiency is determined by 
comparing its estimation using different models. 
The uncertainty is taken as $50\,\%$ of the difference between the CDM and the MEPS 
Monte Carlo simulations. 

\item The uncertainty on the trigger efficiency  is 
obtained  by comparing estimates using different monitor triggers (MT). 

\item  The uncertainty on the luminosity measurement is $1.5\,\%$. 

\item The uncertainty of the charged hadron reconstruction is $2\,\%$ per track. For the 
measurement of the \lsf\ to $K^0_s$ ratio the uncertainty caused by the pion track 
appearing in both  decays is assumed to cancel. 
The systematic uncertainty on the ratio $\ksf/h^{\pm}$ is  estimated to be $2.0\,\%$.

\item The uncertainty due to the decay branching ratios is taken as $0.8\% $ for \lsf\
and is negligible for \ksf~\cite{pdg06}.

\end{itemize}

\begin{table}[!hbp]
\centering
\begin{tabular}{|c|c||c|c|c|c|}
\hline
Source & Variation & $\Delta\sigma(K^0_s)$ & $\Delta\sigma(\Lambda)$ & $R(\Lambda/K^0_s)$  & $R(K^0_s/h^\pm)$  \\
 &  & $[\%]$ & $[\%] $ & $[\%] $  & $[\%]$  \\
\hline
\hline
$E_e'$ & $\pm 1\%$ & $^{+3.3}_{-3.5} $ & $^{+2.8}_{-3.1}$ & $-$ & $-$\\
\hline
$\theta_e$ & $\pm 1\,\textrm{ mrad}$ & $\pm 1.4 $ & $\pm 1.5$ & $-$ & $-$ \\
\hline
signal extraction & $\frac{N^{fit} - N^{count}}{N^{fit}}$ & $ \pm 0.6$ & $\pm 1.4$ & $\pm 1.5$ & $ \pm 0.6$  \\
\hline
model & $0.5*\frac{\epsilon_{rec}^{CDM} - \epsilon_{rec}^{MEPS}}{\epsilon_{rec}^{CDM}}$ & $\pm 0.4 $ 
& $\pm 1.2$ & $\pm 1.2$  & $\pm 3.5$ \\
\hline
trigger efficiency & $\frac{\epsilon_{trig}^{MT set1} - \epsilon_{trig}^{MT set2}}{\epsilon_{trig}^{MT set1}}$ 
& $^{+0.4}_{-0.9} $ & $^{+1.0}_{-1.4}$ & $^{+1.1}_{-1.6}$ &$^{+0.4}_{-1.0} $\\
\hline
luminosity & $$ & $\pm 1.5$ & $\pm 1.5$ & $-$  & $-$ \\
\hline
track reco. & $2.0\,\%$ per track & $\pm 4.0$ & $\pm 4.0$ & $\pm 2.0$ & $\pm 2.0$ \\
\hline
branching ratio &  & $\pm0.1$ & $\pm 0.8$ & $\pm 0.8$ & $\pm 0.1$ \\
\hline
\hline
\multicolumn{2}{|c||}{Total systematic uncertainty} & $^{+5.6}_{-5.8}$ &$^{+5.8}_{-6.0}$ &
$^{+3.1}_{-3.3}$ & $^{+4.1}_{-4.2}$ \\
\hline
\end{tabular}
\caption{Systematic sources, 
variations and corresponding relative  errors of the inclusive cross 
sections and of the ratios
of \lsf\ to \ksf\ and \ksf\ to charged hadrons. 
All relative errors are expressed as percentages.}
\label{table:Systematics}
\end{table}

%
The systematic errors due to these uncertainties are estimated
by varying each quantity within its error in the Monte Carlo simulation and
repeating the cross section measurement.
In the cross section calculation, the contributions 
are added in quadrature and included
in the uncertainty shown in the individual bins of the
differential distributions.
In the ratios, the uncertainties on the electron energy scale and polar angle,
as well as the luminosity, cancel. The other sources of uncertainty are assumed
to be uncorrelated and are added in quadrature.


%
\subsection{Inclusive Production Measurements}

The inclusive \ksf, \lsf\ and charged hadron $h^{\pm}$ production cross sections $\sigma_{\it vis}$
are measured in the 
kinematic region $2 < Q^2 < 100\GeVSq$ and $0.1 < y < 0.6$,
for the ranges 
 $0.5 < p_T(\ksf,\lsf,h^{\pm} ) < 3.5\,\gev$  and $|\eta(\ksf, \lsf,h^{\pm} )|< 1.3$.

The \ksf\ cross section is found to be 
\begin{equation}
\sigma_{vis}(ep \rightarrow e \ksf X) = 21.18 \pm 0.09 
({\rm \textrm {stat.}}) ^{+1.19}_{-1.23} (\textrm {syst.}) \textrm{\, nb}.
\label{eq:xsec-ks}
\end{equation}
The measurement is in agreement with the expectation $21.77\textrm{\, nb}$\,
based on the LO Monte Carlo program DJANGOH, using the CDM approach and 
the default value of $\lambda_s =0.286$.

The cross section for the sum of \lsf\ and \lsa\ baryon production is measured in the
same kinematic region and is found to be
\begin{equation}
\sigma_{vis}(ep \rightarrow e [\lsf + \lsa] X) = 7.88 \pm 0.10 
(\textrm {stat.}) ^{+0.45}_{-0.47} (\textrm{syst.}) \textrm{\, nb},
\label{eq:xsec-la}
\end{equation}
in agreement with the expectation of $7.94\textrm{ \,nb}$
from the DJANGOH calculation.
The individual \lsf\ and \lsa\ production rates 
are measured to be
\begin{eqnarray}
\sigma_{vis}(ep \rightarrow e \lsf X) &=& 3.96 \pm 0.06 (\textrm {stat.}) 
^{+0.23}_{-0.24} (\textrm{syst.}) \textrm{\, nb}, \\
\label{eq:xsec-la2}
\sigma_{vis}(ep \rightarrow e \lsa X) &=& 3.94 \pm 0.07 (\textrm {stat.}) 
^{+0.23}_{-0.24} (\textrm{syst.}) \textrm{ \,nb},
\end{eqnarray}
and are therefore found to be consistent 
with each other within the statistical precision.
The measurements are also in agreement with the DJANGOH 
prediction of $3.97 \textrm{ \,nb}$.
The systematic errors are fully correlated.

The inclusive ratio of strange baryon to meson production is determined to be
\begin{equation}
\frac{\sigma_{vis}(ep \rightarrow e [\lsf + \lsa] X)}{\sigma_{vis}(ep \rightarrow e \ksf X)}
 = 0.372 \pm 0.005 (\textrm {stat.})   ^{+0.011}_{-0.012} (\textrm{syst.}),
\label{eq:xsec-ks-la}
\end{equation}

in agreement with the prediction of $0.365$ from the DJANGOH calculation.

 The ratio of cross sections of \ksf\ mesons to charged hadrons $h^{\pm}$   
is found to be
\begin{equation}
\frac{\sigma_{vis}(ep \rightarrow e \ksf  X)}{\sigma_{vis}(ep \rightarrow 
e h^{\pm} X)} = 0.0645 \pm 0.0002(\textrm {stat.})  ^{+0.0019}_{-0.0020} (\textrm{syst.}),
\label{eq:xsec-ks-h}
\end{equation}

in agreement with the DJANGOH prediction  of $0.0638$ based on MEPS  with \lambdas$= 0.22$.
Similar values of $0.05 - 0.07$ are obtained for the ratio of the average $\ksf$ 
multiplicity over the average charged pion multiplicity in
$e^{+}e{^-}$ annihilation events at centre of mass energies from
$10$ to $200 \gev$ \cite{pdg06}.


\subsection{Differential Production Cross Sections}
\label{sec:diffXsec}

Production cross sections and ratios of \ksf, \lsf\ and charged hadrons $h^{\pm}$
are measured in the visible kinematic region differentially 
in the event variables  $Q^2$ and $x$ and in the laboratory frame variables
$p_T$ and $\eta$. 
Differential cross sections are also measured as a function of the 
variables $x_p^{Breit}$ and $p_T^{Breit}$ defined in the Breit frame. 
The results are bin-averaged and no bin-centre corrections are applied.
The distributions are shown in figures~\ref{fig:ks-ds-lab} 
to~\ref{fig:ks-charged-ratio-lab}
and are compared with the predictions. The numerical values are also listed in 
tables~\ref{table:K0CrossSection} to~\ref{table:RhCrossSection}.

\subsubsection{Discussion of ${\bf K_s^0}$ and ${\bf \Lambda}$ Results}

\begin{figure}
\begin{center}
\includegraphics[width=79mm]{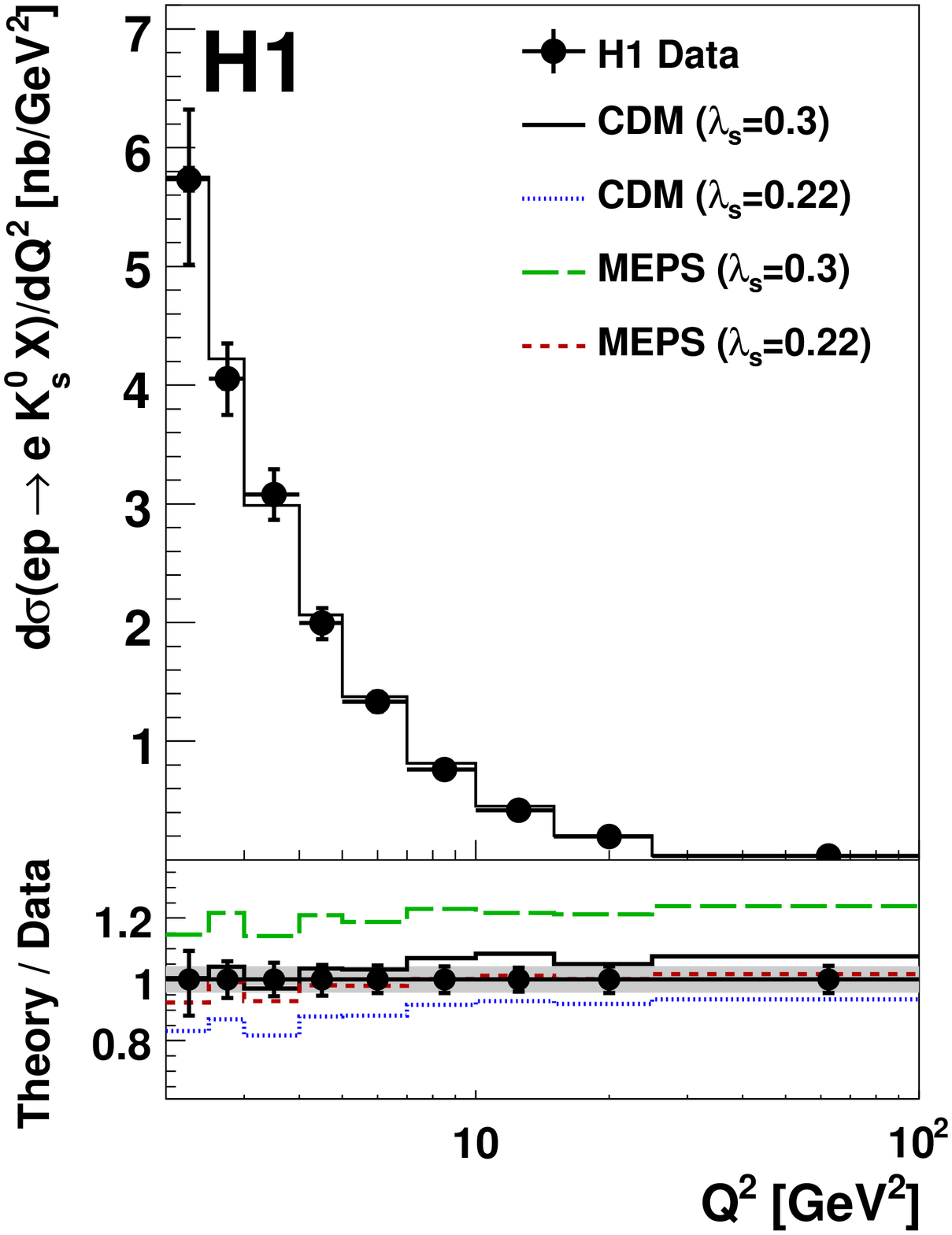}
\includegraphics[width=79mm]{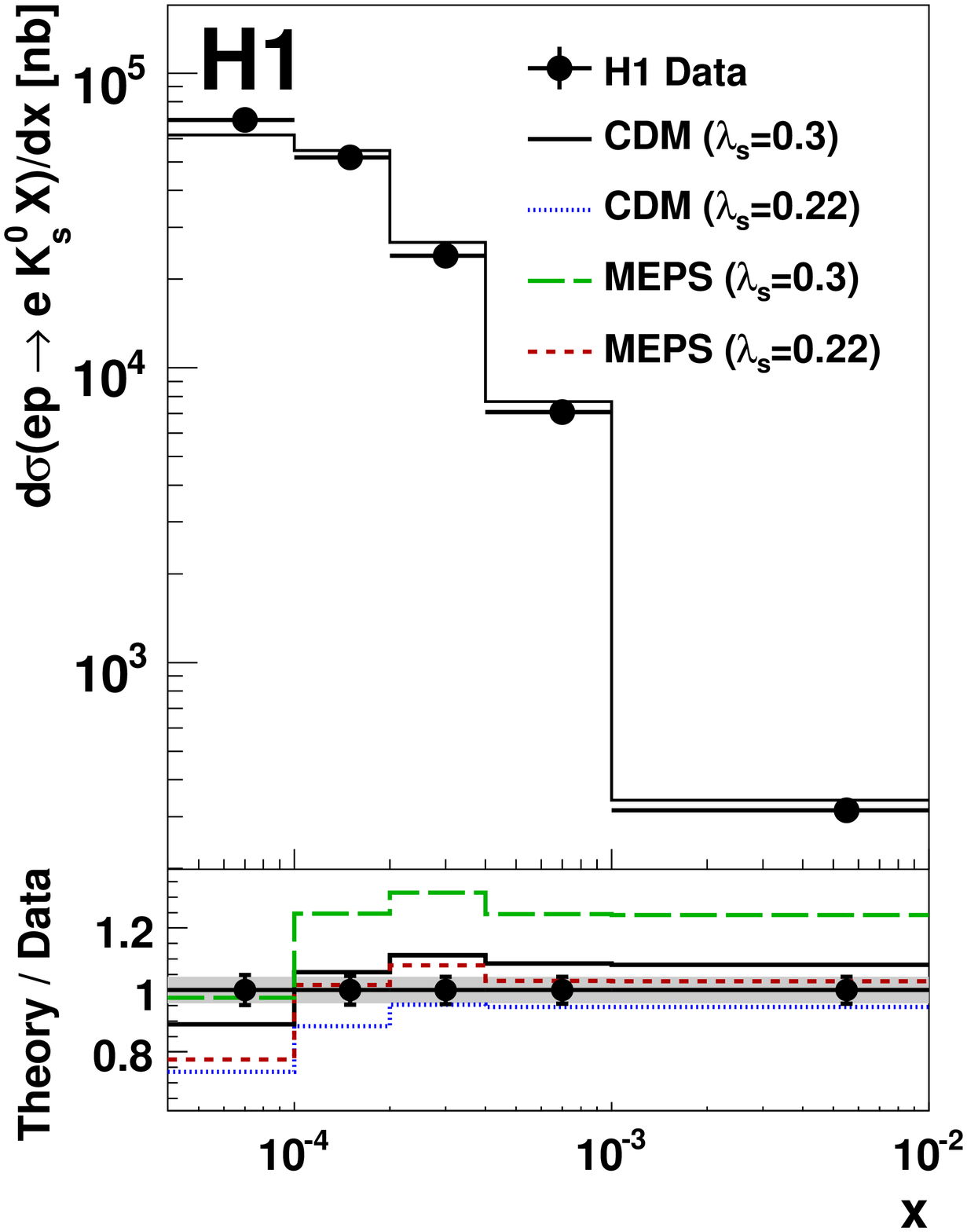}\\
\includegraphics[width=79mm]{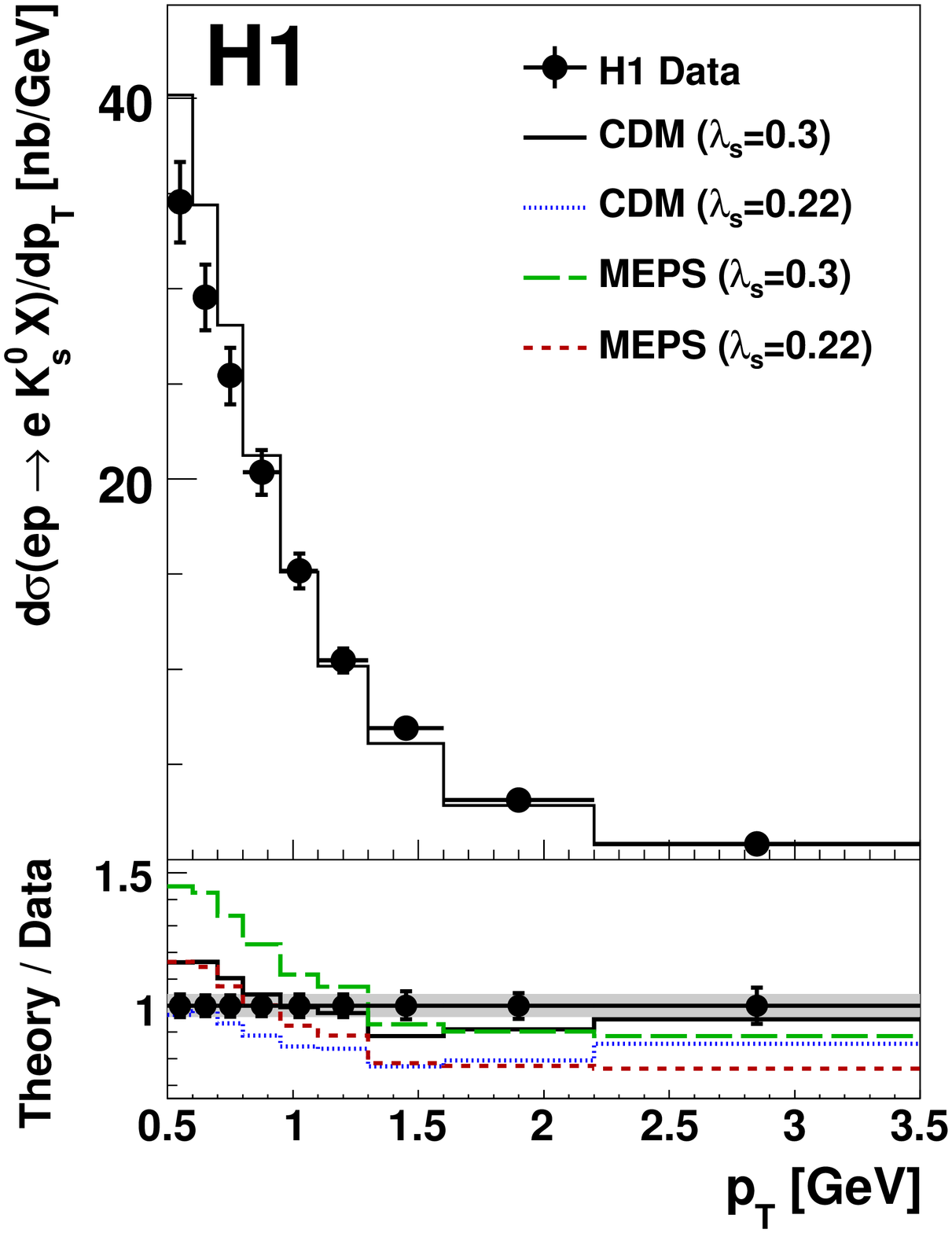}
\includegraphics[width=79mm]{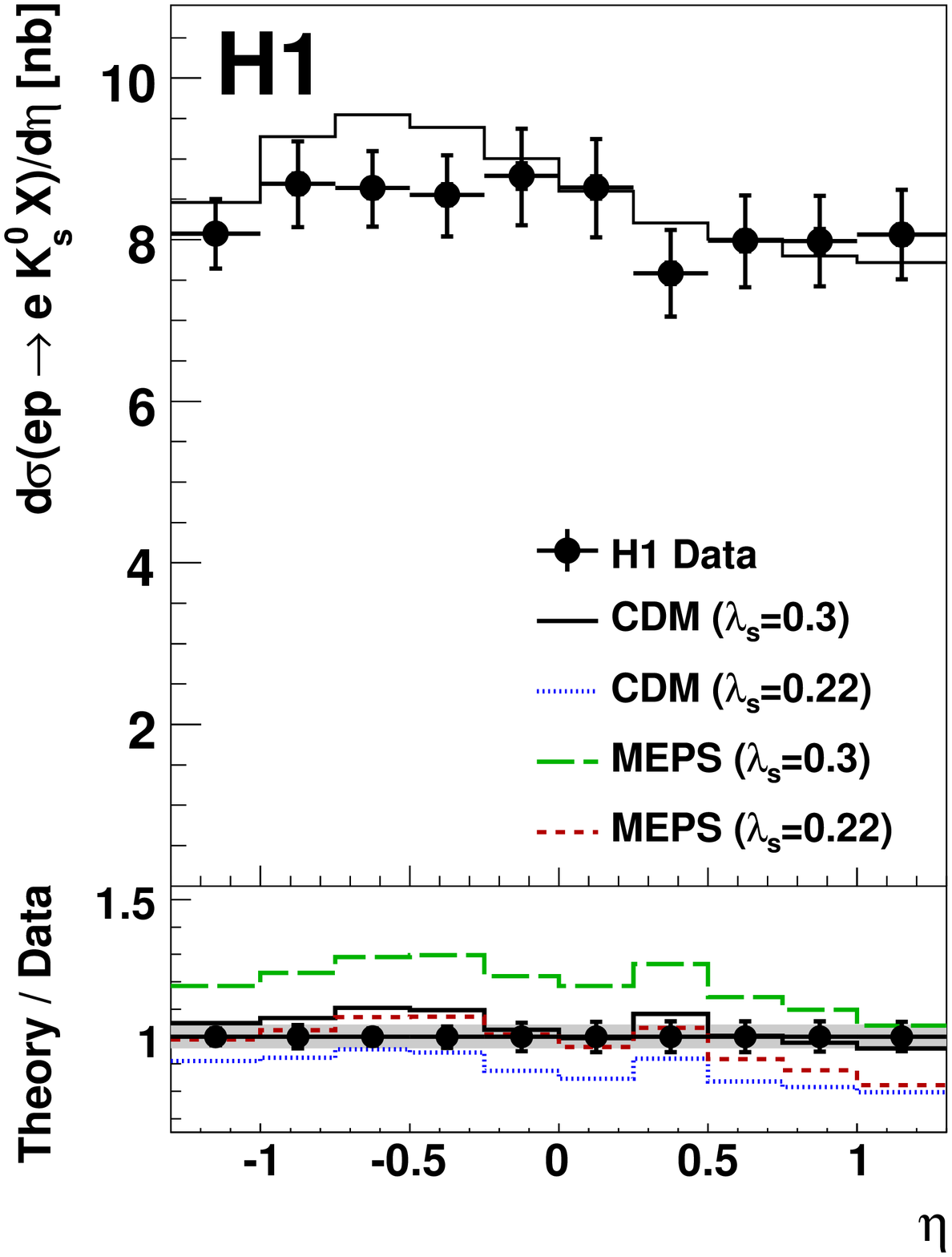}\\
\caption{The differential production cross sections for 
\ksf\ in the laboratory frame as a function of the
a) photon virtuality squared $Q^2$, 
b) Bjorken scaling variable $x$,
c) \ksf\ transverse momentum $p_T$ 
and d) \ksf\ pseudorapidity $\eta$.
The inner (outer) error bars show the statistical (total) errors.
On the bottom of  each figure, the ``Theory/Data'' ratios are shown
for different LO Monte Carlo predictions (see text).
For comparison, the data points are put to one and
only uncorrelated errors are shown; the correlated systematic errors are indicated by the
grey band.
}
\label{fig:ks-ds-lab}
\end{center}
\begin{picture}(0,0)
   \put(17,177){\bfseries a)}
   \put(97,177){\bfseries b)}
   \put(17,77){\bfseries c)}
   \put(97,77){\bfseries d)}
   \put(62,241){\LARGE e\,p $\rightarrow$ e\,K$^0_s$\,X}
\end{picture}
\end{figure}

\begin{figure}
\begin{center}
\includegraphics[width=79mm]{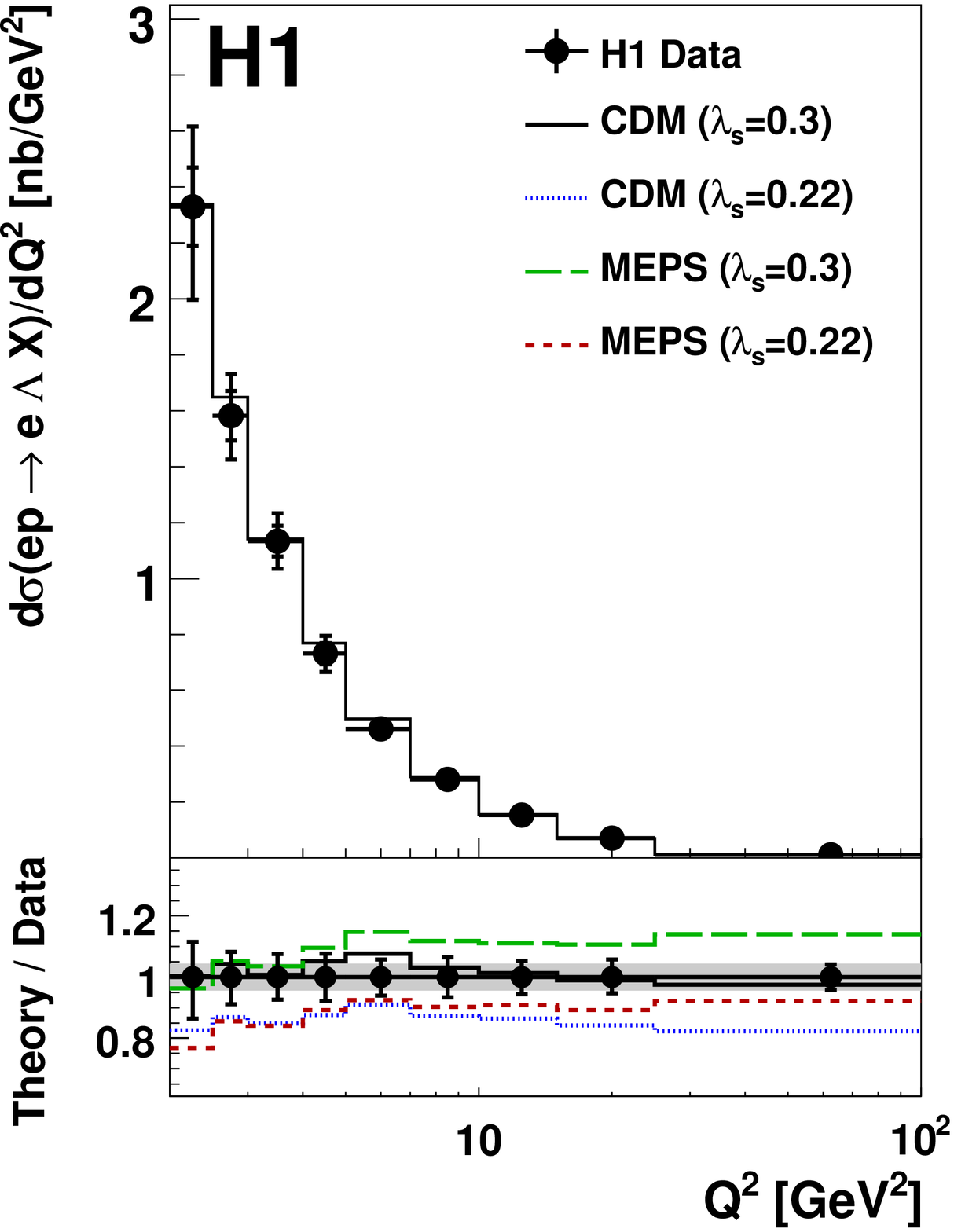}
\includegraphics[width=79mm]{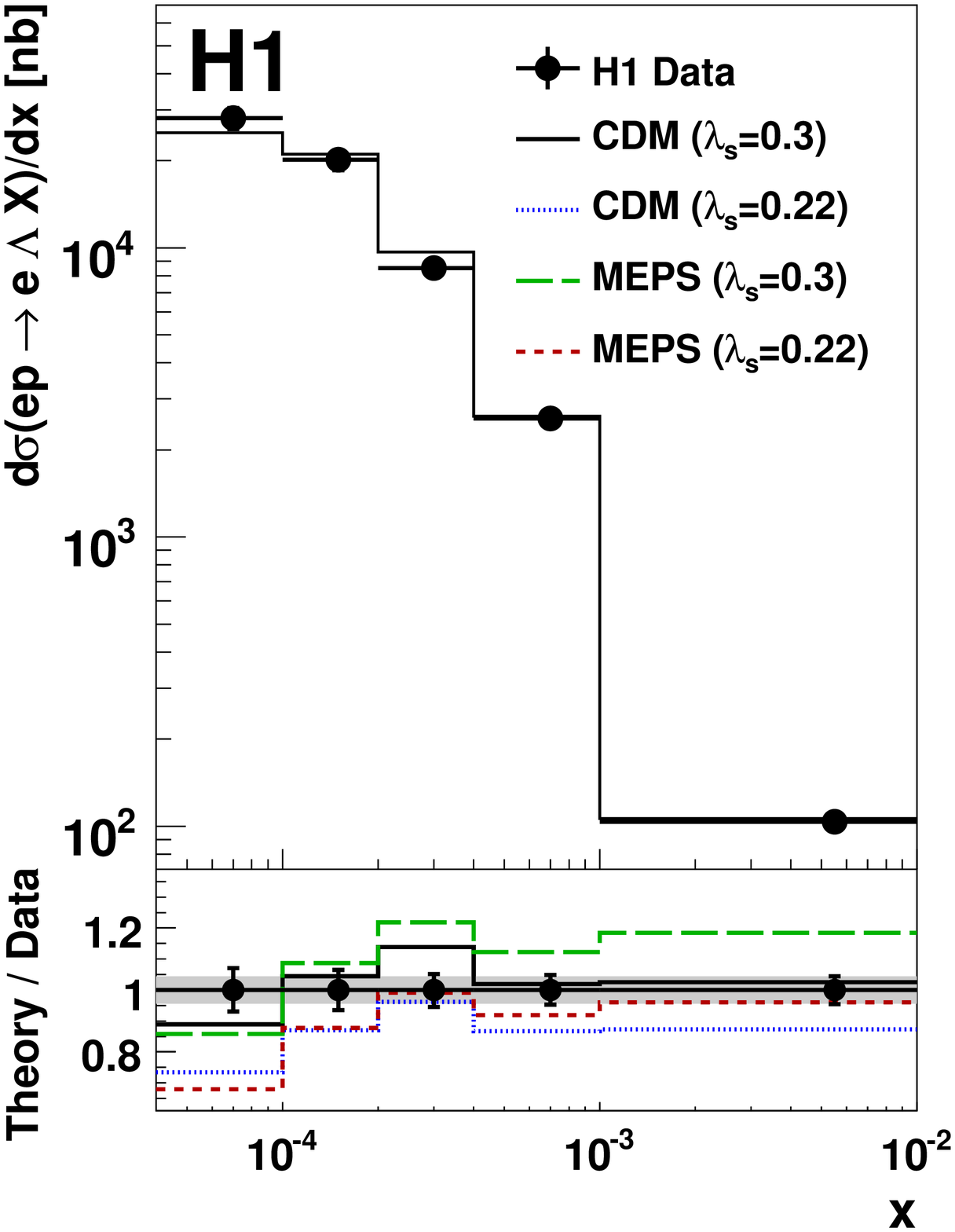}\\
\includegraphics[width=79mm]{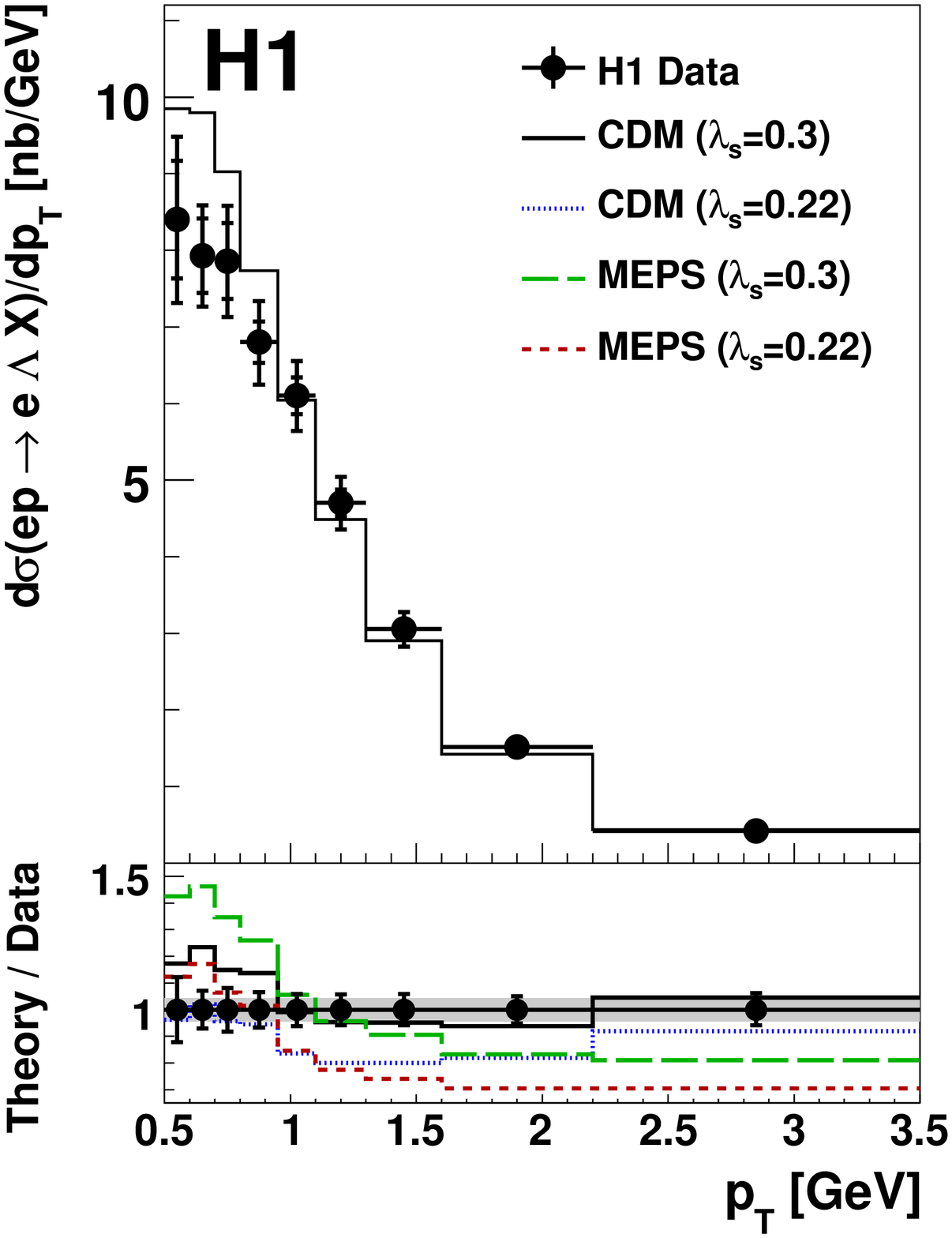}
\includegraphics[width=79mm]{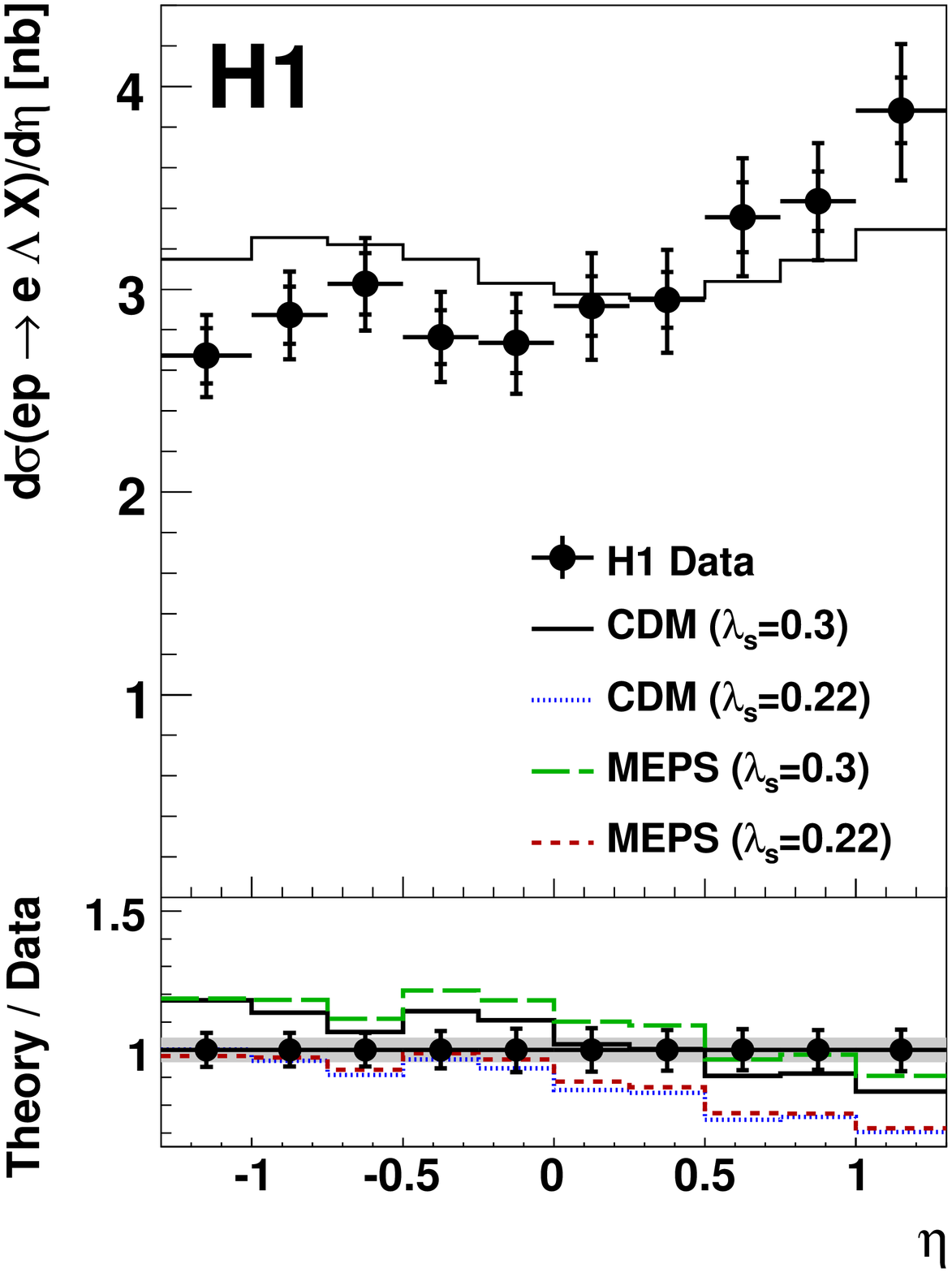}\\
\caption{The differential production cross sections for 
\lsf\ in the laboratory frame as a function of the
a) photon virtuality squared \qsq,
b) Bjorken scaling variable $x$, 
c) \lsf\ transverse momentum $p_T$
and d) \lsf\ pseudorapidity $\eta$. 
More details in the caption of figure~\ref{fig:ks-ds-lab}.
}
\label{fig:la-ds-lab}
\end{center}
\begin{picture}(0,0)
   \put(17,162){\bfseries a)}
   \put(97,162){\bfseries b)}
   \put(17,62){\bfseries c)}
   \put(97,62){\bfseries d)}
   \put(62,227){\LARGE e\,p $\rightarrow$ e\,$\Lambda$\,X}
\end{picture}
\end{figure}

\begin{figure}
\begin{center}
\includegraphics[width=79mm]{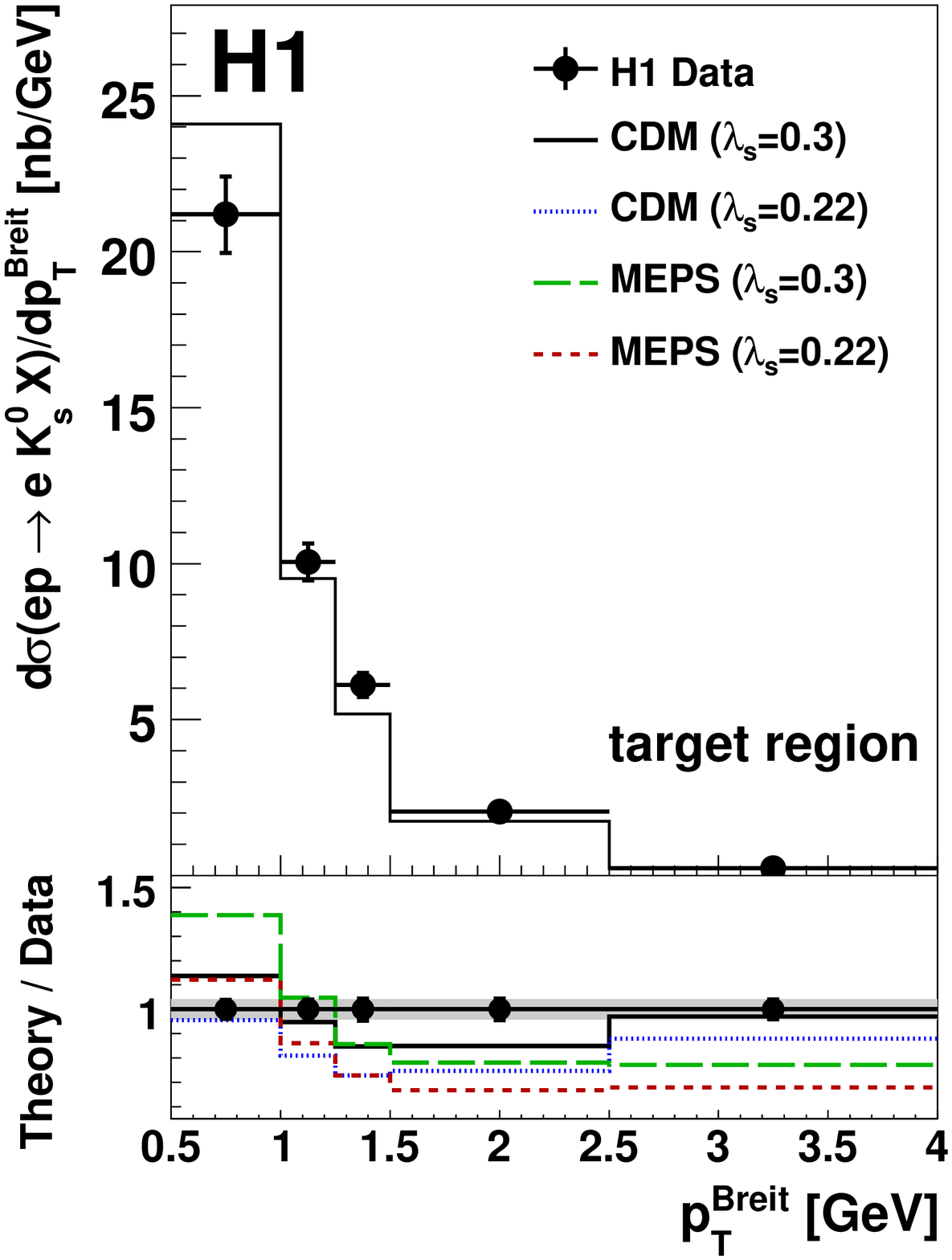}
\includegraphics[width=79mm]{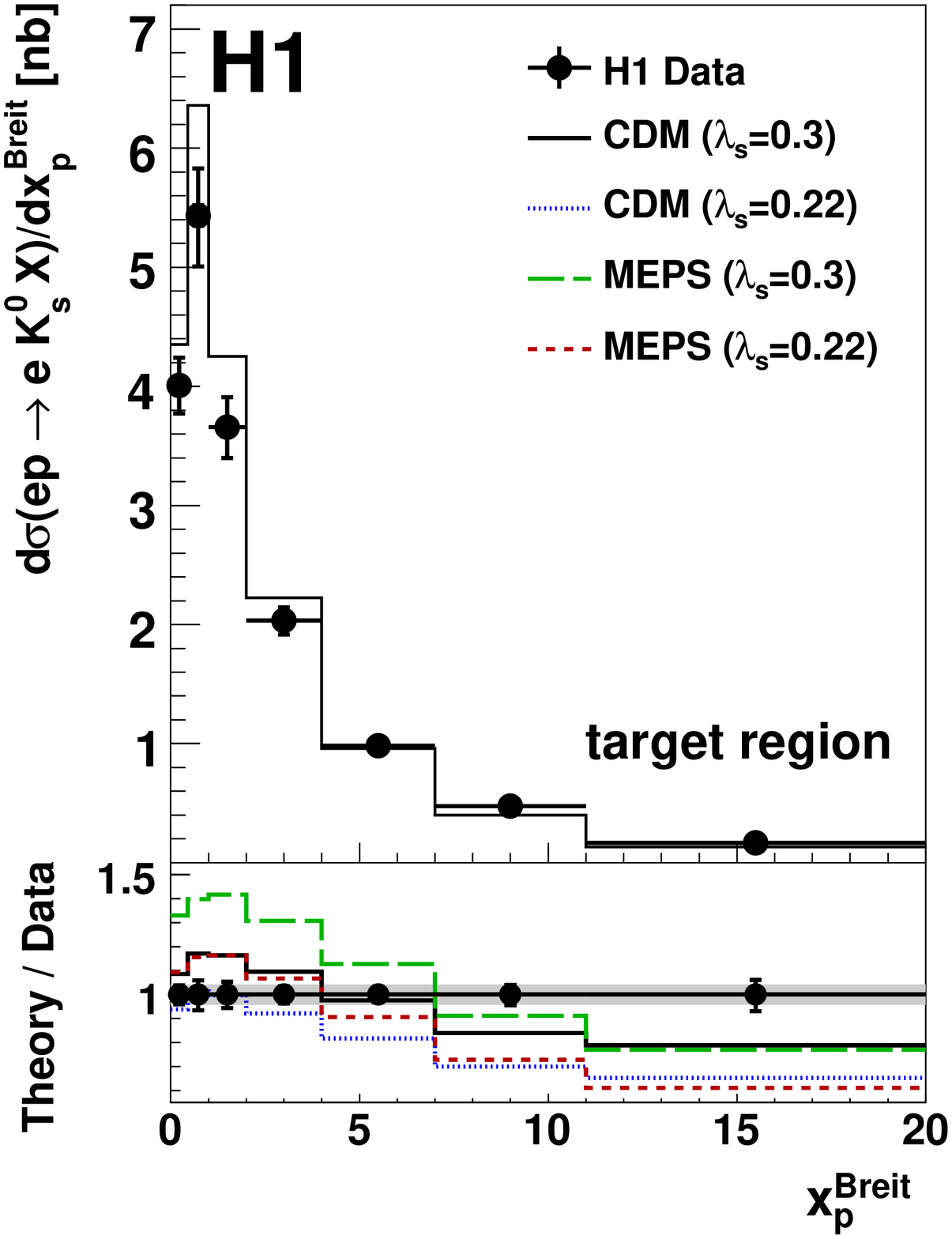}\\
\includegraphics[width=79mm]{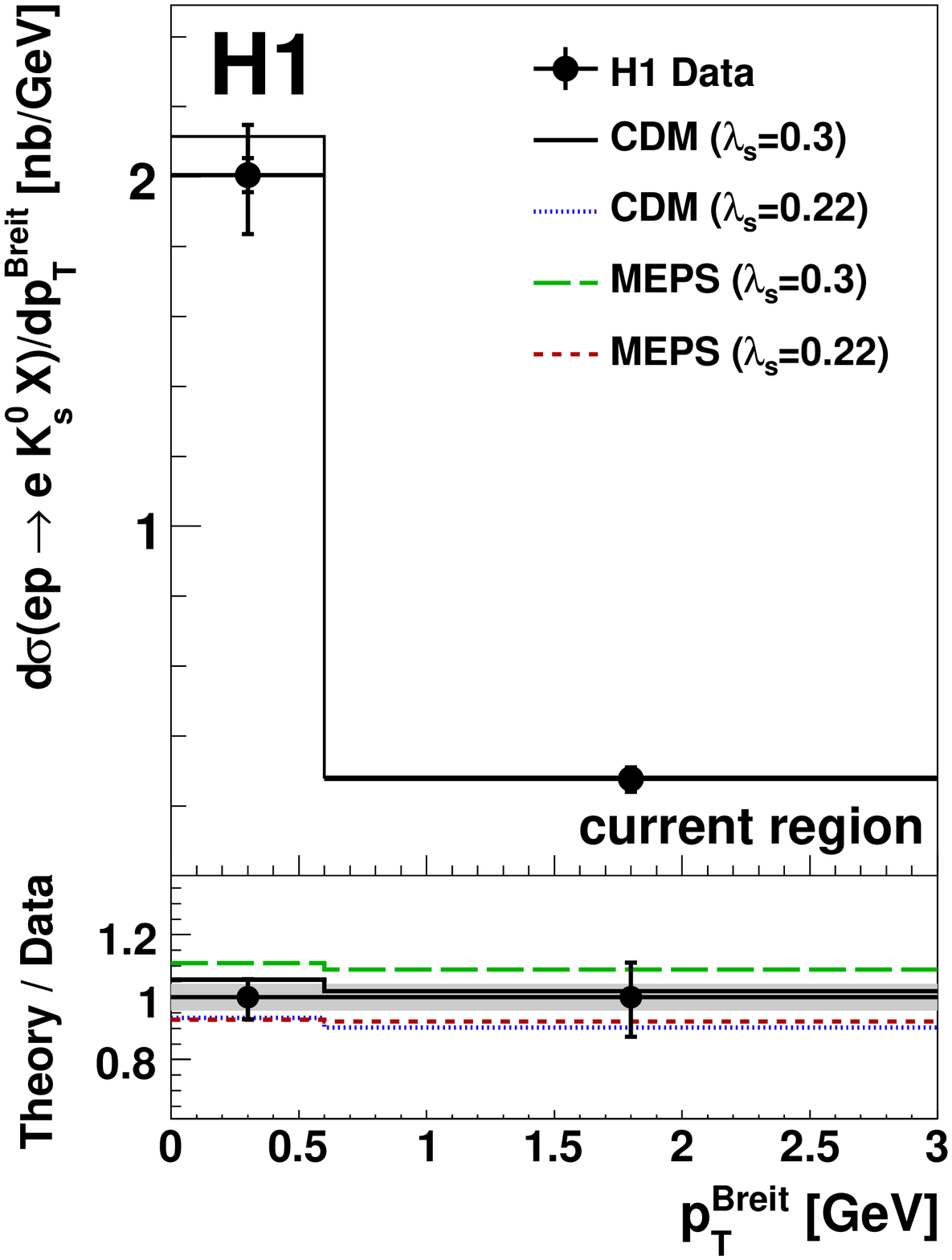}
\includegraphics[width=79mm]{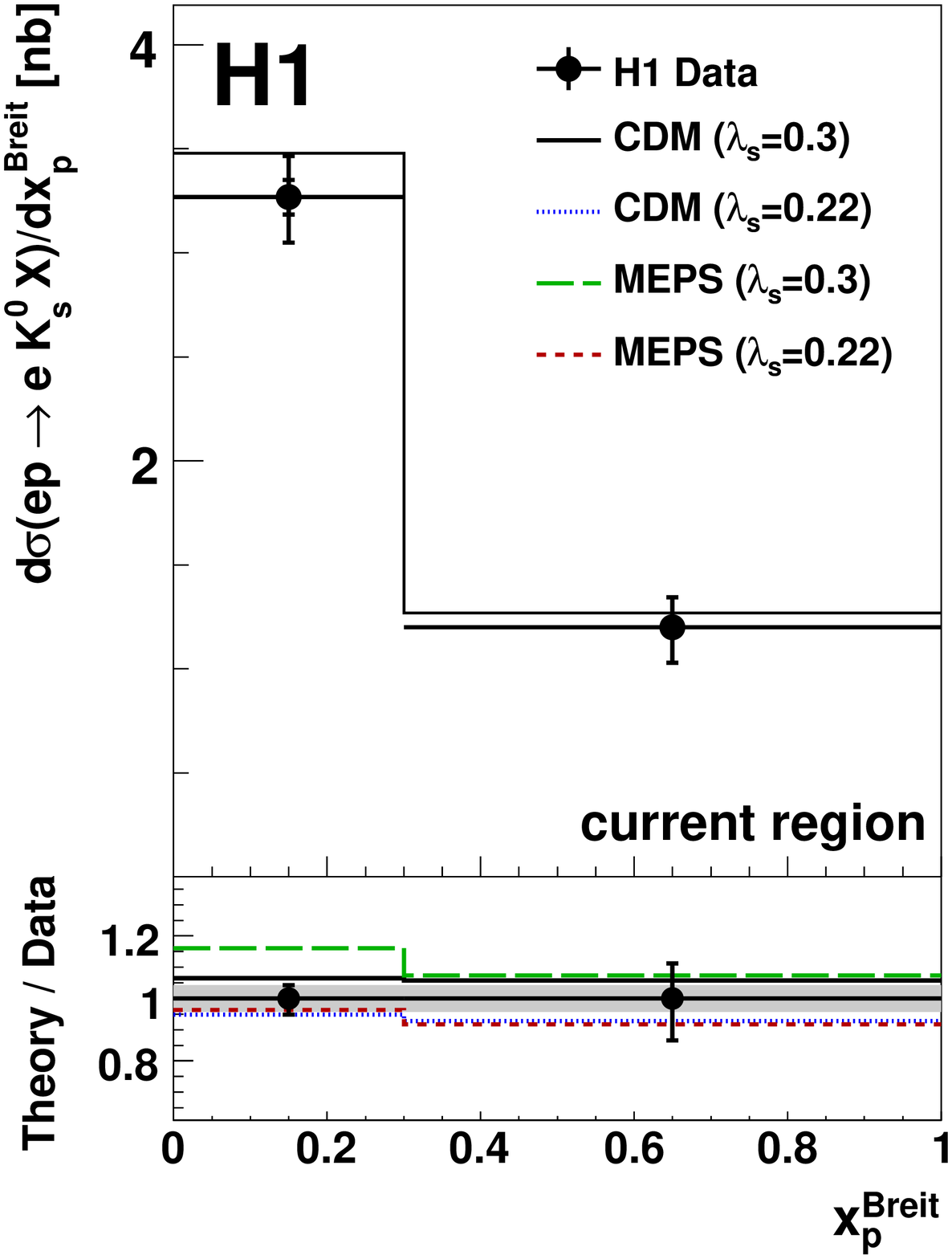}\\

\caption{The differential \ksf\ production cross sections 
in the Breit frame as a function of \ksf\ transverse momentum 
$p_T^{Breit}$ and  momentum fraction $x_p^{Breit}$ in the target 
hemisphere (a, b) and in the current hemisphere (c, d).
More details in the caption of figure~\ref{fig:ks-ds-lab}.
}
\label{fig:ks-ds-breit}
\end{center}
\begin{picture}(0,0)
   \put(17,162){\bfseries a)}
   \put(97,162){\bfseries b)}
   \put(17,62){\bfseries c)}
   \put(97,62){\bfseries d)}
   \put(42,227){\LARGE e\,p $\rightarrow$ e\,K$^0_s$\,X (Breit frame)}
\end{picture}
\end{figure}

\begin{figure}
\begin{center}
\includegraphics[width=79mm]{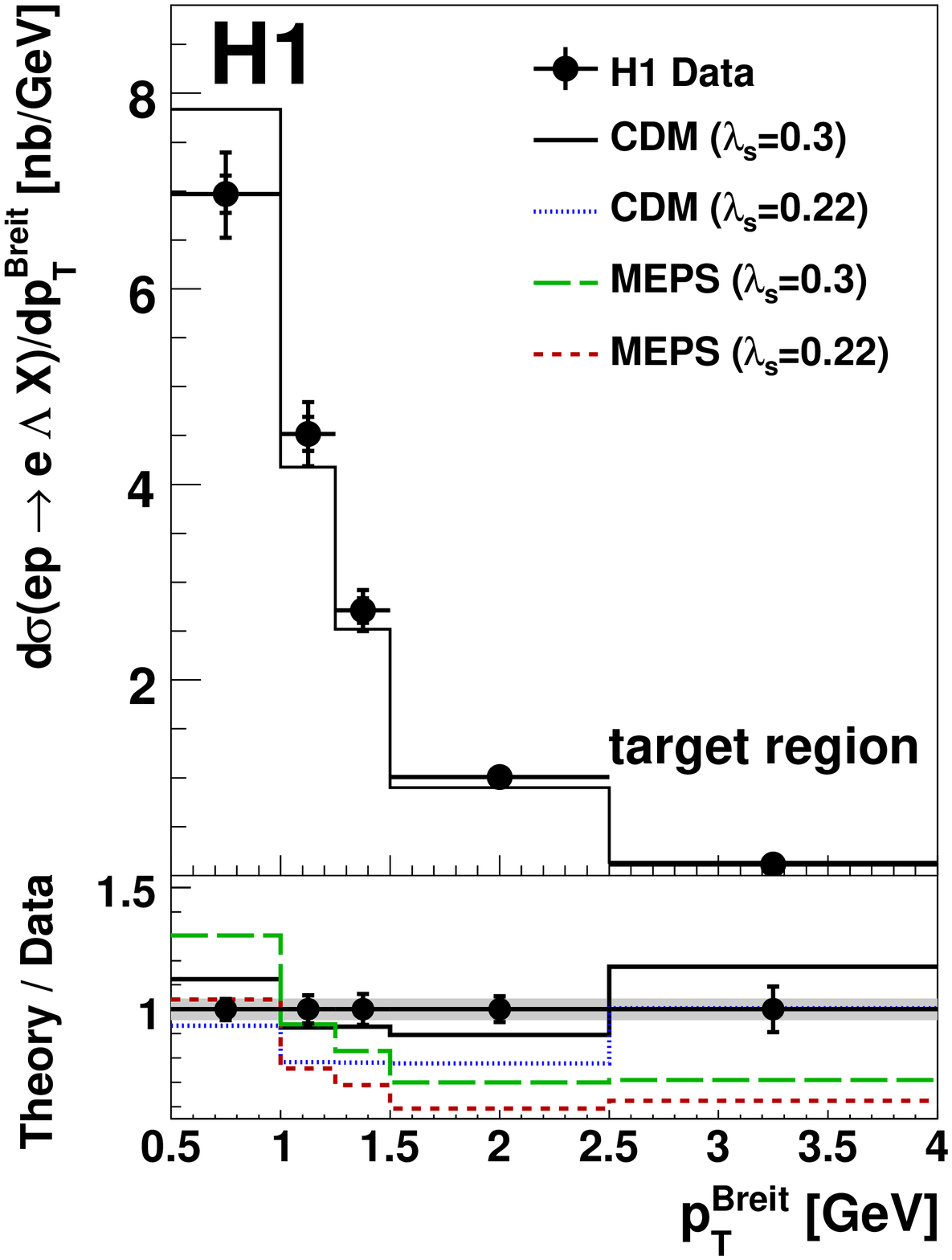}
\includegraphics[width=79mm]{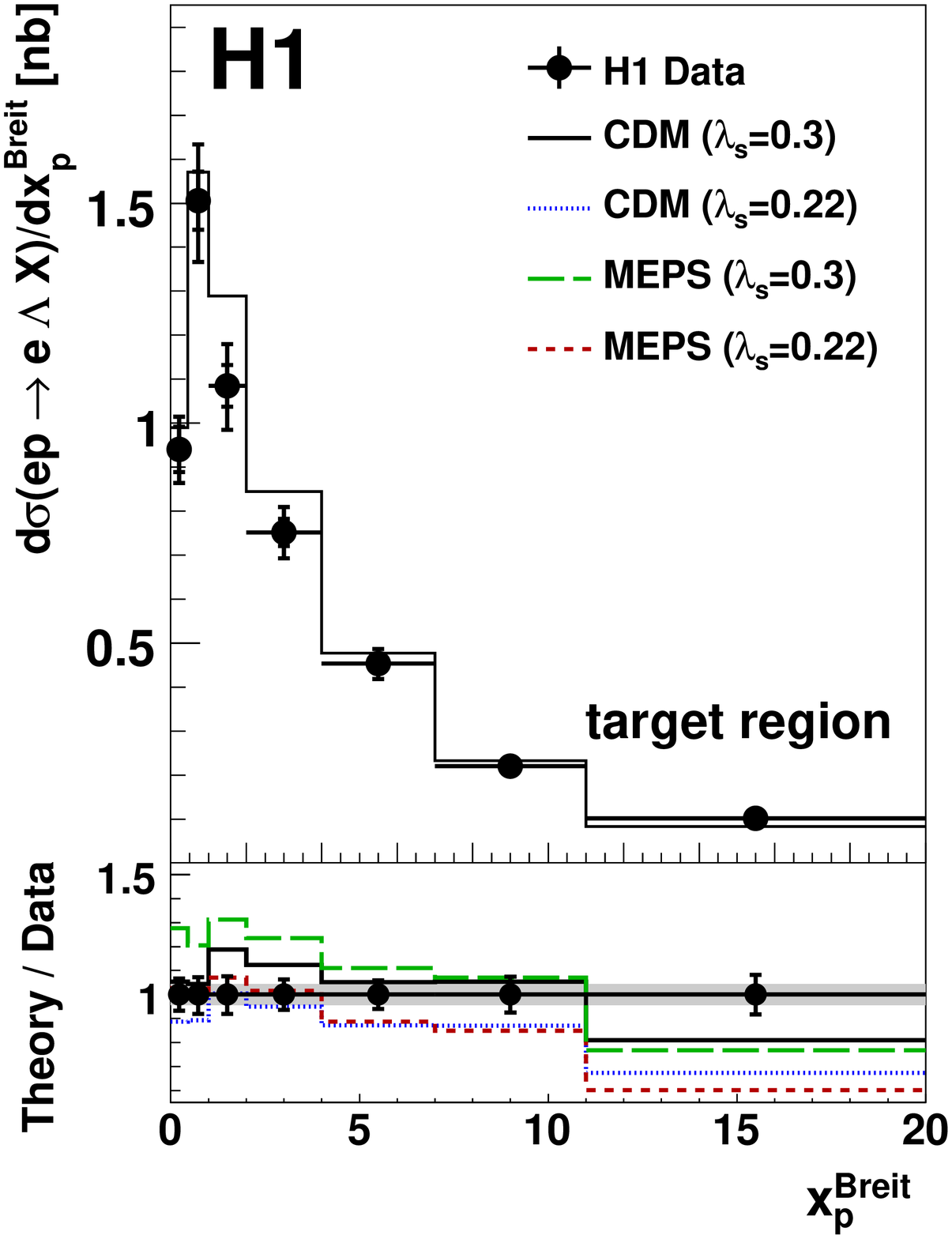}\\
\includegraphics[width=79mm]{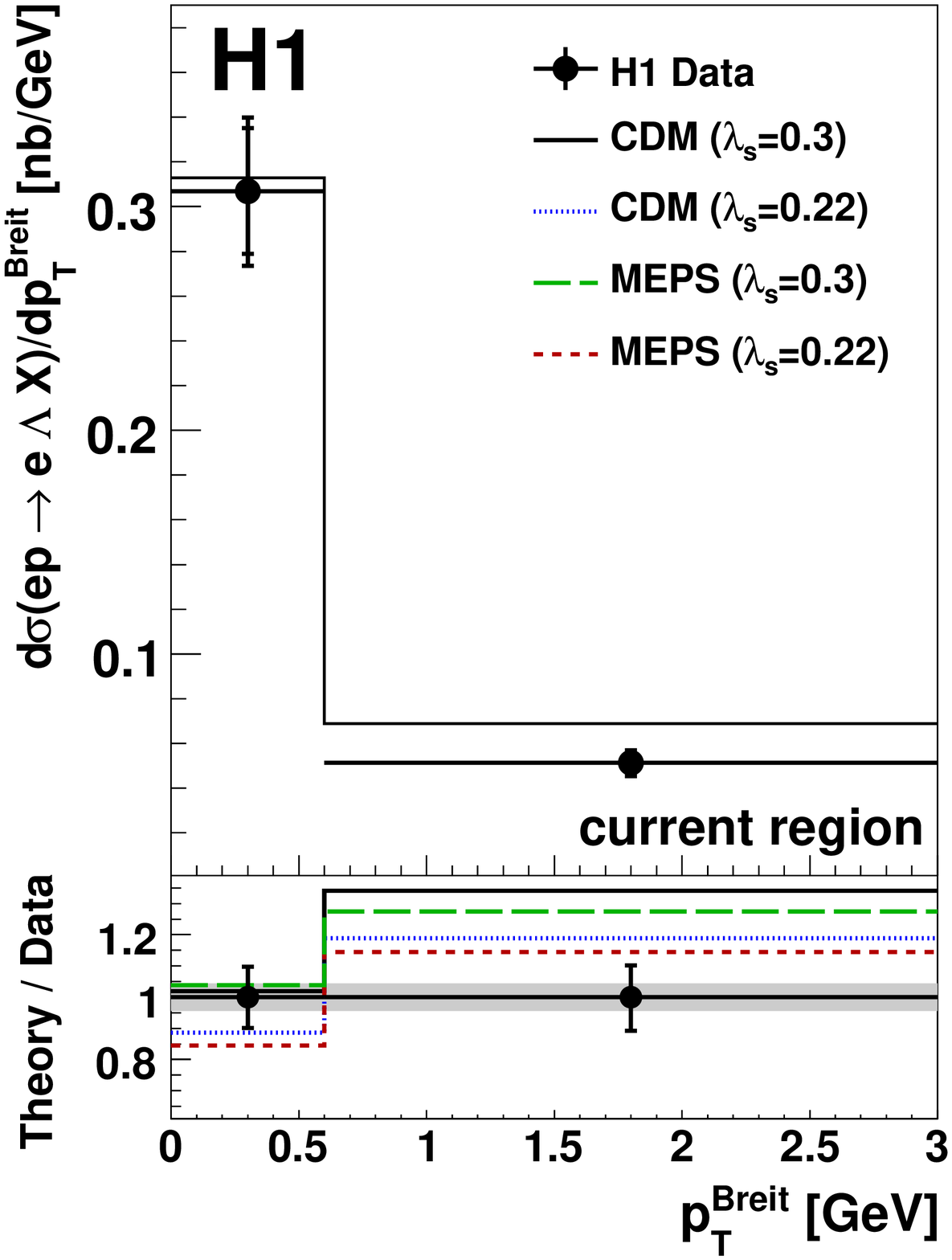}
\includegraphics[width=79mm]{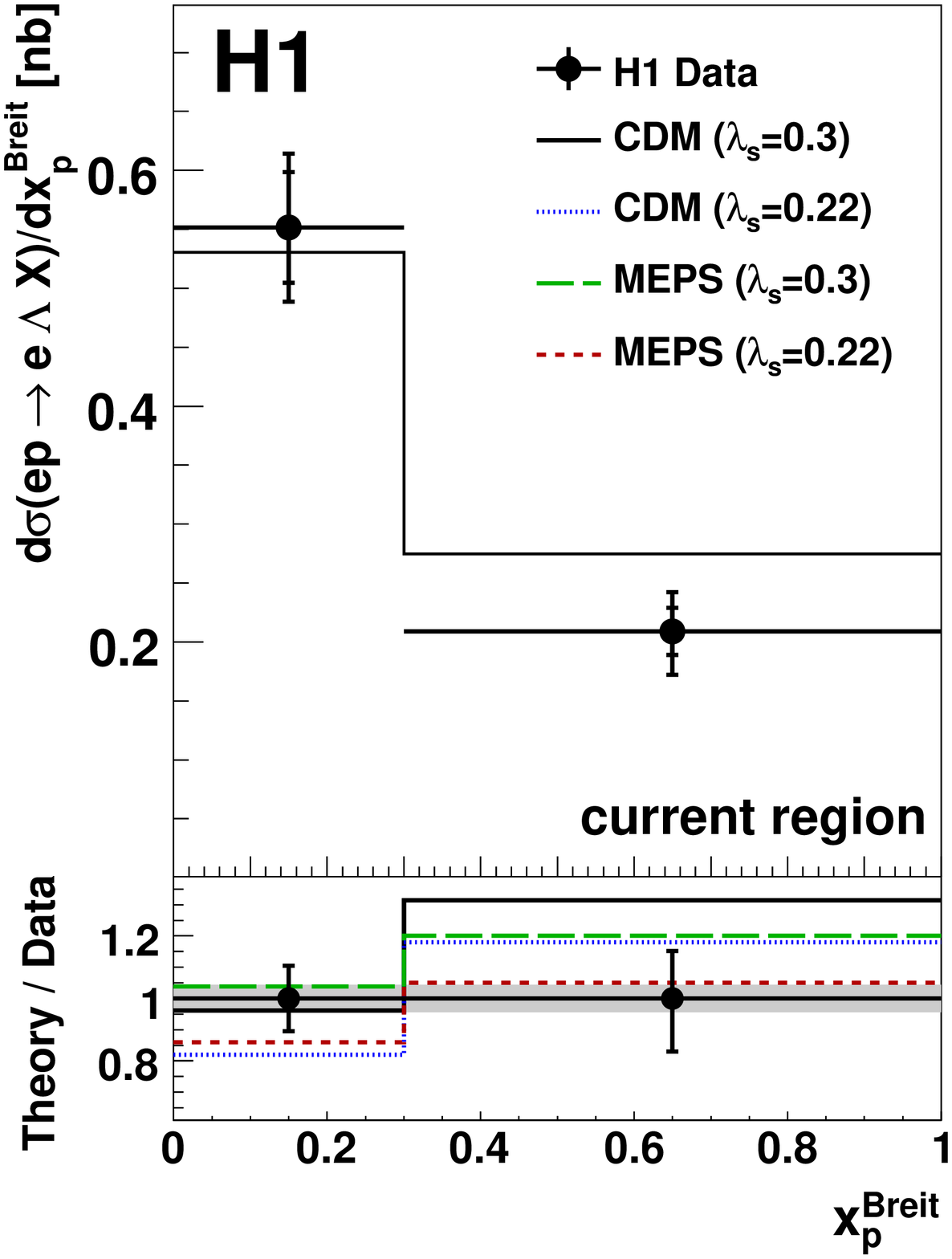}\\
\caption{The differential production cross sections for 
the \lsf\  baryons   
measured in the Breit frame as a function of \lsf\ transverse momentum
$p_T^{Breit}$ and  momentum fraction $x_p^{Breit}$ in the target 
hemisphere (a, b) and in the current hemisphere (c, d).
More details in the caption of figure~\ref{fig:ks-ds-lab}.
}
\label{fig:la-ds-breit}
\end{center}
\begin{picture}(0,0)
   \put(17,167){\bfseries a)}
   \put(97,167){\bfseries b)}
   \put(17,67){\bfseries c)}
   \put(97,67){\bfseries d)}
   \put(42,232){\LARGE e\,p $\rightarrow$ e\,$\Lambda$\,X (Breit frame)}
\end{picture}
\end{figure}

The measured differential cross sections of \ksf\ and  \lsf\ 
production are shown 
in figures~\ref{fig:ks-ds-lab} to~\ref{fig:la-ds-breit} and listed in 
tables~\ref{table:K0CrossSection} to~\ref{table:L0Breit}. 
The cross sections decrease  rapidly as a function of  $Q^2$ and $x$,
similarly to the inclusive DIS distributions. 
The cross sections are also observed to fall rapidly with $p_T$.

In the laboratory frame the overall features of the distributions are 
 reproduced by the DJANGOH simulations at the level of $10$ to $20\%$. 
For comparison, the CDM and MEPS model predictions are each
given with two values of the suppression factor
 \lambdas$ = 0.3$ and \lambdas$= 0.22$.
The predictions based on the CDM model with 
\lambdas $ = 0.3$  provide a reasonably good description of the data for
\ksf\ and \lsf\ production.  
The MEPS simulation  produces distributions, which are quite similar in shape 
to the CDM model predictions
but with a different normalisation in the case of \ksf\ production,
where a lower value of \lambdas $ = 0.22$ describes the data better.
In the case of \lsf\ production, both MEPS and CDM predictions are very similar in shape 
and normalisation 
and $\lambdas = 0.3$ provides a better description of the data.
For these comparisons, 
only the parameter  \lambdas\ is varied to describe the data.
However, in contrast to the \ksf, the  \lsf\  production 
cross sections also depend significantly on the
JETSET parameters that describe diquark and strange diquark creation. 



The cross sections measured as a function of $x_p^{Breit}$ and $p_T^{Breit}$ 
in the Breit frame are shown 
in figures~\ref{fig:ks-ds-breit} and~\ref{fig:la-ds-breit}
and listed in tables~\ref{table:K0Breit} and~\ref{table:L0Breit}, 
for both the target and the current region.
The cross section values in the target regions are about one order
of magnitude higher than in the current region.
They are generally  well described by both the 
MEPS and CDM model predictions. 
The predicted momentum distributions tend to be softer than in the data.
However, in the current region the sensitivity to \lambdas\ is clearly reduced
with respect to the laboratory frame or the target region. This
is due to both larger errors and an
increased fraction of strangeness produced in perturbative processes, which 
contributes up to about 50$\%$ (compared to about 25$\%$ in the target hemisphere).

\begin{figure}
\begin{center}
\includegraphics[width=79mm]{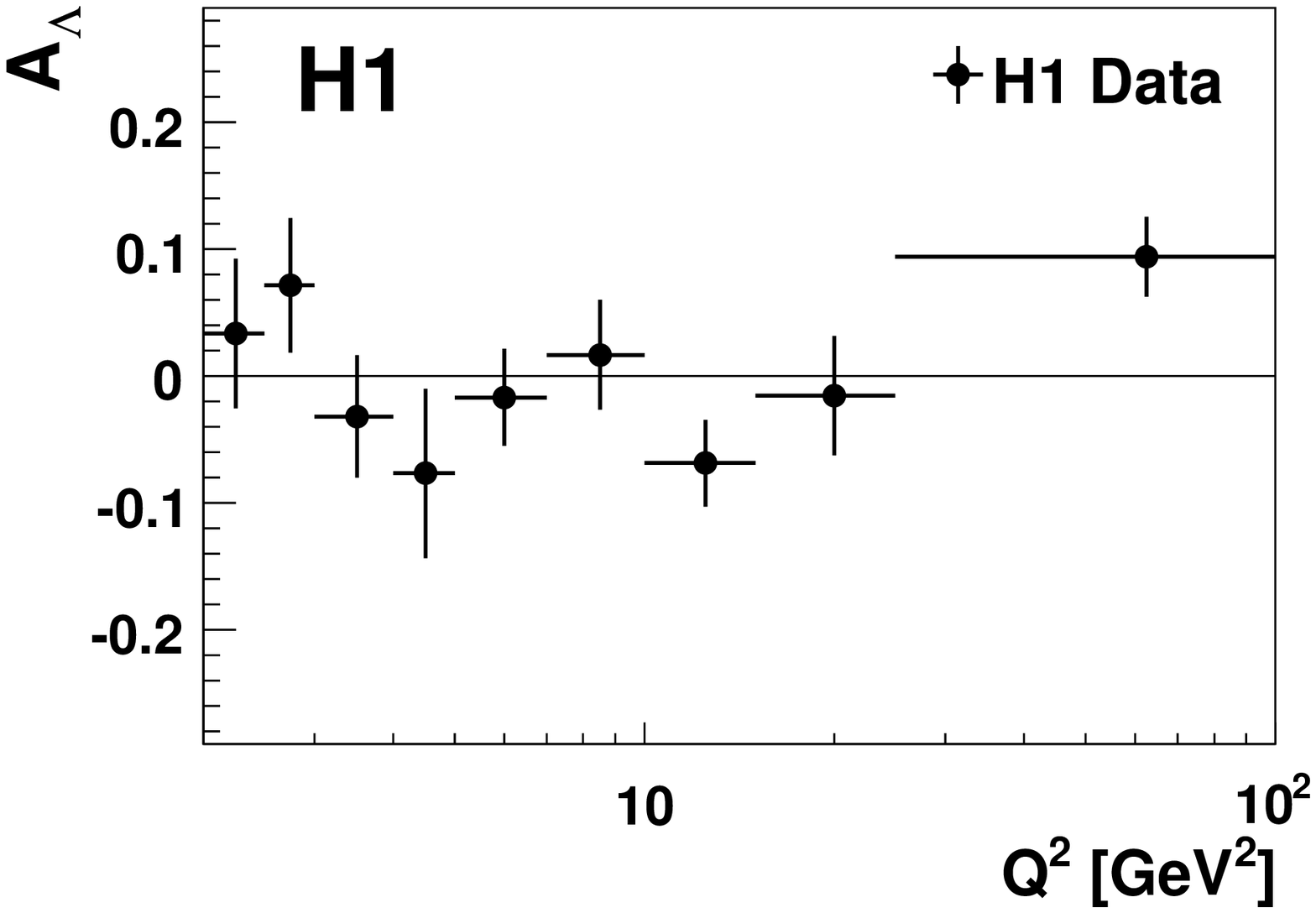}
\includegraphics[width=79mm]{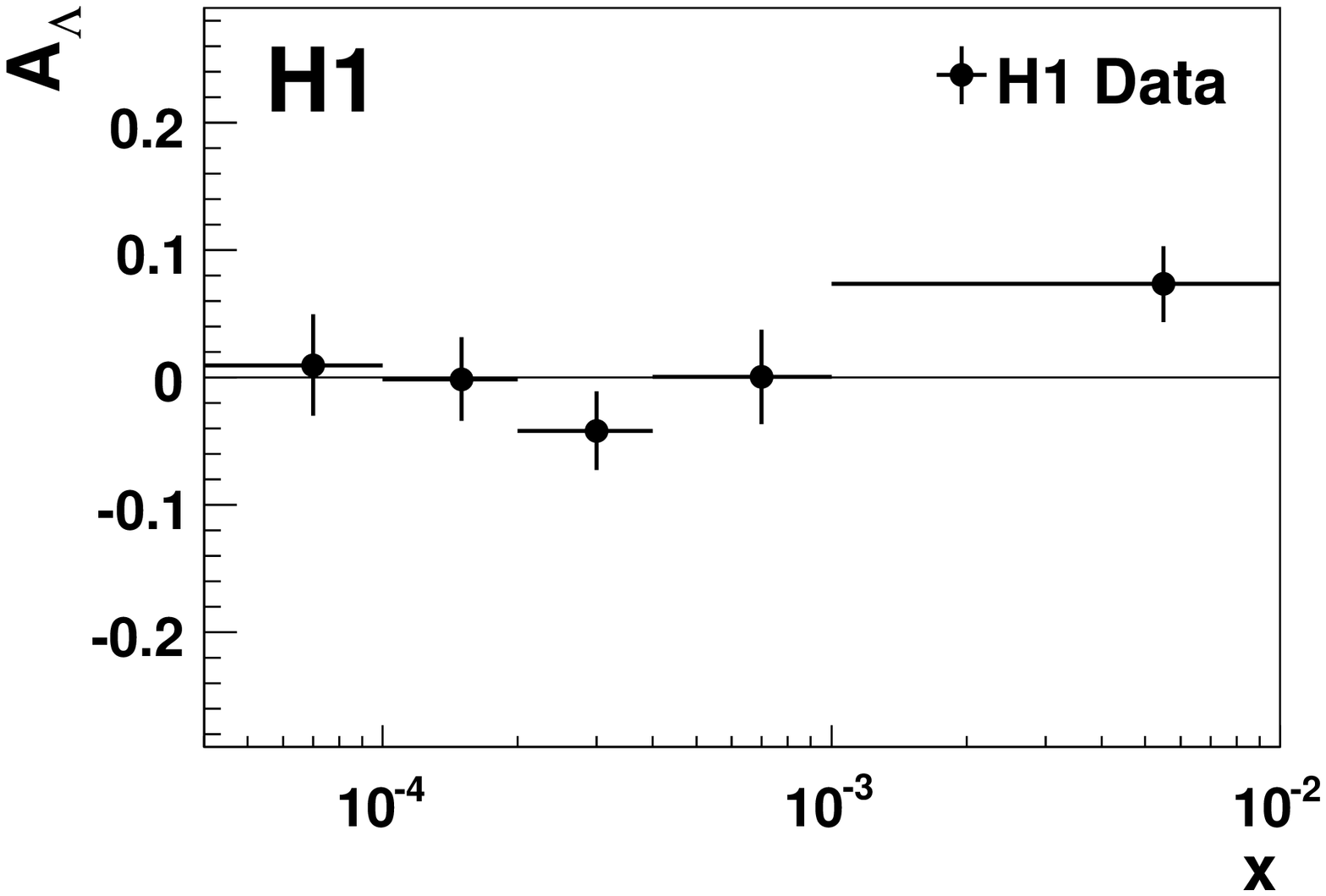}\\
\includegraphics[width=79mm]{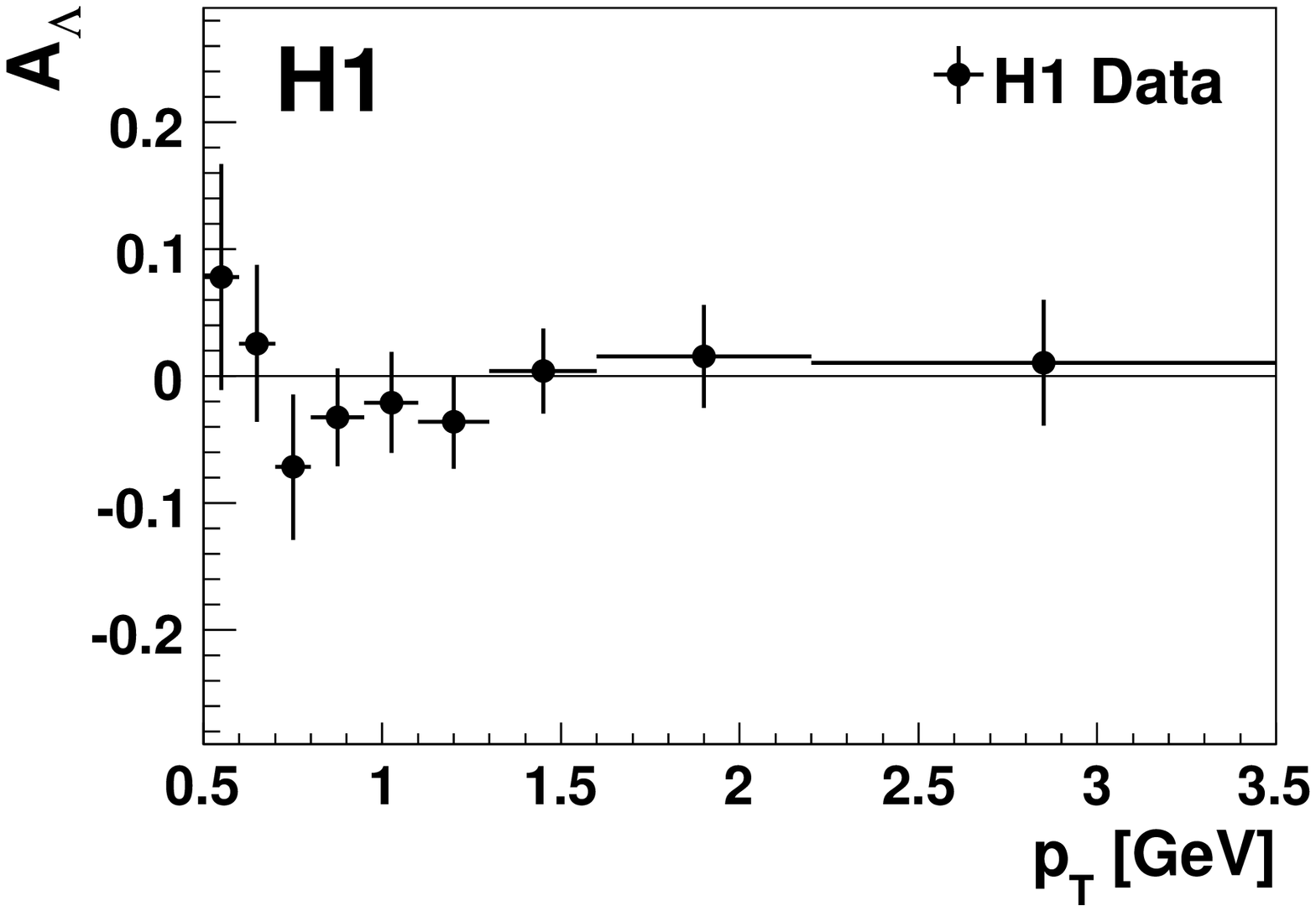}
\includegraphics[width=79mm]{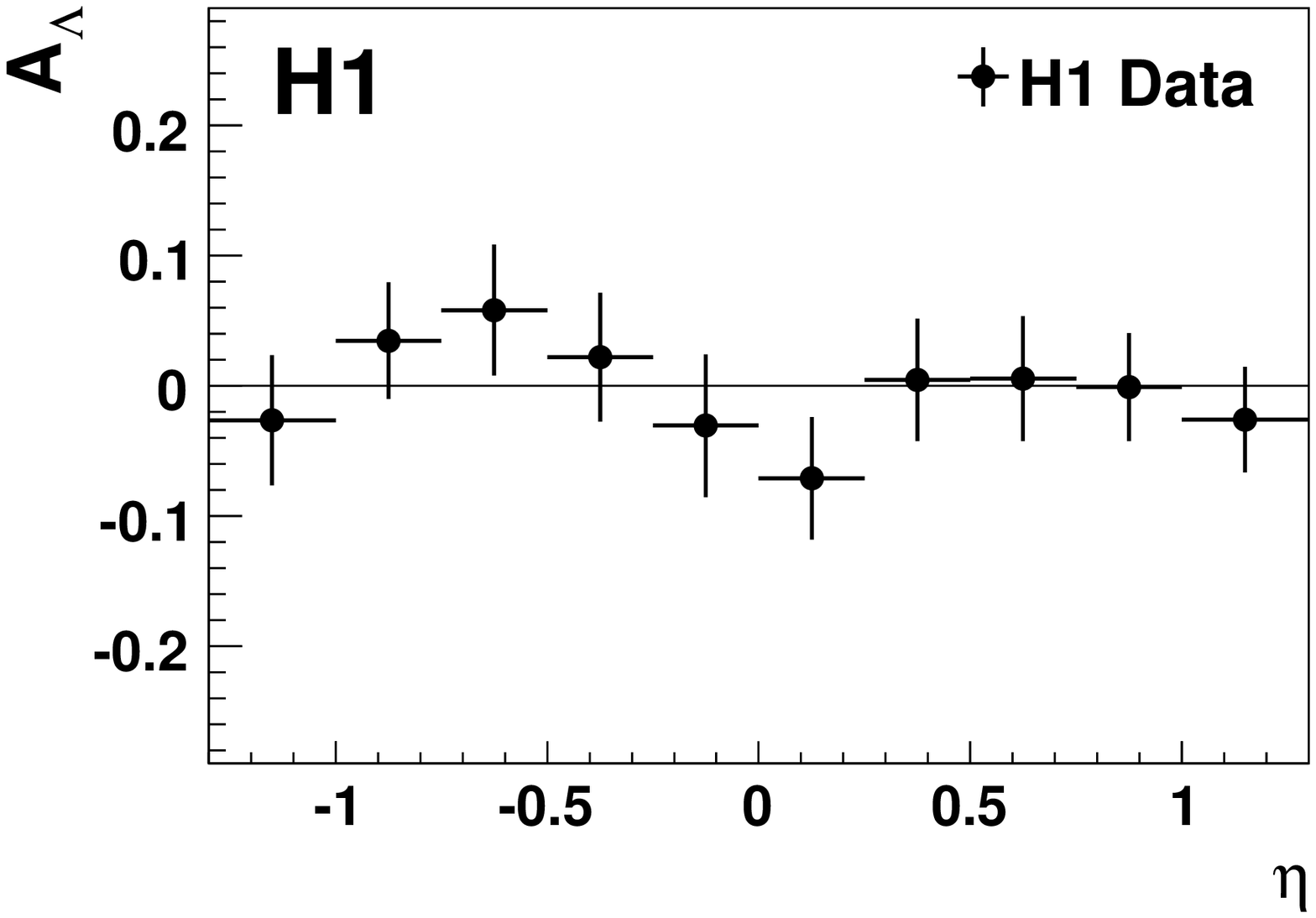}\\
\caption{The asymmetry $A_{\Lambda}$ of the differential 
production cross sections of the \lsf\ and \lsa\ baryons
in the laboratory frame as a function of the
a) photon virtuality squared \qsq,  
b) Bjorken scaling variable $x$, 
c) transverse momentum $p_T$
and  d) pseudorapidity $\eta$.
The asymmetry is defined as 
$A_{\Lambda} = [ \sigma_{vis}(ep \rightarrow e \lsf X) - 
   \sigma_{vis}(ep \rightarrow e \lsa  X) ] /
 [ \sigma_{vis}(ep \rightarrow e \lsf  X) + \sigma_{vis}(ep \rightarrow e \lsa X) ]$.
The error bars show the statistical uncertainty.
}
\label{fig:la-al-lab}
\end{center}
\begin{picture}(0,0)
   \put(17.5,105){\bfseries a)}
   \put(98,105){\bfseries b)}
   \put(17.5,51){\bfseries c)}
   \put(98,51){\bfseries d)}
   \put(62,148){\LARGE $\Lambda - \overline{\Lambda}$ Asymmetry}
\end{picture}
\end{figure}

\begin{figure}
\begin{center}
\includegraphics[width=79mm]{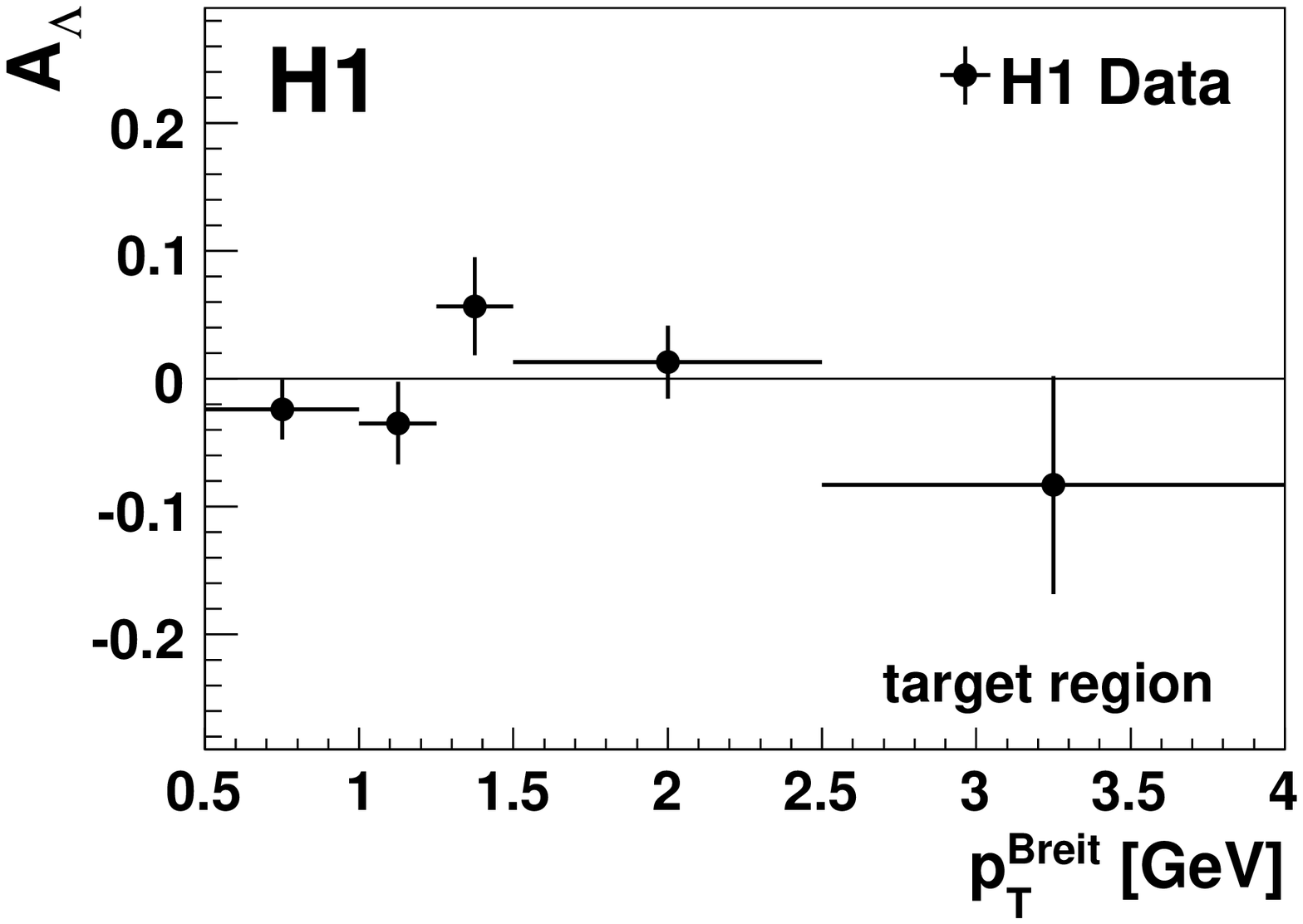}
\includegraphics[width=79mm]{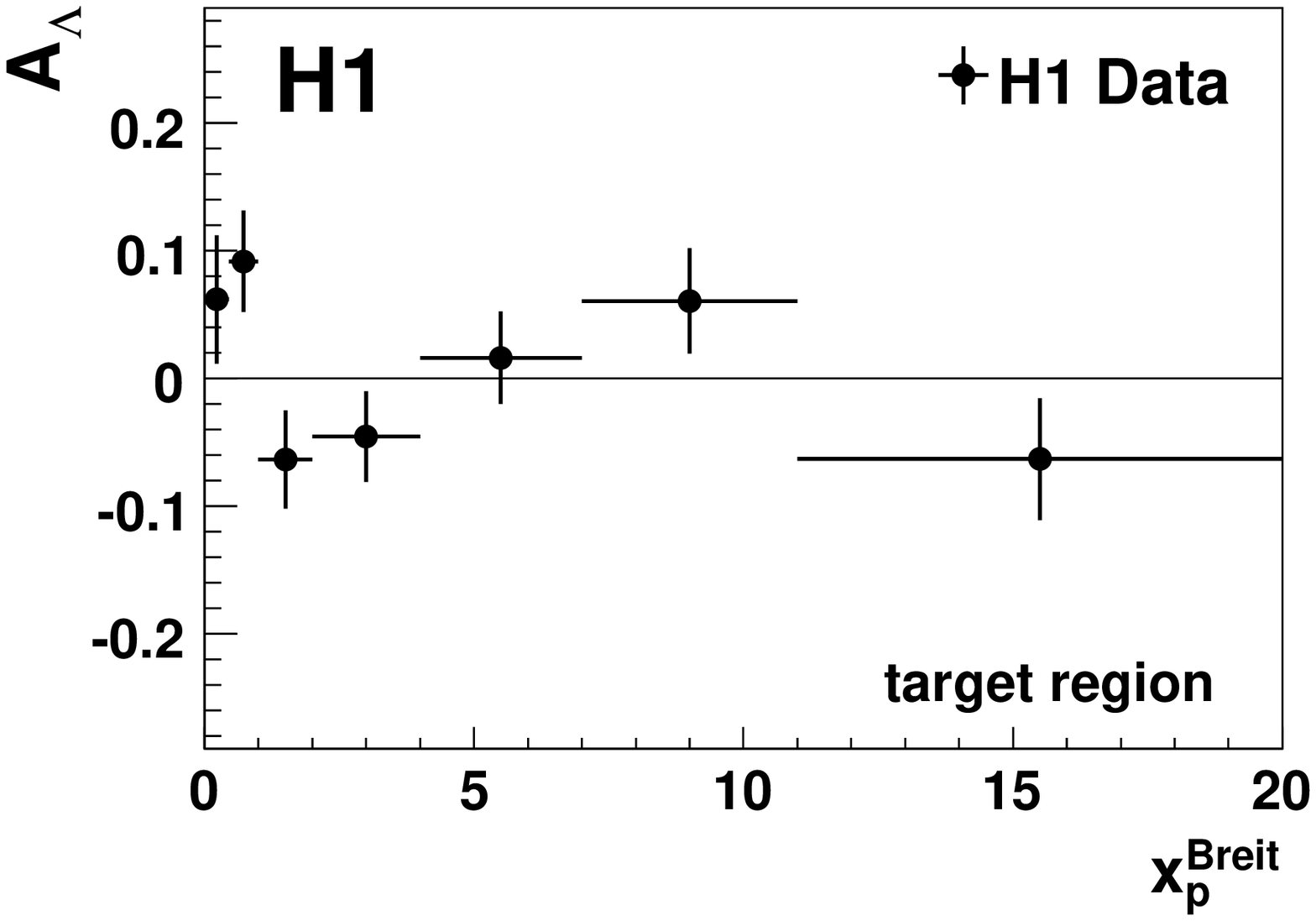}\\
\includegraphics[width=79mm]{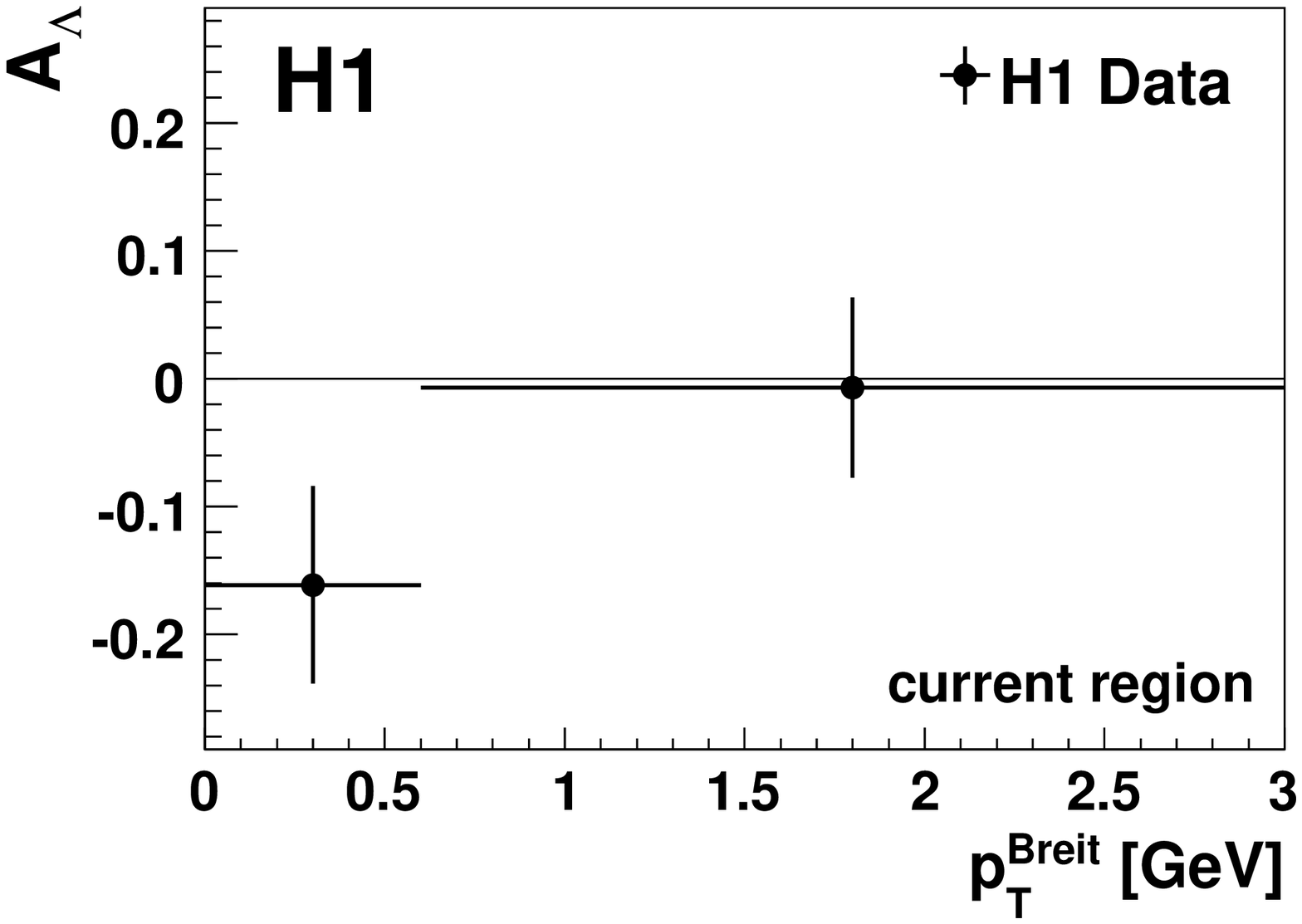}
\includegraphics[width=79mm]{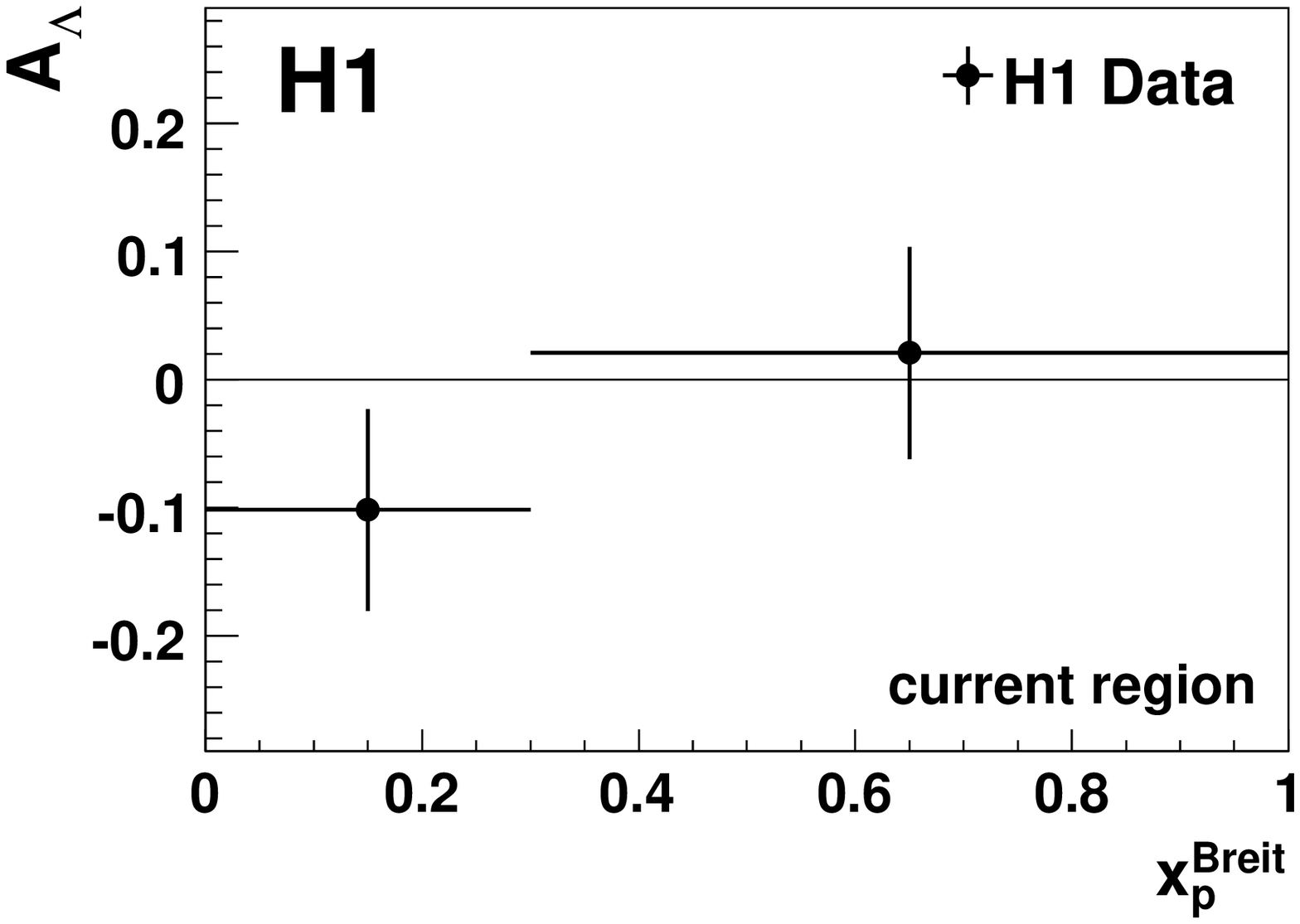}\\
\caption{The asymmetry $A_{\Lambda}$ of the differential 
production cross sections of the \lsf\ and \lsa\ baryons
measured in the Breit frame as a function of 
 transverse momentum  $p_T^{Breit}$ and
 momentum fraction  $x_p^{Breit}$ in the target hemisphere 
(a, b)
and in the current hemisphere (c, d).
The asymmetry is defined as 
$A_{\Lambda} = [ \sigma_{vis}(ep \rightarrow e \lsf X) - 
   \sigma_{vis}(ep \rightarrow e \lsa  X) ] /
 [ \sigma_{vis}(ep \rightarrow e \lsf  X) + \sigma_{vis}(ep \rightarrow e \lsa X) ]$.
The error bars show the statistical uncertainty.
}
\label{fig:la-al-breit}
\end{center}
\begin{picture}(0,0)
   \put(17.5,105){\bfseries a)}
   \put(98,105){\bfseries b)}
   \put(17.5,51){\bfseries c)}
   \put(98,51){\bfseries d)}
   \put(36,148){\LARGE $\Lambda - \overline{\Lambda}$ Asymmetry (Breit frame)}
\end{picture}
\end{figure}

%
To test the mechanism  of baryon number transfer, 
the asymmetry in the production of \lsf\ with respect to \lsa\ is
measured by the variable
\begin{equation}
A_{\Lambda} = \frac{ \sigma_{vis}(ep \rightarrow e \lsf X) - 
   \sigma_{vis}(ep \rightarrow e \lsa  X) }{
  \sigma_{vis}(ep \rightarrow e \lsf  X) + \sigma_{vis}(ep \rightarrow e \lsa X) }.
\end{equation}
A significant \lsf\ - \lsa\ asymmetry $A_{\Lambda}\neq 0$ would indicate a  transfer of the baryon 
number from the proton beam to the final state strange particles.
The measured distributions of $A_{\Lambda}$ in the laboratory and Breit frames are shown in figures~\ref{fig:la-al-lab}
and \ref{fig:la-al-breit}, respectively.
All distributions are observed to be compatible with zero within
errors.  
Thus, no evidence of baryon number transfer is visible in 
the measured \lsf/\lsa\ data.

\begin{figure}
\begin{center}
\includegraphics[width=79mm]{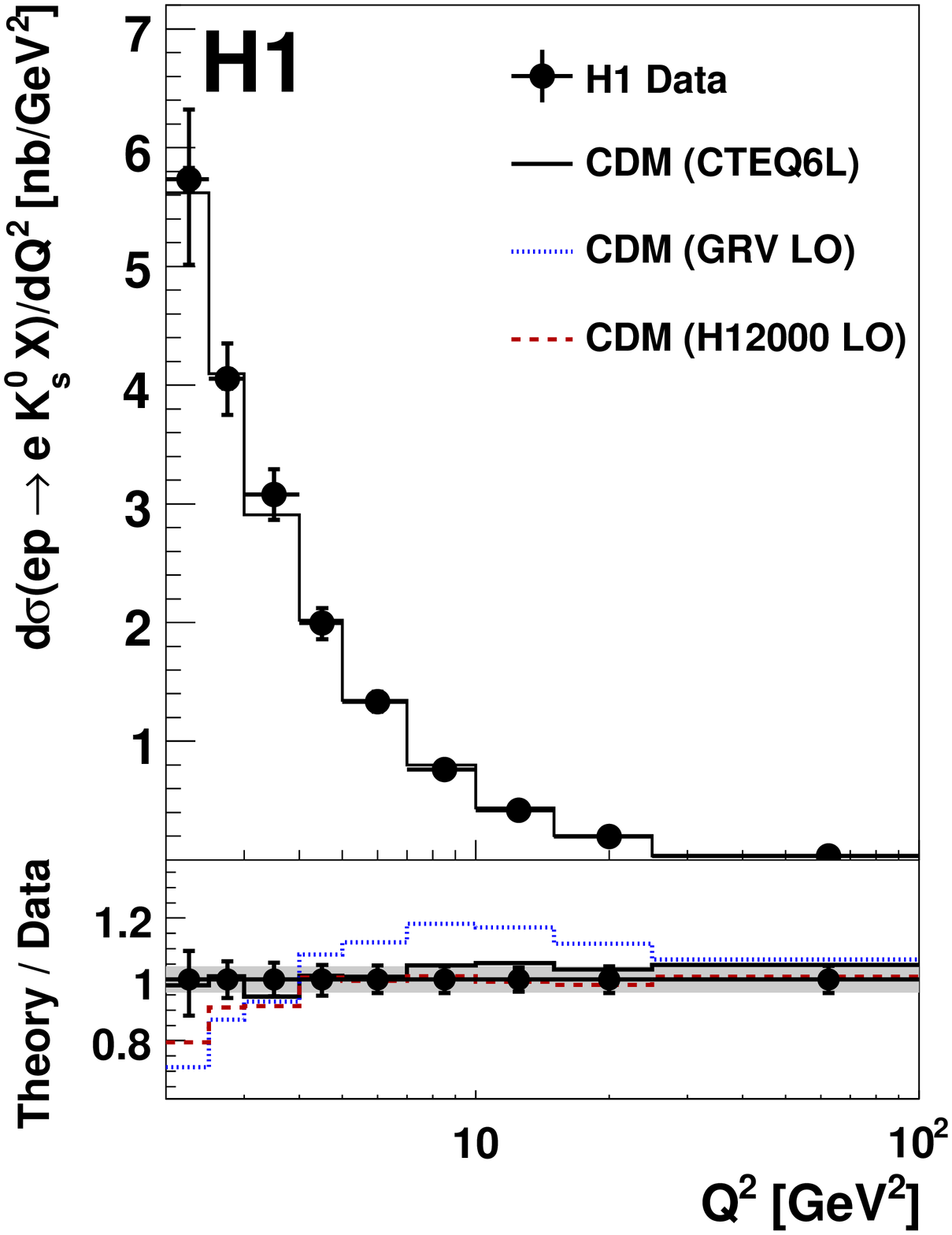}
\includegraphics[width=79mm]{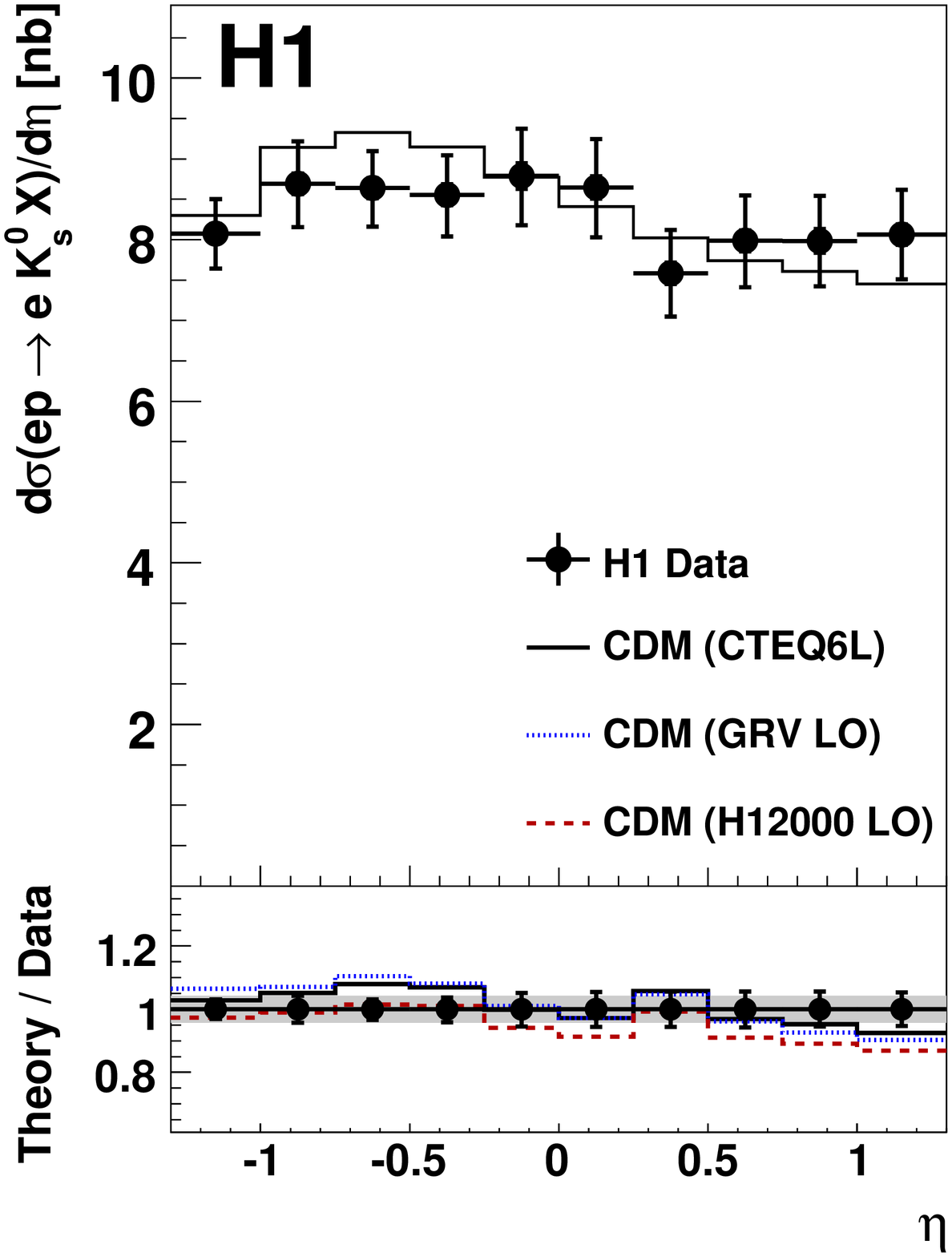}\\
\vspace{1cm}
\includegraphics[width=79mm]{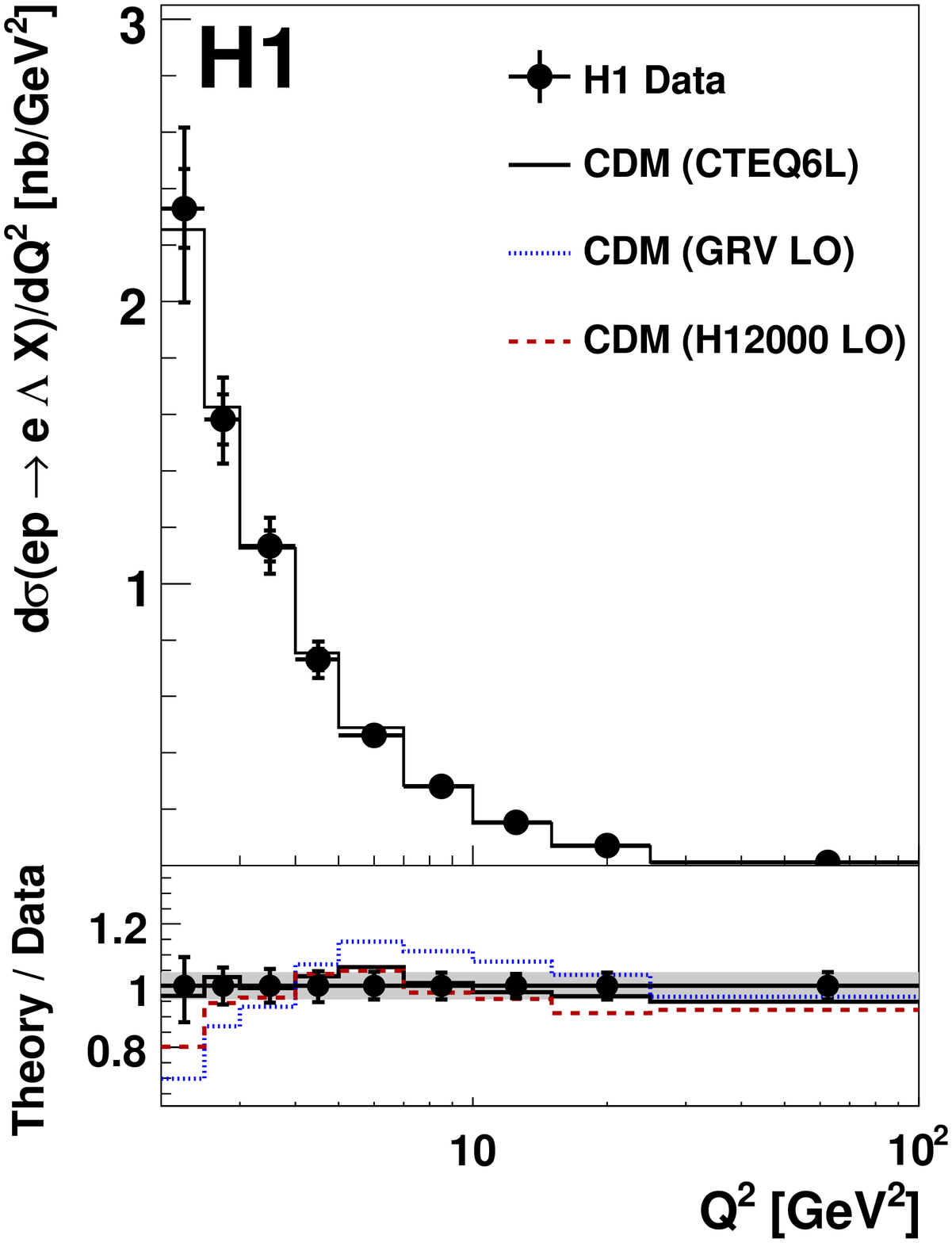}
\includegraphics[width=79mm]{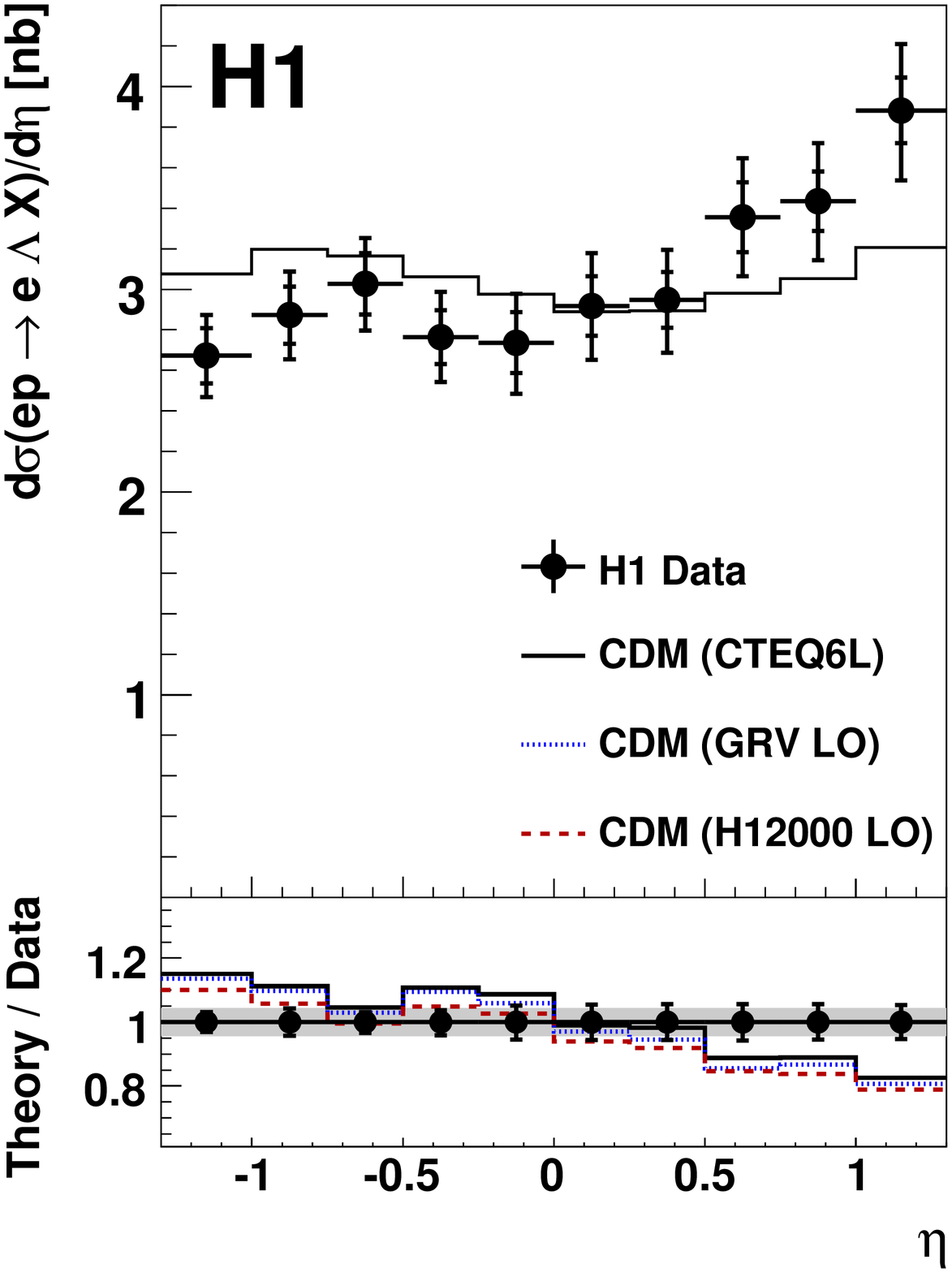}\\
\caption{The differential production cross sections  
 in the laboratory frame as a function of the
event variable \qsq and pseudorapidity $\eta$ for the \ksf\ (a, b) and \lsf\ (c, d).
Overlaid are CDM predictions for $\lambdas=0.286$ 
using three different proton PDFs: 
CTEQ6L, GRV-94 (LO) and H1 2000 LO.
More details in the caption of figure~\ref{fig:ks-ds-lab}.
}
\label{fig:ks-la-pdf}
\end{center}
\begin{picture}(0,0)
   \put(17,176){\bfseries a)}
   \put(97,176){\bfseries b)}
   \put(17,66){\bfseries c)}
   \put(97,66){\bfseries d)}
   \put(62,242){\LARGE e\,p $\rightarrow$ e\,K$^0_s$\,X}
   \put(62,133){\LARGE e\,p $\rightarrow$ e\,$\Lambda$\,X}
\end{picture}
\end{figure}

%
%
In order to test for possible dependencies of strange hadron production
on the proton parton density functions,
the measured distributions are compared with different
PDF parametrisations.
Figure~\ref{fig:ks-la-pdf} shows
the differential cross sections for \ksf\ and \lsf\ production
compared with the CDM predictions using the 
CTEQ6L~\cite{CTEQ6}, H12000LO~\cite{Adloff:2003uh} 
and GRV LO~\cite{Gluck:1994uf} parametrisations and 
$\lambda_s=0.286$.
The predictions of the \qsq\-dependence of the cross section are notably different for different
PDFs for both the 
\ksf\ and the \lsf. The $p_T$ distributions indicate only a slight 
dependence while the $\eta$ distributions do not exhibit any PDF dependence.
The small discrepancy in the forward direction is not resolved by
different PDF parametrisations. Similar results are obtained in the Breit frame.
%

\subsubsection{Ratios of Production Cross Sections}

\begin{figure}
\begin{center}
\includegraphics[width=79mm]{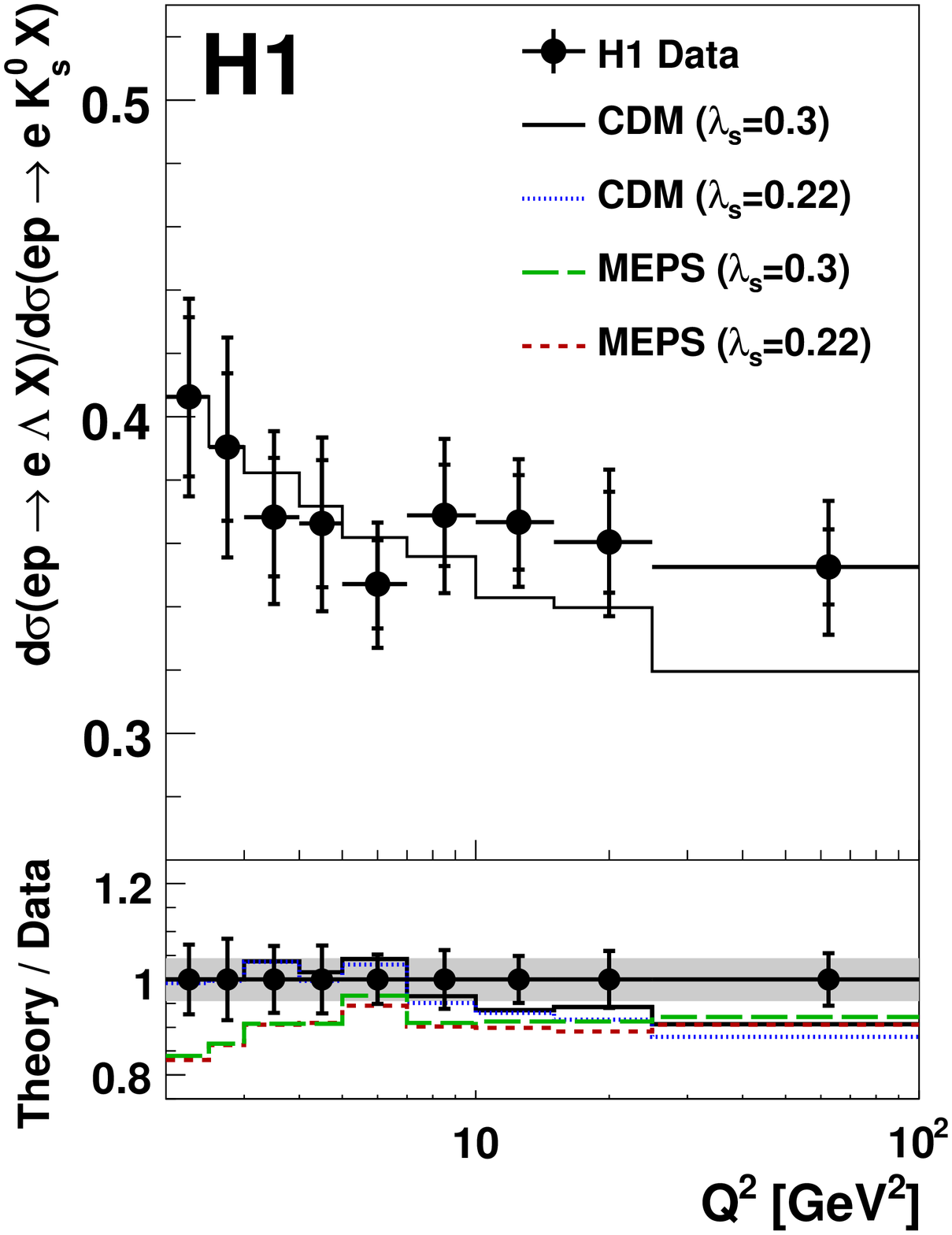}
\includegraphics[width=79mm]{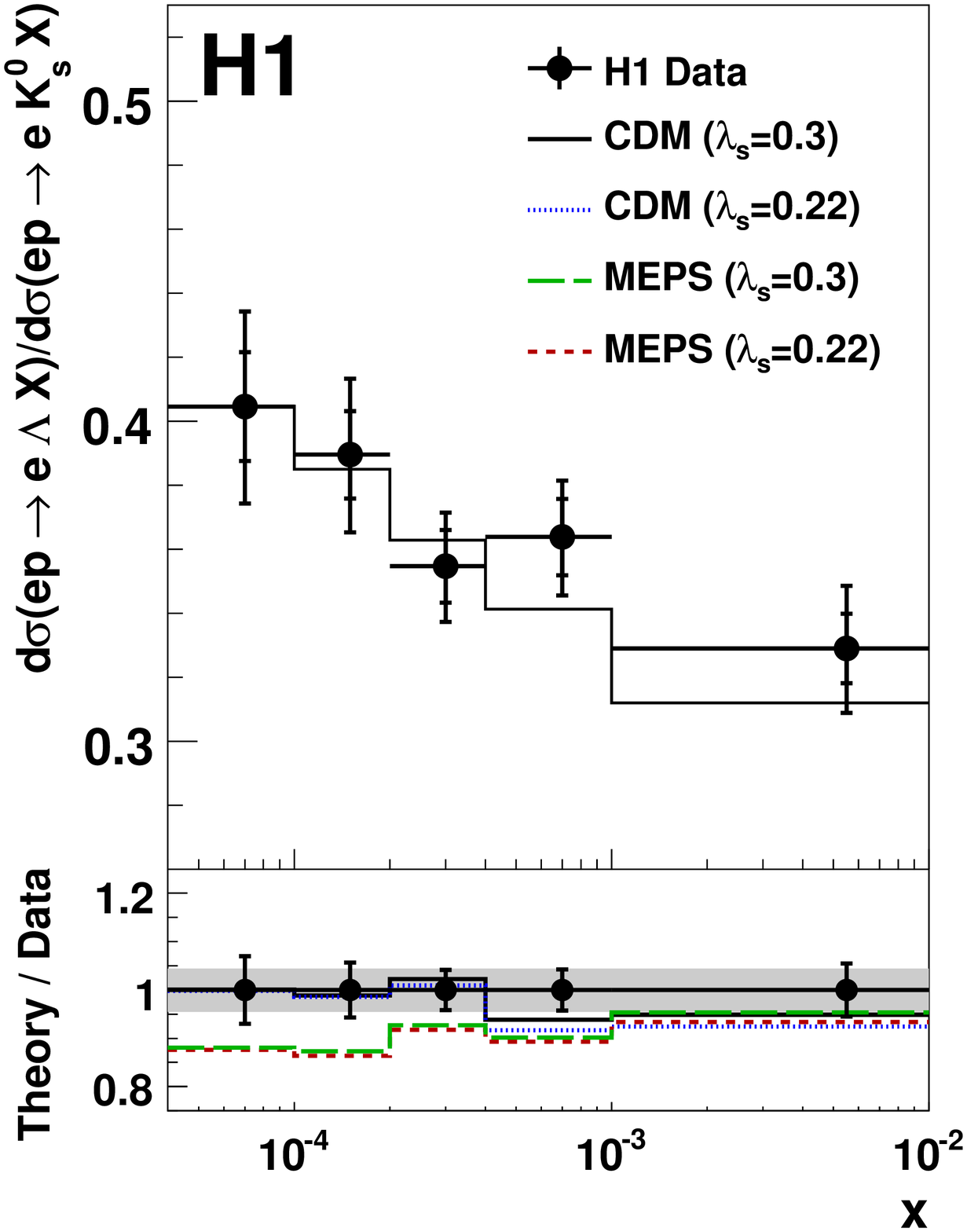}\\
\includegraphics[width=79mm]{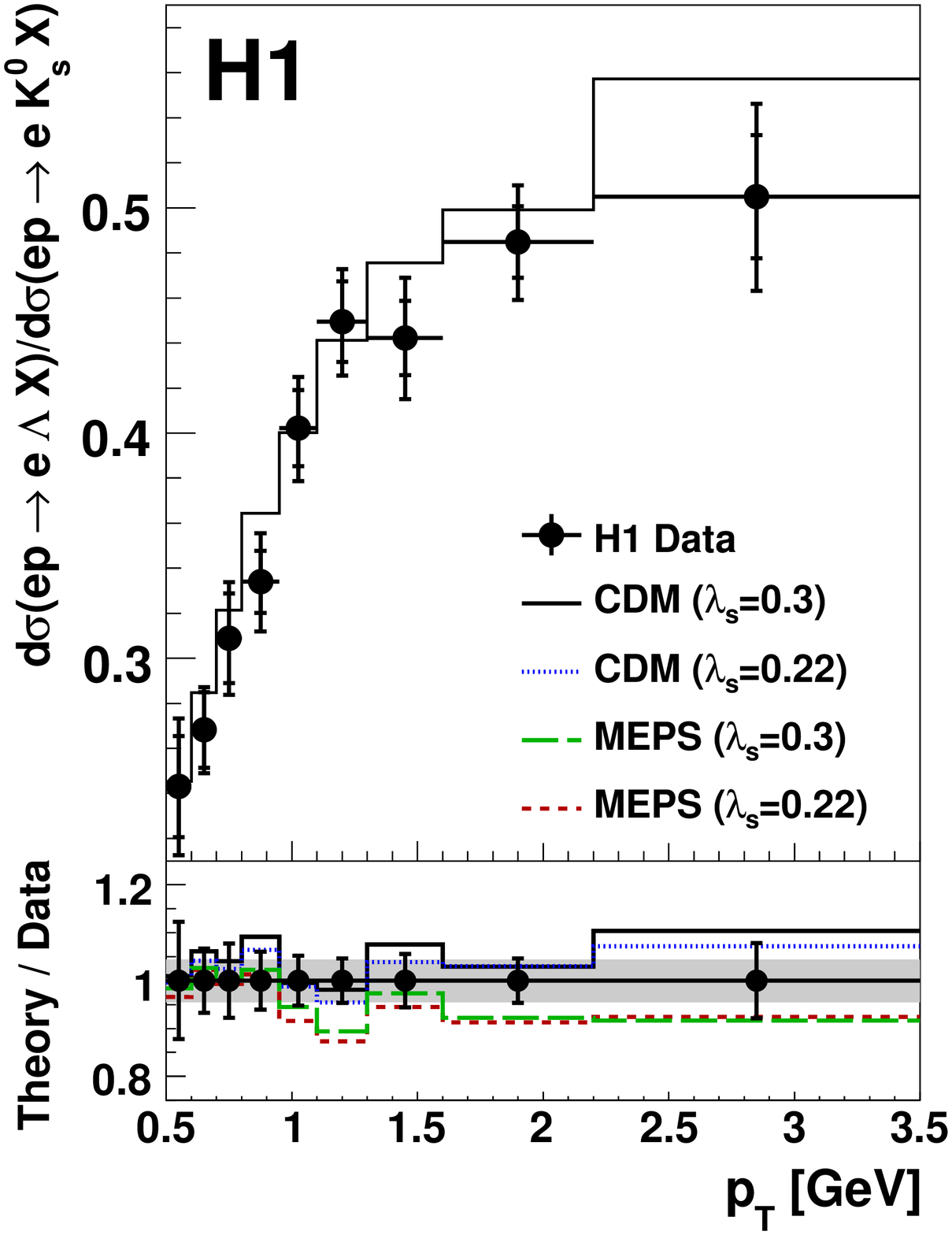}
\includegraphics[width=79mm]{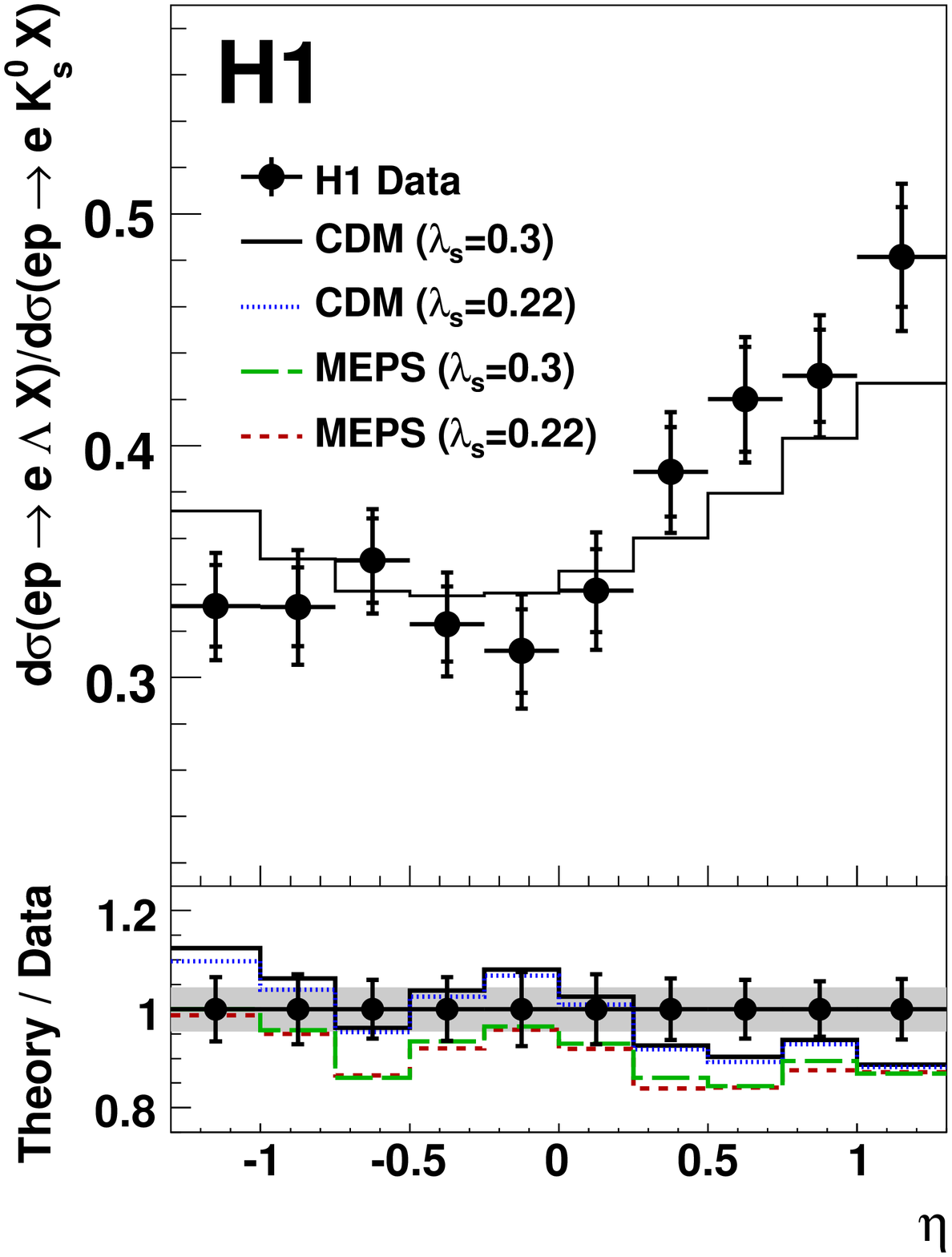}\\
\caption{The ratio of the differential production cross sections for 
\lsf\ baryons and \ksf\ mesons
in the laboratory frame as a function of the
a) photon virtuality squared \qsq,
b) Bjorken scaling variable $x$,
c) transverse momentum $p_T$
and d) pseudorapidity $\eta$. 
More details in the caption of figure~\ref{fig:ks-ds-lab}.
}
\label{fig:la-ks-ratio-lab}
\end{center}
\begin{picture}(0,0)
   \put(17,166){\bfseries a)}
   \put(97,166){\bfseries b)}
   \put(17,66){\bfseries c)}
   \put(97,66){\bfseries d)}
   \put(40,232){\LARGE e\,p $\rightarrow$ e\,$\Lambda$\,X / e\,p $\rightarrow$ e\,K$^0_s$\,X}
\end{picture}
\end{figure}

\begin{figure}
\begin{center}
\includegraphics[width=79mm]{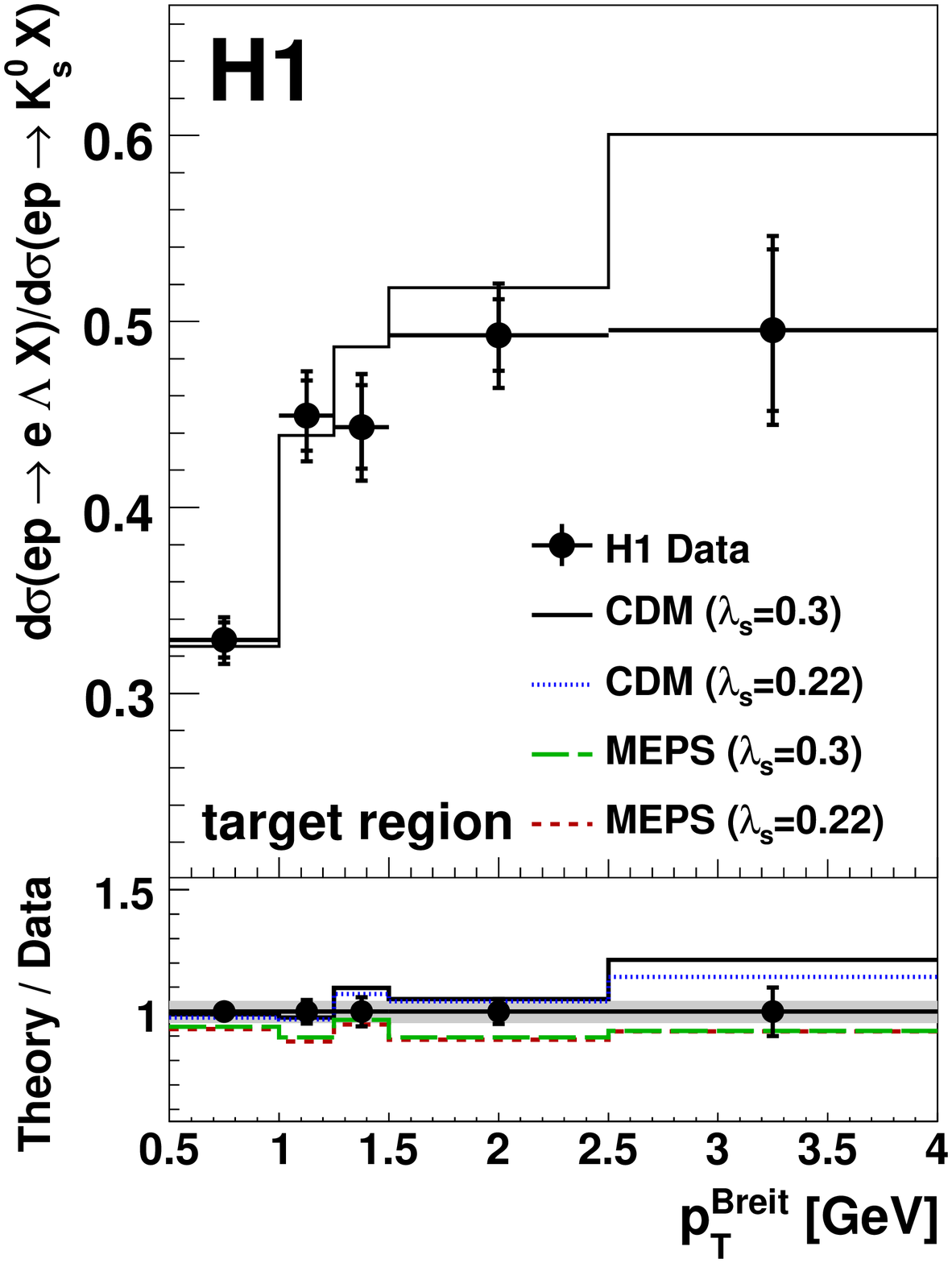}
\includegraphics[width=79mm]{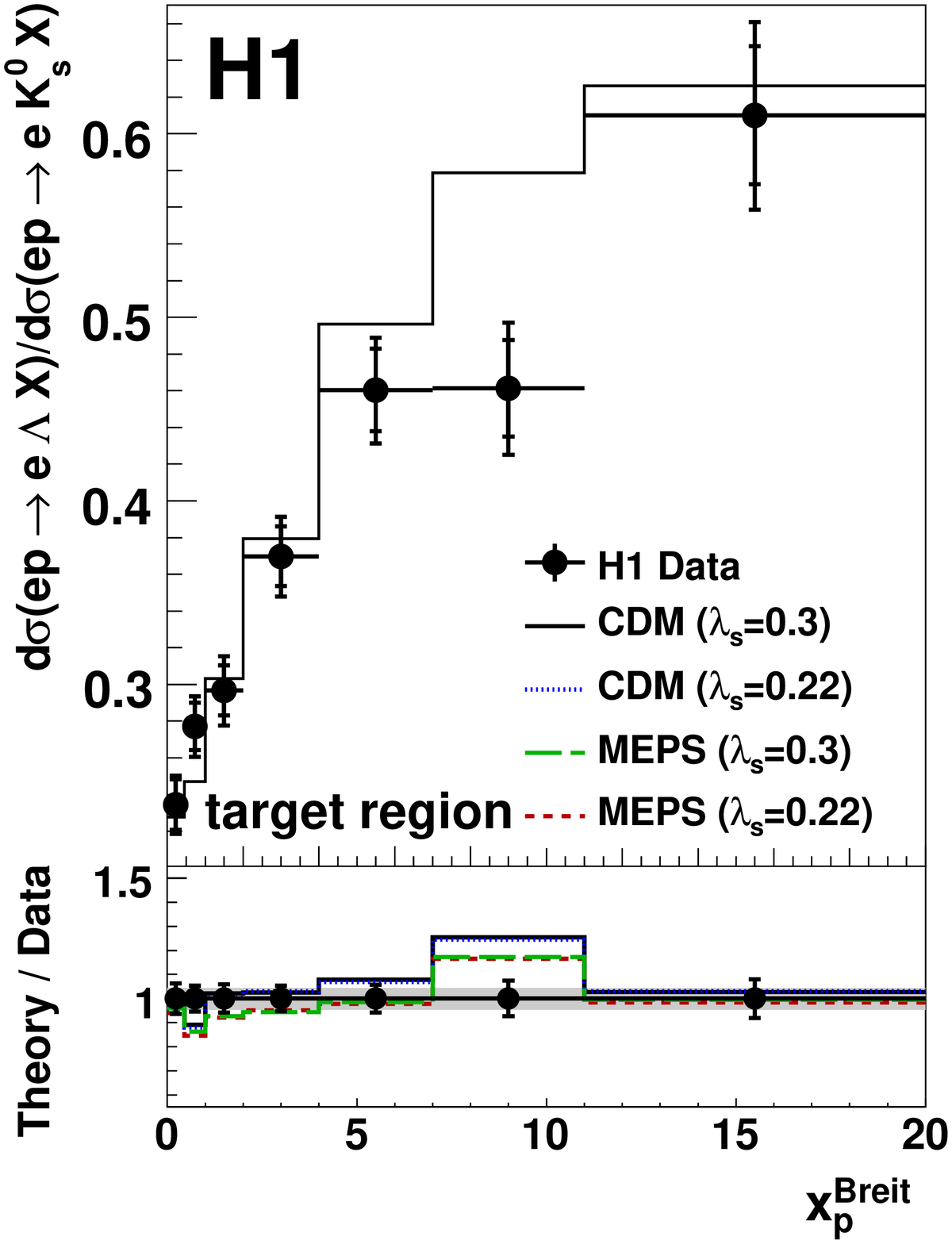}\\
\includegraphics[width=79mm]{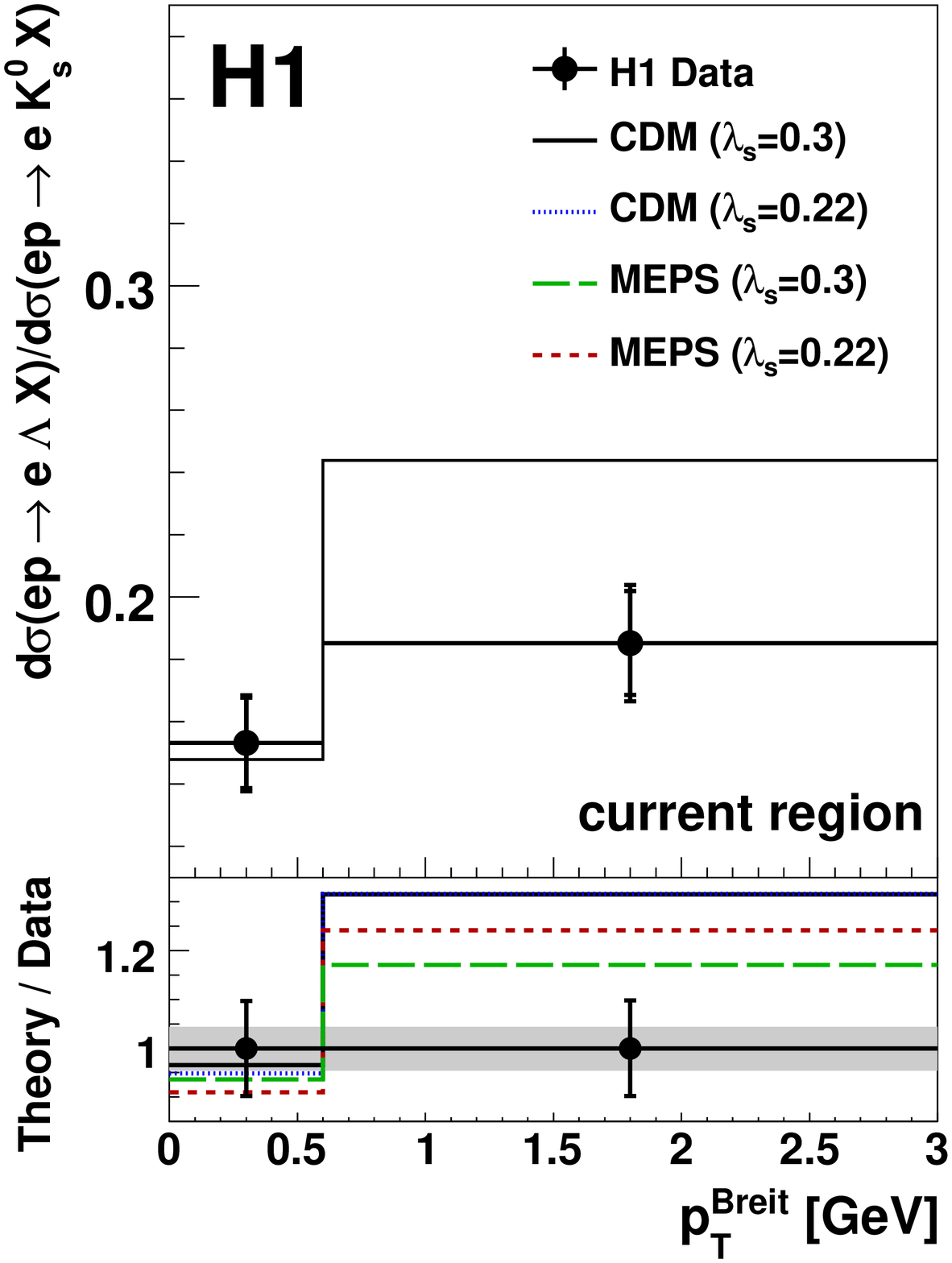}
\includegraphics[width=79mm]{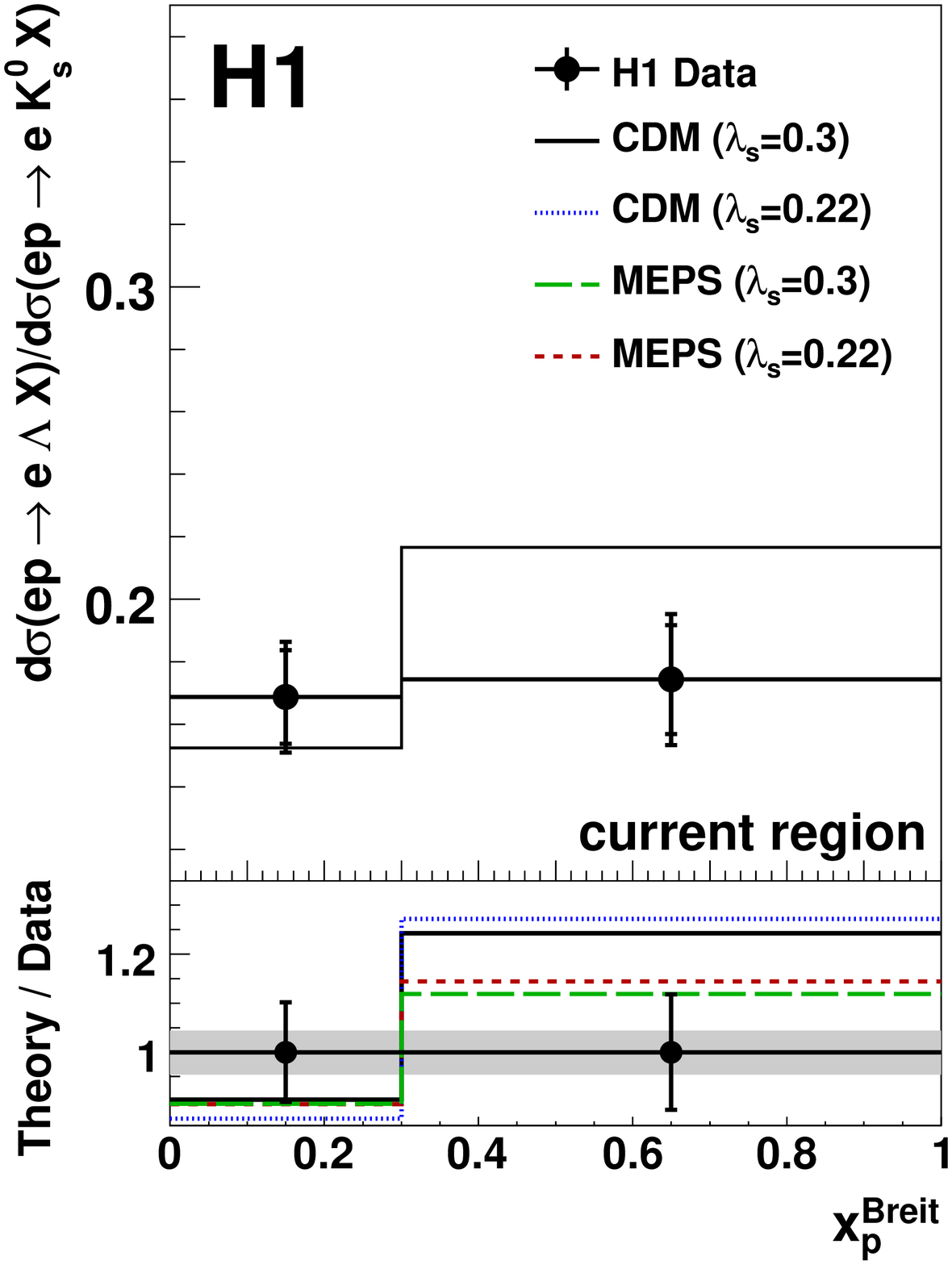}\\
\caption{The ratio of the differential production cross sections for 
\lsf\ baryons and \ksf\ mesons
in the Breit frame as a function of transverse momentum
$p_T^{Breit}$ and momentum fraction $x_p^{Breit}$ in the target 
hemisphere (a, b) and in the current hemisphere (c, d).
More details in the caption of figure~\ref{fig:ks-ds-lab}.
}
\label{fig:la-ks-ratio-breit}
\end{center}
\begin{picture}(0,0)
   \put(17,217){\bfseries a)}
   \put(97,217){\bfseries b)}
   \put(17,117){\bfseries c)}
   \put(97,117){\bfseries d)}
   \put(20,232){\LARGE e\,p $\rightarrow$ e\,$\Lambda$\,X / e\,p $\rightarrow$ e\,K$^0_s$\,X (Breit frame)}
\end{picture}
\end{figure}

Different aspects of baryon production within the
fragmentation models can be tested with reduced theoretical 
uncertainties by studying the ratio of the differential cross sections 
for \lsf\ baryons and \ksf\ mesons
$R(\lsf/\ksf) = {d\sigma(ep \rightarrow e \lsf  X)}$ / ${d\sigma(ep \rightarrow e \ksf X})$.
The measurements
are shown in figures~\ref{fig:la-ks-ratio-lab} and~\ref{fig:la-ks-ratio-breit} 
and listed in tables~\ref{table:RLKlab} and~\ref{table:RLKBreit}.
The CDM implementation provides a reasonably good description of the data
in the laboratory frame (figure~\ref{fig:la-ks-ratio-lab}),
although systematic deviations are seen at high \qsq and 
in the shape of the $\eta$ distribution, whereas
the MEPS predictions clearly underestimate the data.
The model predictions are not sensitive to \lambdas, as expected. 

The dependence of $R(\lsf/\ksf)$ on $p_T^{Breit}$ and $x_p^{Breit}$ 
(figure~\ref{fig:la-ks-ratio-breit})
are reasonably well described in both the target and current hemispheres.  
The predictions are 
almost independent of the model implementation (CDM, MEPS)
and the \lambdas\ values used. 

\begin{figure}
\begin{center}
\includegraphics[width=79mm]{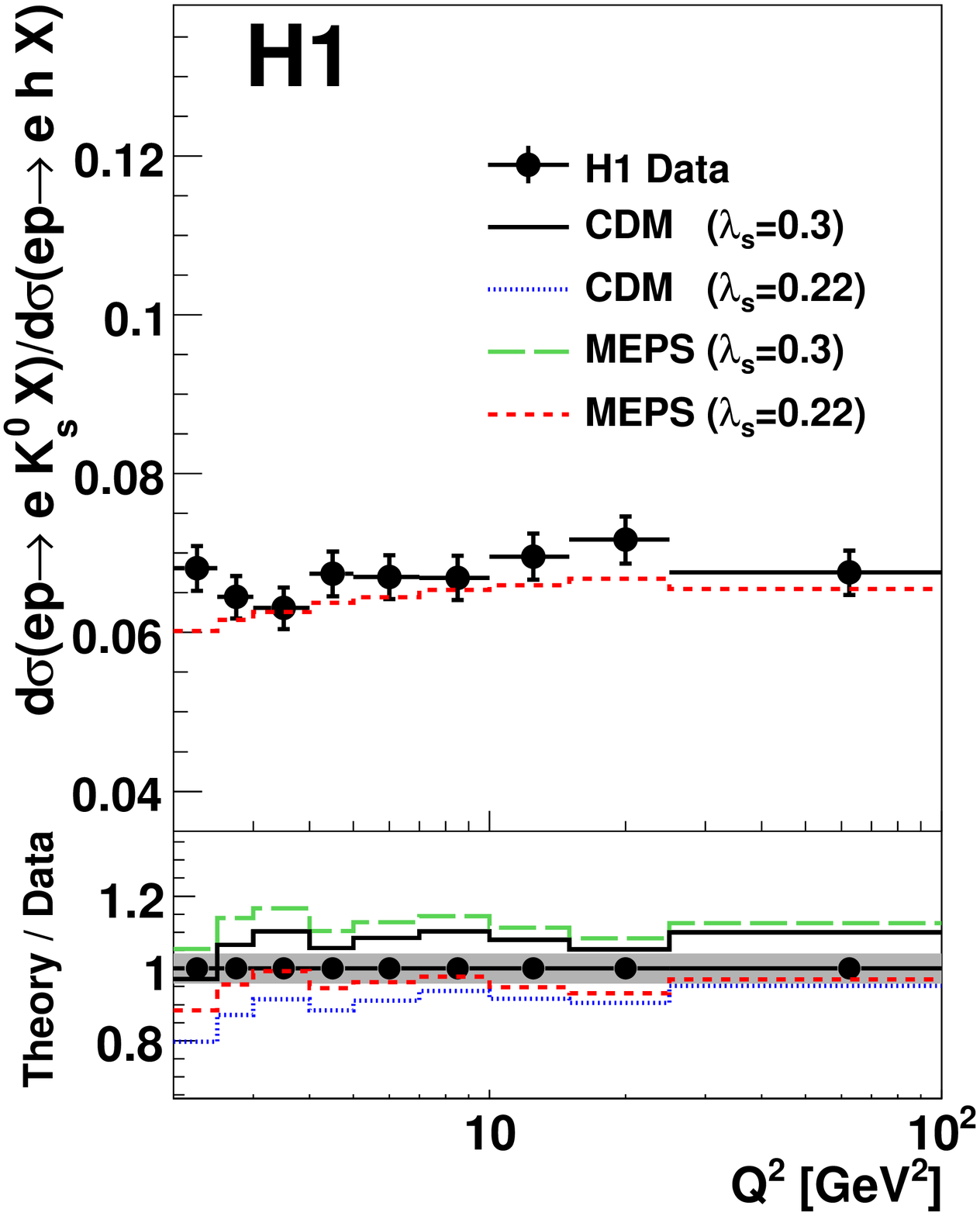}
\includegraphics[width=79mm]{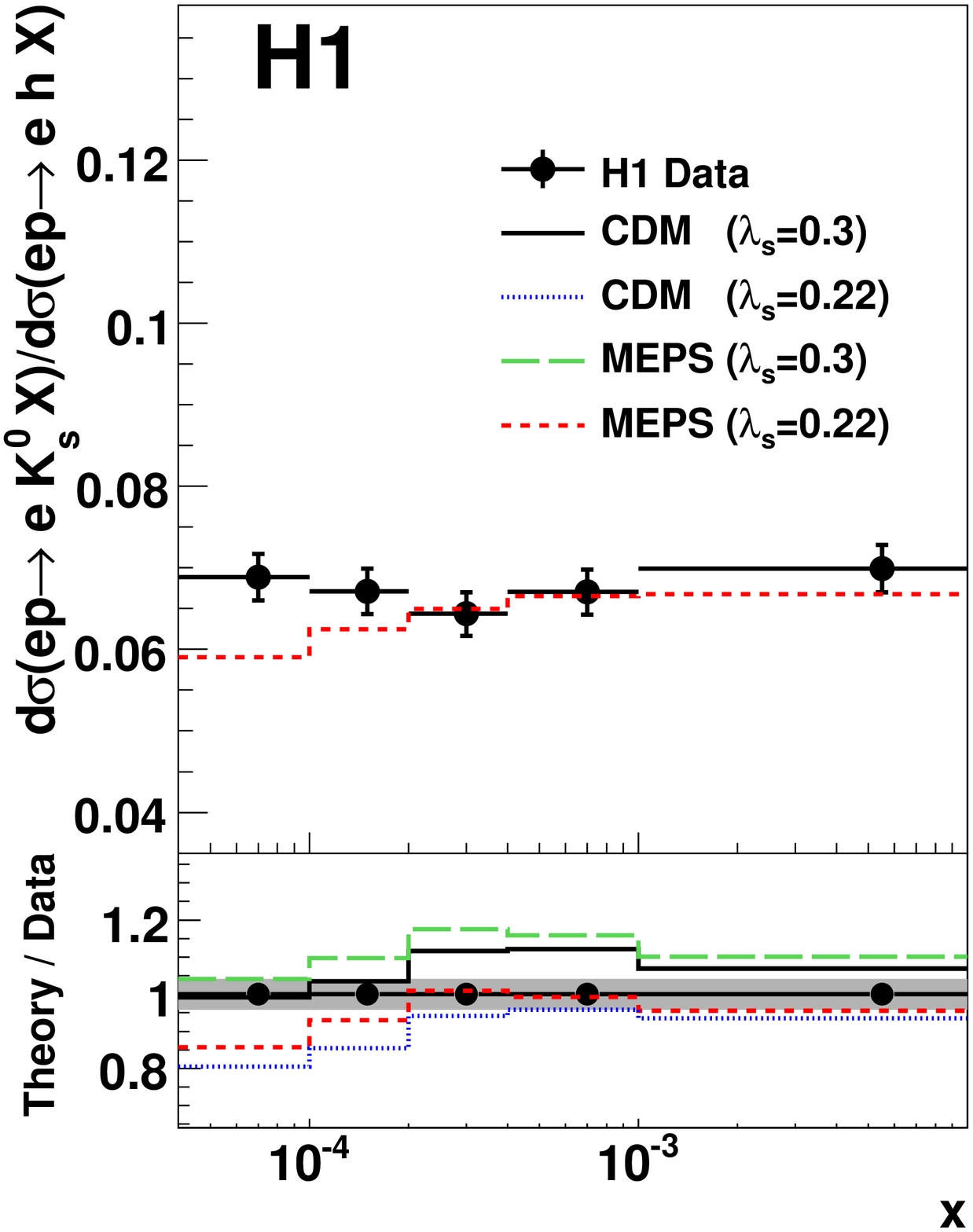}\\
\includegraphics[width=79mm]{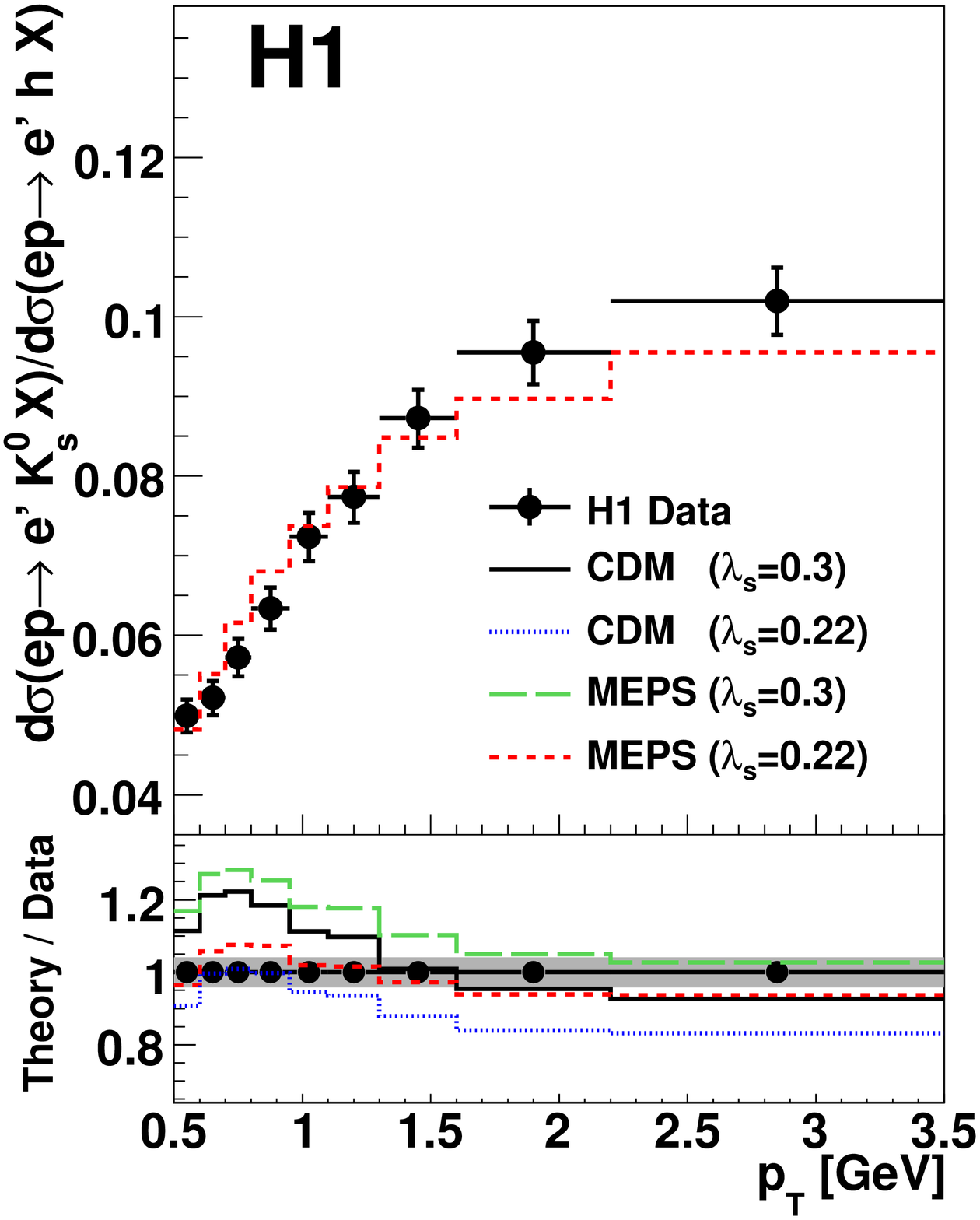}
\includegraphics[width=79mm]{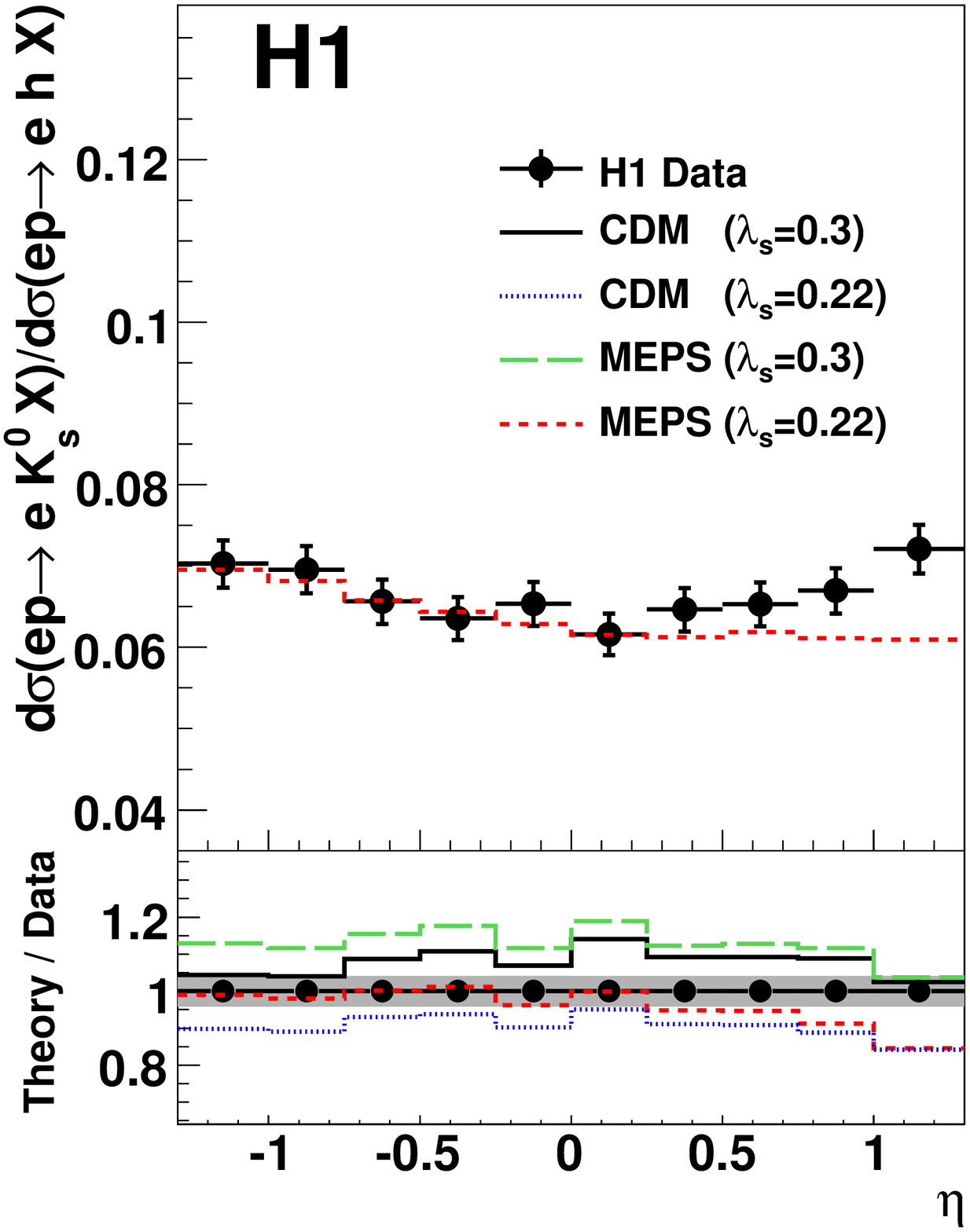}\\
\caption{The ratio of the differential production cross sections for 
\ksf\ mesons and charged hadrons
in the laboratory frame as a function of the
a) photon virtuality squared \qsq,
b) Bjorken scaling variable $x$,
c) transverse momentum $p_T$
and d) pseudorapidity $\eta$.
More details in the caption of figure~\ref{fig:ks-ds-lab}. 
}
\label{fig:ks-charged-ratio-lab}
\end{center}
\begin{picture}(0,0)
   \put(17,170){\bfseries a)}
   \put(97,170){\bfseries b)}
   \put(17,71){\bfseries c)}
   \put(97,71){\bfseries d)}
   \put(40,234){\LARGE e\,p $\rightarrow$ e\,K$^0_s$\,X / e\,p $\rightarrow$ e\,h$^{\pm}$\,X}
\end{picture}
\end{figure}

The ratio of differential production cross sections for $\ksf$ mesons 
and charged hadrons, denoted by 
$R(\ksf/h^{\pm}) = {d\sigma(ep \rightarrow e \ksf  X)}$ / ${d\sigma(ep \rightarrow e 
h^{\pm} X})$,
is equivalent to the ratio of the average multiplicities of $\ksf$ and charged hadrons.
In contrast to inclusive \ksf\ production, the correlation of $R(\ksf/h^{\pm})$ 
to the parameter \lambdas\ is expected to be less model dependent.
By taking the ratio, inadequacies of the model description of the
partonic final states and in particular of the dependence on the proton
structure function should cancel to a large extent.
  The ratio $R(\ksf/h^{\pm})$ is shown in
figure~\ref{fig:ks-charged-ratio-lab} and listed in table~\ref{table:RhCrossSection} 
as a function of 
$\qsq,  x$, $p_T$ and $\eta$.
The ratio strongly rises with increasing $p_T$ and remains approximately constant 
as a function of all the other variables.
This $p_T$ dependence of $R(\ksf/h^{\pm})$ reflects a
general kinematic feature (heavier particles receive the
larger fraction of the system momentum) and can also be observed in the
 \lsf/\ksf\ ratio in figure~\ref{fig:la-ks-ratio-lab}.

Also shown with the data are the CDM and MEPS model predictions for
two values of $\lambda_s$ ($0.22$ and $0.3$). 
Overall, no single prediction is able to fully describe the shapes
of all $R(\ksf/h^{\pm})$  distributions, failing in particular
in the low $p_T$, low $x$ and large positive $\eta$ regions.
The $p_T$ spectrum of $R(\ksf/h^{\pm})$ is found to be harder in the data,
consistent with the conclusions derived from 
the cross section measurement.

The shapes of the ratios $R(\ksf/h^{\pm})$ are reasonably described by
both CDM and MEPS model predictions. 
However, there is a difference in normalisation between the two
models.
The CDM prediction with $\lambdas=0.3$ is in better agreement with the data 
at low $Q^2$,
whereas at high $Q^2$ a value of $\lambdas=0.22$ is preferred, 
as observed in the ZEUS data~\cite{ZEUS-K0-2}. 
In contrast, the MEPS model predictions prefer a lower value of $\lambdas = 0.22$
over the full phase space.

A comparison of the predictions, applying different settings for the
diquark-quark suppression factors ($\lambda_{qq}, \lambda_{sq}$) shows 
the expected behaviour.
In general, no changes are visible in the shapes of the differential distributions, 
however some differences are present in the absolute normalisation.
The \ksf\ distributions are not affected, as expected, and both the \lsf\ and the
ratio $R(\lsf/\ksf)$ show the anticipated correlations to the suppression 
factors ($\lambda_{qq},\lambda_{sq}$).
These predicted effects are mostly independent of the choice of the
$\lambdas$ value, used for the simulation, and indicate
that the ``ALEPH-tune'' from $e^+e^-$ collisions also describes the
overall features of the data in $ep$ collisions, supporting the universality
of strangeness production.

%

\section{Conclusions}
The production cross sections and ratios of the production of \ksf, \lsf\ and
charged hadrons $h^{\pm}$ are  measured inclusively and also differentially 
as a function of the DIS  variables and of the final state particle variables 
in the visible kinematic region, defined by
 $2 < Q^2 < 100\GeVSq$, $0.1 < y < 0.6$,
 $0.5 < p_T(\ksf,\lsf,h^{\pm} ) < 3.5\,\gev$  and $|\eta(\ksf, \lsf,h^{\pm} )|< 1.3$.

The measured total cross sections and their ratios are in agreement 
with the predictions based on DJANGOH.
The overall features of the various differential distributions are   
reasonably well reproduced by both simulations, based either on the CDM or
the MEPS approach,
when applying model parameters obtained from  $e^+e^-$ data at LEP.
However, predictions based on a single value of {\lambdas} fail 
to describe the details of the distributions
in various regions of the phase space, in particular
in the low $p_T$, low $x$ and large positive $\eta$ regions.
The production of \ksf\ and \lsf\ particles, as measured in the Breit frame, is in general 
described by both CDM and MEPS predictions.  
The measurement of the asymmetry in the production of \lsf\ with respect to \lsa,
which is found to be consistent with zero within errors, does not
support the hypothesis of baryon number transfer.
  
The \lsf\ to \ksf\ cross section  ratio  is  better described by the CDM 
prediction and is nearly independent of the {\lambdas} value,
whereas for the 
\ksf\  to charged hadrons cross section ratio 
the MEPS model with   $\lambdas = 0.22$ is in better agreement with the data.
%

\section*{Acknowledgements}

We are grateful to the HERA machine group whose outstanding
efforts have made this experiment possible. 
We thank the engineers and technicians for their work in constructing and
maintaining the H1 detector, our funding agencies for 
financial support, the
DESY technical staff for continual assistance
and the DESY directorate for support and for the
hospitality which they extend to the non DESY 
members of the collaboration.

\clearpage

%
%
%
%

\clearpage


\newpage


\newpage

\clearpage
\newpage

\clearpage
\newpage

\clearpage

\newpage

\clearpage

\newpage
\clearpage

\newpage

\clearpage

\newpage
\clearpage

\newpage

\clearpage

\newpage

\clearpage

\newpage
\clearpage

\newpage

\clearpage


\newpage

\renewcommand{\arraystretch}{1.2}

\begin{table}[htbp]
\footnotesize
\centering
\begin{tabular}{|r@{}c@{}l||cccc|}
\hline
\multicolumn{7}{|c|}{$\mathbf{ep \rightarrow e\,K^0_s\,X}$} \\
\hline
\hline
%
\multicolumn{3}{|c||}{$Q^{2}$} & ${\rm d}\sigma/{\rm d}Q^2$ & stat. & syst. ($+$) & syst. ($-$) \\
\multicolumn{3}{|c||}{${\rm [GeV^2]}$} & \multicolumn{4}{c|}{[${\rm nb/GeV^2}$]} \\
\hline
2$\;\;$&--$\;\;$&2.5  &    5.73  &    0.10  &    0.58  &    0.71  \\
2.5$\;\;$&--$\;\;$&3  &    4.05  &    0.08  &    0.29  &    0.29  \\
3$\;\;$&--$\;\;$&4  &3.08  &    0.05  &    0.21  & 0.21   \\
4$\;\;$&--$\;\;$&5  &    2.00  &    0.03  &    0.12  & 0.13   \\
5$\;\;$&--$\;\;$&7  &    1.332  &    0.018  &    0.082  & 0.082  \\
7$\;\;$&--$\;\;$&10  &    0.764  &    0.011  &    0.045  & 0.047   \\
10$\;\;$&--$\;\;$&15  &    0.417  &    0.006  &    0.023  & 0.024  \\
15$\;\;$&--$\;\;$&25  &    0.197  &    0.003  &    0.012  &  0.012  \\
25$\;\;$&--$\;\;$&100  &    0.0340  &    0.0004  &    0.0020  &  0.0021  \\
\hline
\hline
\multicolumn{3}{|c||}{$x$} & ${\rm d}\sigma/{\rm d}x$ & stat. & syst. ($+$) & syst. ($-$)\\
\multicolumn{3}{|c||}{} & \multicolumn{4}{c|}{[$\mu$b]}  \\
\hline
 0.00004$\;\;$& -- &$\;\;$0.0001   & 69.4  & 1.0 & 4.4 & 4.4  \\
 0.0001$\;\;$& -- &$\;\;$0.0002  &  51.7  & 0.6 &  3.2 &  3.3  \\
 0.0002$\;\;$& -- &$\;\;$0.0004  &  24.0  &  0.3 &  1.4 &  1.5  \\
 0.0004$\;\;$& -- &$\;\;$0.001  &  7.07  & 0.07 &  0.43 &  0.43  \\
 0.001$\;\;$& -- &$\;\;$0.01  &  0.315  &  0.004 & 0.019 &  0.019  \\
\hline
\hline
\multicolumn{3}{|c||}{$p_{T}$} & ${\rm d}\sigma/{\rm d}p_T$ & stat. & syst. ($+$)  & syst. ($-$) \\
\multicolumn{3}{|c||}{[GeV]} & \multicolumn{4}{c|}{[nb/GeV]}  \\
\hline
0.5$\;\;$& -- &$\;\;$0.6  & 34.6  &  0.5  &  2.0  &  2.1  \\
0.6$\;\;$& -- &$\;\;$0.7  & 29.6  &  0.4  &  1.7  &  1.7  \\
0.7$\;\;$& -- &$\;\;$0.8  & 25.5  &  0.4  &  1.4  &  1.5  \\
0.8$\;\;$& -- &$\;\;$0.9  & 20.4  &  0.3  &  1.1  &  1.2  \\
0.9$\;\;$& -- &$\;\;$1.1  & 15.2  &  0.2  &  0.9  &  0.9  \\
1.1$\;\;$& -- &$\;\;$1.3  & 10.46 &  0.14 &  0.61 &  0.63  \\
1.3$\;\;$& -- &$\;\;$1.6  & 6.91  &  0.10 &  0.46 &  0.46  \\
1.6$\;\;$& -- &$\;\;$2.2  & 3.13  &  0.04 &  0.20 &  0.20  \\
2.2$\;\;$& -- &$\;\;$3.5  & 0.83  &  0.02 &  0.06 &  0.06  \\
\hline
\hline
\multicolumn{3}{|c||}{$\eta$} & ${\rm d}\sigma/{\rm d}\eta$ & stat. & syst. ($+$) & syst. ($-$) \\
\multicolumn{3}{|c||}{} & \multicolumn{4}{c|}{[nb]}  \\
\hline
-1.3$\;\;$&  -- &$\;\;$-1     &      8.08  &      0.12  &      0.41  &      0.42  \\
-1$\;\;$&    -- &$\;\;$-0.75  &      8.69  &      0.13  &      0.51  &      0.52  \\
-0.75$\;\;$& -- &$\;\;$-0.5   &      8.64  &      0.12  &      0.44  &      0.46  \\
-0.5$\;\;$&  -- &$\;\;$-0.25  &      8.56  &      0.13  &      0.47  &      0.50  \\
-0.25$\;\;$& -- &$\;\;$0        &      8.79  &      0.16  &      0.56  &      0.59  \\
0$\;\;$&       -- &$\;\;$0.25     &      8.65  &      0.14  &      0.58  &      0.60  \\
0.25$\;\;$&    -- &$\;\;$0.5      &      7.58  &      0.13  &      0.52  &      0.52  \\
0.5$\;\;$&     -- &$\;\;$0.75     &      7.99  &      0.13  &      0.55  &      0.56  \\
0.75$\;\;$&    -- &$\;\;$1        &      7.98  &      0.15  &      0.54  &      0.54  \\
1$\;\;$&       -- &$\;\;$1.3      &      8.06  &      0.13  &      0.54  &      0.54  \\
\hline
\end{tabular}
 \caption{The differential $K^0_s$ cross-section values   
as a function of $Q^2$, $x$, $p_T$ and $\eta$ in the
visible region defined by 
$2 < Q^2 < 100\GeVSq$ and $0.1 < y < 0.6$.
The bin ranges, the bin averaged cross section values,
the statistical and the positive and negative systematic uncertainties are listed.}
\label{table:K0CrossSection}
\end{table}

\clearpage
\newpage

\begin{table}[htbp]
\footnotesize
\centering
\begin{tabular}{|r@{}c@{}l||cccc|}
\hline
\multicolumn{7}{|c|}{$\mathbf{ep \rightarrow e\,\Lambda\,X}$} \\
\hline
\hline
%
\multicolumn{3}{|c||}{$Q^{2}$} & ${\rm d}\sigma/{\rm d}Q^2$ & stat. & syst. ($+$) & syst. ($-$) \\
\multicolumn{3}{|c||}{${\rm [GeV^2]}$} & \multicolumn{4}{c|}{[${\rm nb/GeV^2}$]} \\
\hline
  2 $\;\;$&  -- &$\;\;$ 2.5  &    2.33  &    0.14  &    0.25  &    0.30  \\
  2.5 $\;\;$&  -- &$\;\;$ 3  &    1.58  &    0.09  &    0.12  &    0.13  \\
  3 $\;\;$&  -- &$\;\;$ 4  &    1.13  &    0.05  &    0.08  &    0.08  \\
  4 $\;\;$&  -- &$\;\;$ 5  &    0.73  &    0.04  &    0.05  &    0.05  \\
  5 $\;\;$&  -- &$\;\;$ 7  &    0.462  &    0.018  &    0.028  &    0.030  \\
  7 $\;\;$&  -- &$\;\;$ 10  &   0.282  &    0.012  &    0.019  &    0.020  \\
  10$\;\;$&  -- &$\;\;$ 15  &    0.153  &    0.006  &    0.009  &   0.009  \\
  15$\;\;$&  -- &$\;\;$ 25  &    0.071  &    0.003  &    0.004  &    0.004  \\
  25$\;\;$&  -- &$\;\;$ 100  &    0.0120  &    0.0004  &    0.0006  &    0.0006  \\
\hline
\hline
\multicolumn{3}{|c||}{$x$} & ${\rm d}\sigma/{\rm d}x$ & stat. & syst. ($+$) & syst. ($-$)\\
\multicolumn{3}{|c||}{} & \multicolumn{4}{|c|}{[$\mu$b]}  \\
\hline
0.00004 $\;\;$&  -- &$\;\;$0.0001  & 28.1   &  1.1  & 2.0  &  2.0    \\
0.0001$\;\;$&  -- &$\;\;$0.0002  & 20.1  &  0.7  & 1.4 &  1.4    \\
0.0002$\;\;$&  -- &$\;\;$0.0004  & 8.5  &0.3  &  0.5 &  0.5    \\
0.0004$\;\;$&  -- &$\;\;$0.001  & 2.57  &0.08 &  0.15&  0.15    \\
0.001$\;\;$&  -- &$\;\;$0.01  & 0.104  & 0.003  &  0.006 &  0.006    \\
\hline
\hline
\multicolumn{3}{|c||}{$p_{T}$} & ${\rm d}\sigma/{\rm d}p_T$ & stat. & syst. ($+$)  & syst. ($-$) \\
\multicolumn{3}{|c||}{[GeV]} & \multicolumn{4}{|c|}{[nb/GeV]}  \\
\hline
  0.5$\;\;$& -- &$\;\;$  0.6  &      8.4  &      0.8  &      0.8  &      0.8  \\
  0.6$\;\;$& -- &$\;\;$ 0.7  &      8.0  &      0.5  &      0.4  &      0.5  \\
  0.7$\;\;$& -- &$\;\;$ 0.8  &      7.9  &      0.5  &      0.5  &      0.5  \\
  0.8$\;\;$& -- &$\;\;$ 0.9  &      6.8  &      0.3  &      0.5  &      0.5  \\
  0.9$\;\;$& -- &$\;\;$ 1.1  &      6.1  &      0.2  &      0.4  &      0.4  \\
  1.1$\;\;$& -- &$\;\;$ 1.3  &      4.70  &      0.18  &      0.29  &      0.30  \\
  1.3$\;\;$& -- &$\;\;$ 1.6  &      3.05  &      0.10  &      0.20  &      0.20  \\
  1.6$\;\;$& -- &$\;\;$ 2.2  &      1.52  &      0.05  &      0.09  &      0.09  \\
  2.2$\;\;$& -- &$\;\;$ 3.5  &      0.42  &      0.02  &      0.02  &      0.02  \\
\hline
\hline
\multicolumn{3}{|c||}{$\eta$} & ${\rm d}\sigma/{\rm d}\eta$ & stat. & syst. ($+$) & syst. ($-$) \\
\multicolumn{3}{|c||}{} & \multicolumn{4}{c|}{[nb]}  \\
\hline
 -1.3$\;\;$& -- &$\;\;$-1  &      2.67  &      0.14  &      0.15  &      0.15  \\
  -1$\;\;$& -- &$\;\;$-0.75  &      2.87  &      0.14  &      0.16  &      0.17  \\
  -0.75$\;\;$& -- &$\;\;$-0.5  &      3.03  &      0.15  &      0.17  &      0.17  \\
  -0.5$\;\;$& -- &$\;\;$-0.25  &      2.76  &      0.13  &      0.18  &      0.18  \\
  -0.25$\;\;$& -- &$\;\;$0  &      2.74  &      0.15  &      0.19  &      0.20  \\
  0$\;\;$& -- &$\;\;$0.25  &      2.92  &      0.15  &      0.21  &      0.22  \\
  0.25$\;\;$& -- &$\;\;$0.5  &      2.95  &      0.14  &      0.21  &      0.22  \\
  0.5$\;\;$& -- &$\;\;$0.75  &      3.36  &      0.17  &      0.23  &      0.23  \\
  0.75$\;\;$& -- &$\;\;$ 1  &      3.43  &      0.15  &      0.25  &      0.25  \\
  1$\;\;$& -- &$\;\;$ 1.3  &      3.88  &      0.16  &      0.28  &      0.31  \\
\hline
\end{tabular}
\caption{The differential \lsf\ cross-section values  
 as a function of $Q^2$, $x$, $p_T$ and $\eta$.
 More details in caption of table~\ref{table:K0CrossSection}.}
\label{table:L0CrossSection}
\end{table}

\clearpage
\newpage

\begin{table}[htbp]
\footnotesize
\centering
\begin{tabular}{|r@{}c@{}l||cccc|}
\hline
\multicolumn{7}{|c|}{$\mathbf{ep \rightarrow e\,K^0_s\,X}$} \\
\hline
\hline
\multicolumn{3}{|c||}{$p_{T}^{Breit}$ target} & ${\rm d}\sigma/{\rm d}p_{T}^{Breit}$ & stat. & syst. ($+$) & syst. ($-$) \\
\multicolumn{3}{|c||}{${\rm [GeV]}$} & \multicolumn{4}{|c|}{[${\rm nb/GeV}$]} \\
\hline
 0.5 $\;\;$& -- &$\;\;$ 1 &   21.20   &    0.19  &     1.20  &     1.23 \\
 1 $\;\;$& -- &$\;\;$ 1.25  &    10.05  &     0.16  &     0.57  &     0.57 \\
 1.25$\;\;$& -- &$\;\;$1.5  &     6.12  &     0.13  &     0.35  &     0.37 \\
 1.5$\;\;$& -- &$\;\;$2.5  &     2.04  &     0.04  &     0.12  &     0.12 \\
 2.5$\;\;$& -- &$\;\;$4  &     0.230  &     0.008  &     0.011  &     0.011 \\
\hline
\hline
\multicolumn{3}{|c||}{$p_{T}^{Breit}$ current} & ${\rm d}\sigma/{\rm d}p_{T}^{Breit}$ & stat. & syst. ($+$) & syst. ($-$) \\
\multicolumn{3}{|c||}{${\rm [GeV]}$} & \multicolumn{4}{|c|}{[${\rm nb/GeV}$]} \\
\hline
 0$\;\;$& -- &$\;\;$ 0.6  &     2.00  &     0.05  &     0.14  &     0.16 \\
 0.6$\;\;$& -- &$\;\;$ 3  &     0.277  &     0.009  &     0.031  &     0.036 \\
\hline
\hline
\multicolumn{3}{|c||}{$x_{p}^{Breit}$ target} & ${\rm d}\sigma/{\rm d}x_{p}^{Breit}$ & stat. & syst. ($+$) & syst. ($-$) \\
\multicolumn{3}{|c||}{} & \multicolumn{4}{|c|}{[${\rm nb}$]} \\
\hline
  0$\;\;$& -- &$\;\;$ 0.45  &     4.01  &     0.08  &     0.22  &     0.23 \\
  0.45$\;\;$& -- &$\;\;$ 1  &     5.43  &     0.09  &     0.39  &     0.42 \\   
  1$\;\;$& -- &$\;\;$ 2  &     3.66  &     0.05  &     0.25  &     0.25 \\   
  2$\;\;$& -- &$\;\;$ 4  &     2.03  &     0.03  &     0.11  &     0.11 \\   
  4$\;\;$& -- &$\;\;$ 7  &     0.984  &     0.016  &     0.05  &     0.05 \\   
  7$\;\;$& -- &$\;\;$ 11  & 	0.478  &     0.011  &     0.026  &     0.028 \\  
  11$\;\;$& -- &$\;\;$ 20  & 	 0.167  & 	0.005  &     0.011  &     0.013 \\ 
\hline
\hline
\multicolumn{3}{|c||}{$x_{p}^{Breit}$ current} & ${\rm d}\sigma/{\rm d}x_{p}^{Breit}$ & stat. & syst. ($+$) & syst. ($-$) \\
\multicolumn{3}{|c||}{} & \multicolumn{4}{|c|}{[${\rm nb}$]} \\
\hline
 0$\;\;$& -- &$\;\;$  0.3  &      3.27  &      0.08  &      0.18  &      0.20 \\
 0.3$\;\;$& -- &$\;\;$ 1  &      1.20  &      0.04  &      0.14  &      0.17 \\
\hline
\end{tabular}
\caption{The differential \ksf\ cross-section values   
as a function of $p_{T}^{Breit}$ and $x_p^{Breit}$ in the target 
and current hemispheres of the Breit frame.
More details in caption of table~\ref{table:K0CrossSection}. }
\label{table:K0Breit}
\end{table}

\clearpage
\newpage

\begin{table}[htbp]
\footnotesize
\centering
\begin{tabular}{|r@{}c@{}l||cccc|}
\hline
\multicolumn{7}{|c|}{$\mathbf{ep \rightarrow e\,\Lambda\,X}$} \\
\hline
\hline
\multicolumn{3}{|c||}{$p_{T}^{Breit}$ target} & ${\rm d}\sigma/{\rm d}p_{T}^{Breit}$ & stat. & syst. ($+$) & syst. ($-$) \\
\multicolumn{3}{|c||}{${\rm [GeV]}$} & \multicolumn{4}{|c|}{[${\rm nb/GeV}$]} \\
\hline
  0.5$\;\;$& -- &$\;\;$1  &     6.97  &     0.19  &     0.38  &     0.40 \\
  1$\;\;$& -- &$\;\;$1.25  &     4.52  &     0.17  &     0.28  &     0.28 \\
  1.25$\;\;$& -- &$\;\;$1.5  &     2.71  &     0.13  &     0.17  &     0.17 \\
  1.5$\;\;$& -- &$\;\;$2.5  &     1.01  &     0.04  &     0.06  &     0.06 \\
  2.5$\;\;$& -- &$\;\;$4  &     0.114  &     0.009  &     0.007  &     0.008 \\
\hline
\hline
\multicolumn{3}{|c||}{$p_{T}^{Breit}$ current} & ${\rm d}\sigma/{\rm d}p_{T}^{Breit}$ & stat. & syst. ($+$) & syst. ($-$) \\
\multicolumn{3}{|c||}{${\rm [GeV]}$} & \multicolumn{4}{|c|}{[${\rm nb/GeV}$]} \\
\hline
\hline
0$\;\;$& -- &$\;\;$ 0.6  &     0.307  &     0.028  &     0.017  &     0.018 \\
0.6$\;\;$& -- &$\;\;$3  &     0.051  &     0.004  &     0.004  &     0.004 \\
\hline
\multicolumn{3}{|c||}{$x_{p}^{Breit}$ target} & ${\rm d}\sigma/{\rm d}x_{p}^{Breit}$ & stat. & syst. ($+$) & syst. ($-$) \\
\multicolumn{3}{|c||}{} & \multicolumn{4}{|c|}{[${\rm nb}$]} \\
\hline
 0$\;\;$& -- &$\;\;$0.45  &     0.94  &     0.05  &     0.05  &     0.06 \\
 0.45$\;\;$& -- &$\;\;$ 1  &     1.51  &     0.07  &     0.11  &     0.12 \\   
 1$\;\;$& -- &$\;\;$ 2  &     1.09  &     0.05  &     0.08  &     0.09 \\   
 2$\;\;$& -- &$\;\;$ 4  &     0.75  &     0.03  &     0.05  &     0.05 \\   
 4$\;\;$& -- &$\;\;$ 7  &     0.45  &     0.02  &     0.03  &     0.03 \\   
 7$\;\;$& -- &$\;\;$ 11  & 	0.220  &     0.011  &     0.015  &     0.016 \\  
 11$\;\;$& -- &$\;\;$ 20  & 	 0.102  & 	0.006  &     0.008  &     0.008 \\ 
\hline
\hline
\multicolumn{3}{|c||}{$x_{p}^{Breit}$ current} & ${\rm d}\sigma/{\rm d}x_{p}^{Breit}$ & stat. & syst. ($+$) & syst. ($-$) \\
\multicolumn{3}{|c||}{} & \multicolumn{4}{|c|}{[${\rm nb}$]} \\
\hline
 0$\;\;$& -- &$\;\;$0.3  &      0.55  &      0.05  &      0.04  &      0.04 \\
 0.3$\;\;$& -- &$\;\;$1  &      0.21  &      0.02  &      0.03  &      0.03 \\
\hline
\end{tabular}
\caption{The differential \lsf\ cross-section values   
as a function of $p_{T}^{Breit}$ and $x_p^{Breit}$ in the target 
and current hemispheres of the Breit frame.
More details in caption of table~\ref{table:K0CrossSection}. }
\label{table:L0Breit}
\end{table}

\clearpage
\newpage

\begin{table}[htbp]
\footnotesize
\centering
\begin{tabular}{|r@{}c@{}l||cccc|}
\hline
\multicolumn{7}{|c|}{$\mathbf{ R(\Lambda/K^0_s) }$} \\
\hline
\hline
%
\multicolumn{3}{|c||}{$Q^{2}$} & R($\Lambda/K^0_s$) & stat. & syst. ($+$) & syst. ($-$) \\
\multicolumn{3}{|c||}{${\rm [GeV^2]}$} & \multicolumn{4}{|c|}{} \\
\hline
  2$\;\;$&  -- &$\;\;$ 2.5  &     0.406  &     0.025  &     0.018  &     0.019  \\
  2.5$\;\;$&  -- &$\;\;$ 3  &     0.390  &     0.020  &     0.030  &     0.030  \\
  3$\;\;$&  -- &$\;\;$ 4  &     0.368  &     0.019  &     0.020  &     0.020  \\
  4$\;\;$&  -- &$\;\;$ 5  &     0.366  &     0.020  &     0.018  &     0.019  \\
  5$\;\;$&  -- &$\;\;$ 7  &     0.347  &     0.014  &     0.014  &     0.014  \\
  7$\;\;$&  -- &$\;\;$ 10  &     0.369  &     0.016  &     0.018  &     0.019  \\
  10$\;\;$&  -- &$\;\;$ 15  &     0.367  &     0.015  &     0.013  &     0.014  \\
  15$\;\;$&  -- &$\;\;$ 25  &     0.360  &     0.016  &     0.017  &     0.017  \\
  25$\;\;$&  -- &$\;\;$ 100  &     0.353  &     0.012  &     0.017  &     0.018  \\
\hline
\hline
\multicolumn{3}{|c||}{$x$} & R($\Lambda/K^0_s$) & stat. & syst. ($+$) & syst. ($-$)\\
\hline
 0.00004  $\;\;$&  -- &$\;\;$  0.0001 &     0.405 &     0.017 &     0.024  &     0.025  \\
 0.0001   $\;\;$&  -- &$\;\;$  0.0002 &     0.390 &     0.014 &     0.019  &     0.020  \\
 0.0002   $\;\;$&  -- &$\;\;$  0.0004 &     0.355 &     0.011 &     0.012  &     0.013  \\
 0.0004   $\;\;$&  -- &$\;\;$  0.001 &     0.364 &     0.012 &     0.013  &     0.014  \\
 0.001    $\;\;$&  -- &$\;\;$   0.01 &     0.329 &     0.011 &     0.016  &     0.017  \\
\hline
\hline
\multicolumn{3}{|c||}{$p_{T}$} & R($\Lambda/K^0_s$) & stat. & syst. ($+$)  & syst. ($-$) \\
\multicolumn{3}{|c||}{[GeV]} & \multicolumn{4}{|c|}{}  \\
\hline
  0.5$\;\;$& -- &$\;\;$ 0.6  &     0.24  &     0.02  &     0.02  &     0.02  \\
  0.6$\;\;$& -- &$\;\;$ 0.7  &     0.268  &     0.017  &     0.009  &     0.010  \\
  0.7$\;\;$& -- &$\;\;$ 0.8  &     0.309  &     0.020  &     0.015  &     0.016  \\
  0.8$\;\;$& -- &$\;\;$ 0.9  &     0.334  &     0.014  &     0.016  &     0.017  \\
  0.9$\;\;$& -- &$\;\;$ 1.1  &     0.402  &     0.017  &     0.015  &     0.016  \\
  1.1$\;\;$& -- &$\;\;$ 1.3  &     0.450  &     0.018  &     0.015  &     0.016  \\
  1.3$\;\;$& -- &$\;\;$ 1.6  &     0.442  &     0.016  &     0.021  &     0.022  \\
  1.6$\;\;$& -- &$\;\;$ 2.2  &     0.485  &     0.016  &     0.019  &     0.020  \\
  2.2$\;\;$& -- &$\;\;$ 3.5  &     0.505  &     0.027  &     0.031  &     0.032  \\
\hline
\hline
\multicolumn{3}{|c||}{$\eta$} & R($\Lambda/K^0_s$) & stat. & syst. ($+$) & syst. ($-$) \\
\hline
 -1.3$\;\;$& -- &$\;\;$-1  &     0.331  &     0.018  &     0.015  &     0.015  \\
  -1$\;\;$& -- &$\;\;$-0.75  &     0.330  &     0.017  &     0.018  &     0.018  \\
  -0.75$\;\;$& -- &$\;\;$-0.5  &     0.350  &     0.018  &     0.013  &     0.014  \\
  -0.5$\;\;$& -- &$\;\;$-0.25  &     0.323  &     0.016  &     0.015  &     0.016  \\
  -0.25$\;\;$& -- &$\;\;$0  &     0.311  &     0.018  &     0.017  &     0.017  \\
  0$\;\;$& -- &$\;\;$0.25  &     0.337  &     0.018  &     0.018  &     0.018  \\
  0.25$\;\;$& -- &$\;\;$0.5  &     0.389  &     0.019  &     0.017  &     0.018  \\
  0.5$\;\;$& -- &$\;\;$0.75  &     0.420  &     0.023  &     0.014  &     0.015  \\
  0.75$\;\;$& -- &$\;\;$1 &     0.430  &     0.020  &     0.017  &     0.018  \\
  1$\;\;$& -- &$\;\;$1.3  &     0.48  &     0.02  &     0.02  &     0.02  \\
\hline
\end{tabular}
\caption{The values of the ratio R($\Lambda/K^0_s$) of the differential cross-sections
for \lsf\ baryons and \ksf\ mesons
 as a function of $Q^2$, $x$, $p_T$ and $\eta$. More details in caption of 
table~\ref{table:K0CrossSection}.}
\label{table:RLKlab}
\end{table}

\clearpage

\newpage

\begin{table}[htbp]
\footnotesize
\centering
\begin{tabular}{|r@{}c@{}l||cccc|}
\hline
\multicolumn{7}{|c|}{$\mathbf{ R(\Lambda/K^0_s) }$} \\
\hline
\hline
\multicolumn{3}{|c||}{$p_{T}^{Breit}$ target} & R($\Lambda/K^0_s$) & stat. & syst. ($+$) & syst. ($-$) \\
\multicolumn{3}{|c||}{${\rm [GeV]}$} & \multicolumn{4}{|c|}{} \\
\hline
  0.5$\;\;$& -- &$\;\;$1  &     0.329  &     0.009  &     0.008  &     0.009 \\
  1$\;\;$& -- &$\;\;$1.25  &     0.449  &     0.019  &     0.015  &     0.016 \\
  1.25$\;\;$& -- &$\;\;$1.5  &     0.443  &     0.023  &     0.017  &     0.018 \\
  1.5$\;\;$& -- &$\;\;$2.5  &     0.493  &     0.019  &     0.020  &     0.021 \\
  2.5$\;\;$& -- &$\;\;$4  &     0.495  &     0.043  &     0.026  &     0.027 \\
\hline
\hline
\multicolumn{3}{|c||}{$p_{T}^{Breit}$ current} & R($\Lambda/K^0_s$) & stat. & syst. ($+$) & syst. ($-$) \\
\multicolumn{3}{|c||}{${\rm [GeV]}$} & \multicolumn{4}{|c|}{} \\
\hline
  0$\;\;$& -- &$\;\;$0.6  &     0.153  &     0.015  &     0.005  &     0.005 \\
  0.6$\;\;$& -- &$\;\;$3  &     0.185  &     0.017  &     0.008  &     0.008 \\
\hline
\hline
\multicolumn{3}{|c||}{$x_{p}^{Breit}$ target} & R($\Lambda/K^0_s$) & stat. & syst. ($+$) & syst. ($-$) \\
\hline
  0$\;\;$& -- &$\;\;$0.45  &     0.235  &     0.014  &     0.008  &     0.008 \\
  0.45$\;\;$& -- &$\;\;$1  &    0.277  &     0.013  &     0.010  &     0.011 \\   
  1$\;\;$& -- &$\;\;$2  &     0.297  &     0.014  &     0.013  &     0.013 \\   
  2$\;\;$& -- &$\;\;$4  &     0.370  &     0.016  &     0.014  &     0.015 \\   
  4$\;\;$& -- &$\;\;$7  &     0.460  &     0.022  &     0.017  &     0.018 \\   
  7$\;\;$& -- &$\;\;$11  & 	0.46  &     0.03  &     0.02  &     0.03 \\  
  11$\;\;$& -- &$\;\;$20  &	 0.61  & 	0.04  &     0.03  &     0.04 \\ 
\hline
\hline
\multicolumn{3}{|c||}{$x_{p}^{Breit}$ current} & R($\Lambda/K^0_s$) & stat. & syst. ($+$) & syst. ($-$) \\
\hline
 0$\;\;$& -- &$\;\;$0.3  &     0.169  &     0.015  &     0.009  &     0.009 \\
 0.3$\;\;$& -- &$\;\;$1  &     0.174  &     0.017  &     0.012  &     0.012 \\
\hline
\end{tabular}
\caption{The values of the ratio R($\Lambda/K^0_s$) of the differential cross-sections
for \lsf\ baryons and \ksf\ mesons 
 as a function of $p_{T}^{Breit}$ and $x_p^{Breit}$  in the target and current hemispheres
of the Breit frame. More details in caption of table~\ref{table:K0CrossSection}.}
\label{table:RLKBreit}
\end{table}

\clearpage
\newpage


%
\begin{table}[htbp]
\footnotesize
\centering
\begin{tabular}{|r@{}c@{}l||cccc|}
\hline
\multicolumn{7}{|c|}{$\mathbf{ R(K^0_s/h^{\pm}) }$} \\
\hline
\hline
%
\multicolumn{3}{|c||}{$Q^{2}$} & $R(K^0_s/h^{\pm})$ & stat. & syst. ($+$) & syst. ($-$) \\
\multicolumn{3}{|c||}{${\rm [GeV^2]}$} & \multicolumn{4}{|c|}{} \\
\hline
 2.0$\;\;$& -- &$\;\;$2.5 & 0.0681 & 0.0002 & 0.0028 & 0.0029 \\
 2.5$\;\;$& -- &$\;\;$3.0 & 0.0645 & 0.0001 & 0.0026 & 0.0027 \\
 3.0$\;\;$& -- &$\;\;$4.0 & 0.0631 & 0.0002 & 0.0026 & 0.0027 \\
 4.0$\;\;$& -- &$\;\;$5.0 & 0.0674 & 0.0002 & 0.0028 & 0.0028 \\
 5.0$\;\;$& -- &$\;\;$7.0 & 0.0670 & 0.0002 & 0.0027 & 0.0028 \\
 7.0$\;\;$& -- &$\;\;$10.0 & 0.0669 & 0.0002 & 0.0027 & 0.0028 \\
 10.0$\;\;$& -- &$\;\;$15.0 & 0.0696 & 0.0002 & 0.0029 & 0.0029 \\
 15.0$\;\;$& -- &$\;\;$25.0 & 0.0717 & 0.0002 & 0.0029 & 0.0030 \\
 25.0$\;\;$& -- &$\;\;$100.0 & 0.0676 & 0.0002 & 0.0028 & 0.0028 \\
\hline
\hline
\multicolumn{3}{|c||}{$x$} & $R(K^0_s/h^{\pm})$ & stat. & syst. ($+$) & syst. ($-$)\\
\hline
 0.00004$\;\;$& -- &$\;\;$ 0.0001 & 0.0689 & 0.0002 & 0.0028 & 0.0029 \\
 0.0001$\;\;$& -- &$\;\;$ 0.0002  & 0.0671 & 0.0002 & 0.0027 & 0.0028 \\
 0.0002$\;\;$& -- &$\;\;$0.0004 & 0.0644 & 0.0001 & 0.0026 & 0.0027 \\
 0.0004$\;\;$& -- &$\;\;$ 0.001 & 0.0671 & 0.0002 & 0.0027 & 0.0028 \\
 0.001$\;\;$& -- &$\;\;$0.01 & 0.0699 & 0.0002 & 0.0028 & 0.0029 \\
\hline
\hline
\multicolumn{3}{|c||}{$p_{T}$} & $R(K^0_s/h^{\pm})$ & stat. & syst. ($+$)  & syst. ($-$) \\
\multicolumn{3}{|c||}{[GeV]} & \multicolumn{4}{|c|}{}  \\
\hline
 0.5$\;\;$& -- &$\;\;$0.6 & 0.0499 & 0.0001 & 0.0020 & 0.0021 \\
 0.6$\;\;$& -- &$\;\;$0.7 & 0.0522 & 0.0002 & 0.0021 & 0.0022 \\
 0.7$\;\;$& -- &$\;\;$ 0.8 & 0.0572 & 0.0002 & 0.0023 & 0.0024 \\
 0.8$\;\;$& -- &$\;\;$0.9 & 0.0633 & 0.0002 & 0.0026 & 0.0027 \\
 0.9$\;\;$& -- &$\;\;$1.1 & 0.0723 & 0.0002 & 0.0030 & 0.0030 \\
 1.1$\;\;$& -- &$\;\;$1.3 & 0.0773 & 0.0002 & 0.0032 & 0.0032 \\
 1.3$\;\;$& -- &$\;\;$1.6 & 0.0872 & 0.0003 & 0.0036 & 0.0037 \\
 1.6$\;\;$& -- &$\;\;$2.2 & 0.0955 & 0.0003 & 0.0039 & 0.0040 \\
 2.2$\;\;$& -- &$\;\;$3.5 & 0.1020 & 0.0003 & 0.0042 & 0.0043 \\
\hline
\hline
\multicolumn{3}{|c||}{$\eta$} & $R(K^0_s/h^{\pm})$ & stat. & syst. ($+$) & syst. ($-$) \\
\hline
 -1.3$\;\;$& -- &$\;\;$-1  & 0.0700 & 0.0002 & 0.0029 & 0.0030 \\
 -1$\;\;$& -- &$\;\;$-0.75 & 0.0696 & 0.0002 & 0.0029 & 0.0029 \\
 -0.75$\;\;$& -- &$\;\;$-0.5  & 0.0656 & 0.0002 & 0.0027 & 0.0028 \\
 -0.5$\;\;$& -- &$\;\;$-0.25  & 0.0635 & 0.0002 & 0.0026 & 0.0027 \\
 -0.25$\;\;$& -- &$\;\;$ 0 & 0.0654 & 0.0002 & 0.0027 & 0.0027 \\
 0$\;\;$& -- &$\;\;$ 0.25  & 0.0616 & 0.0002 & 0.0025 & 0.0026 \\
 0.25$\;\;$& -- &$\;\;$ 0.5  & 0.0646 & 0.0002 & 0.0027 & 0.0027 \\
 0.5$\;\;$& -- &$\;\;$ 0.75  & 0.0653 & 0.0002 & 0.0027 & 0.0027 \\
 0.75$\;\;$& -- &$\;\;$ 1  & 0.0670 & 0.0002 & 0.0027 & 0.0028 \\
 1$\;\;$& -- &$\;\;$ 1.3  & 0.0721 & 0.0002 & 0.0030 & 0.0030 \\
\hline
\end{tabular}
\caption{The values of the ratio of the differential production cross sections for 
\ksf\ mesons and charged hadrons  as a function of $Q^2$, $x$, $p_T$ and $\eta$. 
More details in caption of table~\ref{table:K0CrossSection}.
}
\label{table:RhCrossSection}
\end{table}


\end{document}